\documentclass[11pt]{article}
\renewcommand{\footnoterule}{%
  \kern -3pt
  \hrule width \textwidth height 0.5pt
  \kern 2pt
}
\usepackage{latexsym}
\usepackage{amssymb}
\usepackage{amsmath}
\usepackage{graphicx}
\usepackage{bm}
\usepackage{enumerate}
\usepackage{sectsty}
\usepackage{subfig}
\usepackage{float}
\usepackage{caption}
\usepackage{framed}
\usepackage{placeins}

\sectionfont{\normalsize}
\subsectionfont{\small}

\newcommand{\qzero}{Q_{0}}
\newcommand{\qp}{Q^{\prime}}
\newcommand{\qpp}{Q^{\prime\prime}}
\newcommand{\qzerop}{\qzero^{\prime}}
\newcommand{\qzeropp}{\qzero^{\prime\prime}}
\newcommand{\qzeroppp}{\qzero^{\prime\prime\prime}}
\newcommand{\qzeropppp}{\qzero^{\prime\prime\prime\prime}}
\newcommand{\qzeroppppp}{\qzero^{\prime\prime\prime\prime\prime}}

\newcommand{\phizero}{\Phi_{0}}
\newcommand{\phip}{\Phi^{\prime}}
\newcommand{\phipp}{\Phi^{\prime\prime}}
\newcommand{\phizerop}{\Phi_{0}^{\prime}}
\newcommand{\phizeropp}{\Phi_{0}^{\prime\prime}}
\newcommand{\phizeroppp}{\Phi_{0}^{\prime\prime\prime}}
\newcommand{\gxx}{1+\left(Q^{\prime}\right)^{2}}
\newcommand{\dalembertian}{\raise-1mm\hbox{\Large$\Box$}} 
\newcommand{\rhot}{\tilde{\rho}}
\newcommand{\pr}{\tilde{p}_{r}}
\newcommand{\pt}{\tilde{p}_{t}}
\newcommand{\ppar}{p_{r}}

\newcommand{\pperp}{p_{t}}
\newcommand{\gtt}{g_{tt}(x)}
\newcommand{\romansubs}{\renewcommand{\theequation}{\theparentequation \roman{equation}}}
\newcommand{\HRule}{\rule{\linewidth}{0.5mm}}

\begin{document}
\title{{\small \bf{ON WORMHOLE THROATS IN ${\bm{f(R)}}$ GRAVITY THEORY}}}
\author {{\small Andrew DeBenedictis\hspace{-0.004cm}\footnote{Department of Physics,\qquad\qquad\qquad\qquad\qquad\qquad\qquad\quad $^\dagger$Department of Physics, \newline $\mbox{\quad\;\; and}$ \hspace{7.6cm} Faculty of Electrical Engineering \newline $\mbox{\quad\;\; The Pacific Institute for the Mathematical Sciences,}$ \hspace{0.852cm}and Computing, \newline $\mbox{\quad\;\; Simon Fraser University,}$\hspace{4.75cm} University of Zagreb, \newline $\mbox{\quad\;\; Burnaby, British Columbia, Canada, V5A 1S6}$ \hspace{1.65cm}Unska 3, HR-10 000 Zagreb, Croatia \newline $\mbox{\quad\;\; adebened@sfu.ca}$\hspace{6.0cm}dubravko.horvat@fer.hr}}
\hspace{0.3cm}{\small Dubravko Horvat{$^\dagger$}}\\
\HRule }
\date{{\small May 5, 2012}}
\maketitle

\begin{abstract}
\noindent We study the existence and properties of wormhole throats in modified $f(R)$ gravity theory. Specifically, we concentrate on the cases where the lapse is not necessarily constant, and hence are not limited to the zero tidal force scenarios. In the class of theories whose actions are generated by Lagrangians of the form $f(R)=\sum \alpha_{n}\,R^{n}$ we find parameters which allow for the existence of energy condition respecting throats, which do not exist in Einstein gravity. We also consider the effect of the modified action on the anisotropy of the models, and find that modified gravity can minimize the amount of anisotropy required to support the existence of a throat. {In both these respects, the sector containing theories with positive $n$ is more promising than the negative $n$ sector in comparison to Einstein gravity alone, with large $n$ being most favorable}.
\end{abstract}
\rule{\linewidth}{0.2mm}
\vspace{-1mm}

\section{{\small INTRODUCTION}}
General relativity is arguably the most successful theory of gravity to date. The theory, aside from being pleasing on grounds such as diffeomorphism invariance, has passed all major tests in the solar system. On scales larger than the solar system, general relativity, supplemented with non-standard matter fields such as dark matter or dark energy, explains the dynamics of the large-scale structure of the universe rather well. 

However, there has long been an interest in extended theories of gravitation, which possess general relativity in some limit (for example, see \cite{ref:faraonibook}, \cite{ref:caplau}.) These extensions are mainly motivated by the desire to eliminate the potentially ``exotic'' non-standard matter fields mentioned \cite{ref:noeg}, \cite{ref:noeg2}, or to mimic some low energy quantum gravity effects which are thought to manifest at very high curvatures \cite{ref:birrellanddavies}, \cite{ref:qeg}. Other motivations are perhaps more academic; we have only been able to study gravity effectively in its weak limit, and hence cannot be certain that Einstein gravity holds in stronger gravitational fields. Therefore it is of interest to study gravitational theories which are diffeomorphism invariant and give Einstein gravity in an appropriate limit, but deviate from Einstein gravity in some way outside of the realm where gravitational effects have commonly been observed. This, admittedly, gives us much freedom regarding the types of theories that may be successful candidates.

Of particular interest are the class of theories known as $f(R)$ gravity. These theories consider gravitational fields generated by actions of the form \cite{ref:buchfr}:
\begin{equation}
 S=\frac{1}{2\kappa}\int\,\sqrt{-g}\,f(R)\,d^{4}x + \int\,\sqrt{-g}\, \mathcal{L}_{\mbox{\tiny{mat}}}(g)\,d^{4}x \,, \label{eq:lag}
\end{equation}
where $f(R)$ is some function of the Ricci scalar, $R$ and $\mathcal{L}_{\mbox{\tiny{mat}}}(g)$ is the matter Lagrangian density. The equations of motion generated via varying this action with respect to the metric, and demanding that this variation vanish. These equations of motion are: 
\begin{equation}
 \tfrac{\partial f(R)}{\partial R} R^{\mu}_{\;\,\nu} -\frac{1}{2} f(R)\,\delta^{\mu}_{\;\,\nu} -\nabla^{\mu}\nabla_{\nu} \tfrac{\partial f(R)}{\partial R} +\delta^{\mu}_{\;\,\nu}\, \dalembertian \tfrac{\partial f(R)}{\partial R} = \kappa\,T^{\mu}_{\;\,\nu}\,, \label{eq:eom}
\end{equation}
where as usual $T^{\mu}_{\;\nu}$ is the stress-energy tensor of the matter field which comes from varying the matter action.

A number of earlier generalizations to Einstein gravity fall within the paradigm of $f(R)$ theories. For example, theories with $f(R)=R+\alpha_{2}R^{2}$ have long been studied (called ``$R$-squared'' gravity. See, for example, \cite{ref:rsq1}-\cite{ref:rsqlast}). The Starobinsky inflationary theory \cite{ref:starob} may be the most popular application of $R$-squared gravity theory. In the opposite regime, where curvature is low, there has been less interest in modifying Einstein gravity. However, it is worth mentioning that $f(R)$ theory with \emph{inverse powers} of the Ricci scalar has been considered as a possible candidate to explain the observed accelerated expansion of the universe \cite{ref:o3}-\cite{ref:o6}. Inverse Ricci terms may also appear in certain sectors of string/M
theory (see \cite{ref:odin}, \cite{ref:gzbr} for example). The vacuum state for inverse-$R$ theories is not Minkowski space-time, but is instead either deSitter or anti-deSitter space-time. Admittedly, these inverse theories are highly constrained from stability considerations or solar system tests but may not be completely ruled out. (See \cite{ref:dolgov}-\cite{ref:faraonisolar} and references therein for a discussion of the restrictions.) For generality we also include this sector in our study. A thorough study of solutions in $R^{n}$ cosmology may be found in \cite{ref:nodint2} as well as \cite{ref:cosdyn} and a complete discussion of $f(R)$ modifications to aid acceleration may be found in \cite{ref:lobo}. Outside of cosmology, and perhaps more relevant here, studies have been performed that have more bearing to stellar physics \cite{ref:frstars1}-\cite{ref:frstars9}. In another popular extension, namely that of $f(T)$ gravity where $T$ is the torsion scalar, such solutions have also been studied \cite{ref:s1}, \cite{ref:s2}.

Given that integer powers of $R$ have arguably generated the most interest in $f(R)$ extensions, we propose here to study the properties of wormhole throats in gravity theories given by
\begin{equation}
f(R)={\underset{{\hspace{-0.05cm}\vspace{-0.1cm}n}}{\sum}} \alpha_{n}\,R^{n}\,, \label{eq:fofr}
\end{equation}
with the $\alpha_{n}$ constants. This is quite general as any $f(R)$ analytic in $R$ may be expanded as such a power series (in the positive $n$ sector) such as exponential gravity \cite{ref:o5}. For completion, we allow for both positive and negative powers of $R$ in the action (hence extending to $f(R)$ expandable in negative powers as well). Of course, this series includes the $n=1$ term (Einstein term) and can include $n=0$ (cosmological constant).

The wormhole is of interest as it may provide a simple model for the space-time foam thought to be manifest at very high energies, where the corrections to Einstein gravity due to higher curvature terms may be important in encompassing some quantum gravity effects. As well, exhaustive studies have been performed on wormhole geometries in Einstein gravity and their properties (mainly in the static cases with spherical and axial symmetry) are now well known in Einstein gravity. For example, it is an interesting property that static wormhole throats necessarily must violate the weak/null energy conditions in Einstein gravity \cite{ref:WEC1}, \cite{ref:hochvis} and, furthermore, must be anisotropic. It has been of much interest in wormhole physics to find either somewhat realistic matter models which can meet these slightly exotic conditions in a way required to support a wormhole throat, or else to consider alternative gravitational theories which may allow for energy condition respecting matter. In the latter vein, due to the complexity of the resulting equations, the zero tidal force class of wormholes is most often studied ($g_{tt}=\mbox{const.}$). Even then, analytic models are hard to come by and one often resorts to numerics. The zero tidal force models are interesting due to their tractability, which allows one to study important properties of these geometries. However, they may not be very realistic from a physics perspective though, as they yield a constant frequency shift in inhomogeneous structures possessing a preferred center and, in the weak-field limit, a constant Newtonian potential.

In this work we study wormhole \emph{throats} and \emph{without} the restriction of zero tidal force, although we do consider the zero tidal force cases as well for completion. The consideration of the throat is in many ways more general than considering the asymptotics, as the most salient features of a wormhole occur in the throat region. For example, in Einstein gravity, the necessary violation of energy conditions in static spherically-symmetric wormholes occurs in the neighborhood of the throat, regardless of asymptotics. In fact, in the Einstein gravity scenarios, if the matter field falls off sufficiently fast, or is patched to a vacuum or other solution, the properties far from the throat do not generally differ greatly from similar systems with trivial topology. As well, it has been argued (for example in \cite{ref:hochvis}) that the global topology is too limited a tool to study wormholes, and a local geometric analysis near the throat is generally more useful in discerning interesting properties of wormholes. In such cases, the throat does not necessarily coincide with the common definition of a wormhole, which relies on global properties, and is viewed as an interesting object in its own right, capable of describing more general scenarios than just wormholes. We therefore now concentrate on the near throat region and study the properties in this vicinity. Admittedly, in some cases, demanding flatness at infinity would place extra restrictions on the properties of the wormhole \cite{ref:GS}, \cite{ref:bronnfr} but as the asymptotics of the universe are not exactly known, and we are interested in the existence of throats only, we will not consider such restrictions here. 

In section 2 we briefly review some background material and present a non-standard coordinate gauge that is more suited for wormhole analysis than the standard spherical coordinate chart. We discuss the mathematical construction of the throat in this chart. This is followed by a study, which is analytic when possible but otherwise numerical, of various scenarios depending on the terms present in the gravitational action. {Of special importance, due to their physical relevance, are the non-zero tidal force models, which we examine in some detail.} We particularly concentrate on the properties of energy conditions as well as the degree of anisotropy. {In general, we find that the parameter space of Einstein gravity supplemented with terms where $n$ is positive is more favorable in both these respects than Einstein gravity alone for a large range of parameters. Supplements with negative $n$ have either a very small parameter space where they improve the energy conditions or else tend to worsen the situation when compared to Einstein gravity alone.} Finally, in section 3, we conclude the study.

\section{{\small THE MODELS}}
As is common in wormhole studies in Einstein gravity, we will utilize here an anisotropic fluid source whose stress-energy tensor is given by:
\begin{equation}
T^{\mu}_{\;\,\nu}=(\rho + \pperp)u^{\mu}u_{\nu} + \pperp\,\delta^{\mu}_{\;\,\nu} + (\ppar - \pperp)s^{\mu}s_{\nu}\,. \label{eq:anisoT}
\end{equation}
Here $\rho$, $\pperp$ and $\ppar$ are the energy density, the perpendicular (to the inhomogeneous direction) pressure, and the parallel pressure respectively as measured in the fluid element's rest frame. The vector $u^{\mu}$ is the fluid 4-velocity and $s^{\mu}$ is a space-like vector orthogonal to $u^{\mu}$. These vectors satisfy:
\begin{equation}
 u^{\mu}u_{\mu}=-1,\;\;\; s^{\mu}s_{\mu}=+1,\;\;\; u^{\mu}s_{\mu}=0\,. \label{eq:fluidvecs}
\end{equation}
Since we are interested in the properties of the actual material generating the gravitational field, we \emph{do not} transform the system into an effective scalar-tensor system but instead keep the geometry and the material properties explicit.

We impose spherical symmetry and hence may write the metric as
\begin{align}
ds^{2}=&-e^{\Phi(r)}\,dt^{2} + e^{\Xi(r)}\,dr^{2} + r^{2}\,d\theta^{2} + r^{2}\sin^{2}(\theta)\,d\varphi^{2}\,, \label{eq:spherechart} \\[0.2cm]
=& -e^{\Phi(r)}\,dt^{2} + \left\{1+ \left[\partial_{r} P(r)\right]^{2} \right\}\,dr^{2} + r^{2}\,d\theta^{2} + r^{2}\sin^{2}(\theta)\,d\varphi^{2}\,, \nonumber
\end{align}
where the function $P(r)$ describes the profile of the three geometry for fixed $\theta$ and $\varphi$ (see below for details). In this chart the Ricci scalar is given by
\begin{align}
 R=&\frac{1}{2r^{2}\left[1+\left(P^{\prime}\right)^{2}\right]^{2}}\Bigl[ 4(P^{\prime})^{4} -4r(P^{\prime})^{2}\Phi^{\prime} -4r\Phi^{\prime} +8rP^{\prime}P^{\prime\prime}+2r^{2}P^{\prime}P^{\prime\prime}\Phi^{\prime} \nonumber \\
&\qquad\qquad\qquad\quad +4(P^{\prime})^{2} -2r^{2}(P^{\prime})^{2}\Phi^{\prime\prime} -r^{2}(P^{\prime})^{2}(\Phi^{\prime})^{2} -2r^{2}\Phi^{\prime\prime} -r^{2}(\Phi^{\prime})^{2}\Bigr]\,, \label{eq:Rscal1}
\end{align}
where the prime denotes differentiation with respect to $r$. Although the above is the most common form of metric system to locally describe spherically symmetric gravitational fields, this coordinate system is not ideal for studies of wormhole throats and hence we use a different chart, which we now present.

\subsection{{\footnotesize WORMHOLES}}
\begin{figure}[!ht]
\begin{framed}
\begin{center}
\vspace{0.0cm}
%\fbox{
\includegraphics[bb=0 0 1205 485, scale=0.325, clip, keepaspectratio=true]
{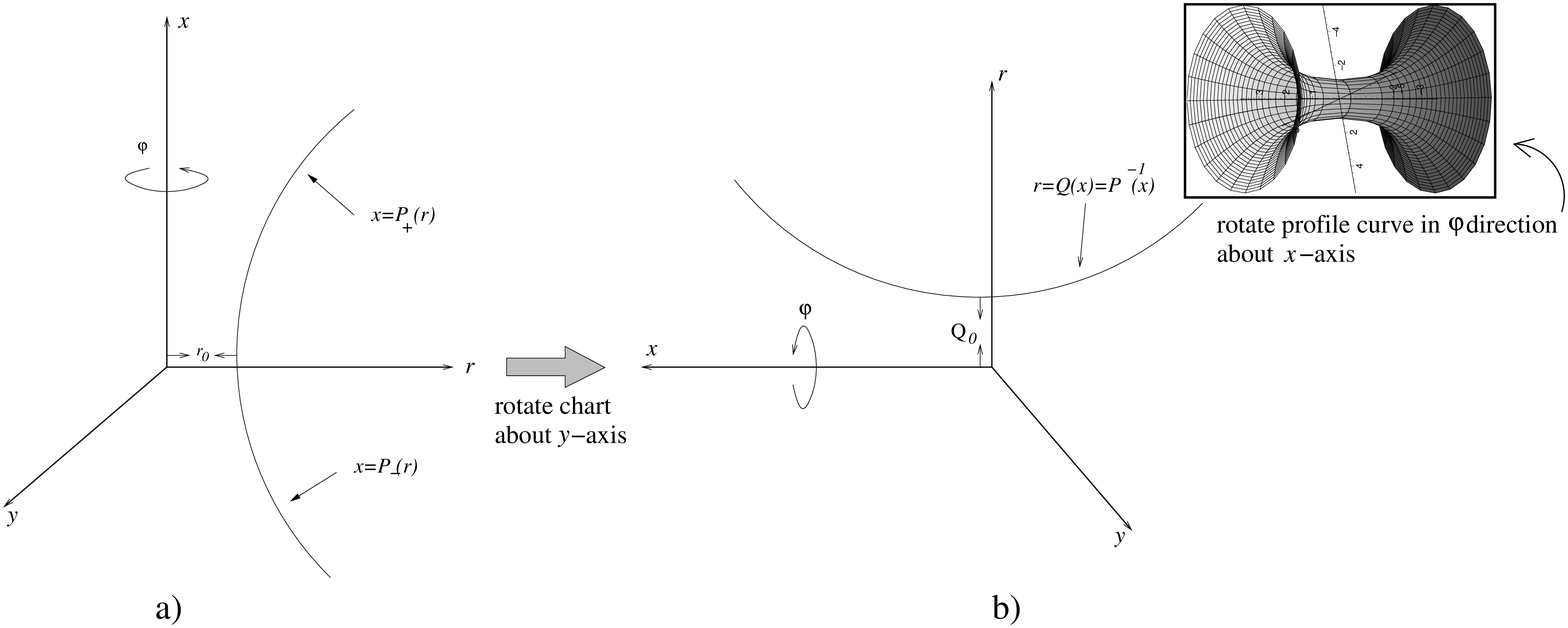}
%}
\end{center}\vspace{-0.1cm}
\caption{\small{a) In the standard spherical chart the wormhole throat is generated by the surface of revolution created by rotating the profile curve $x=P(r)=P_{-}(r) \bigcup P_{+}(r)$ but this leads to a metric singularity at the throat ($r=r_{0}$) where $\partial_{r}P(r) \rightarrow \infty$. A new chart is created in b) via rotating the coordinates by $\pi/2$, and the surface of revolution (inset) is generated by rotating the curve $r=Q(x)=P^{-1}(x)$. This does not have the metric singularity at the throat ($x=0$). Also, a single chart can now cover both regions.}}%
\label{fig:rotatechart}%
\end{framed}
\end{figure}

We consider here static throats, which must be anisotropic in Einstein gravity, in the class of modified gravities given by $f(R)={\underset{{\hspace{-0.05cm}\vspace{-0.1cm}n}}{\sum}} \alpha_{n}\,R^{n}$ for both positive and negative $n$. As mentioned previously, it is common in the literature on wormholes to consider zero tidal force models via the condition $g_{tt}(r)=\mbox{constant}$. However here we relax this restriction and study more general scenarios. As well, although the spherical coordinate chart of (\ref{eq:spherechart}) can be utilized for wormhole throats if care is taken, it is not optimal, as $g_{rr}(r) \rightarrow \infty$ as $r \rightarrow r_{0}$ (see Figure \ref{fig:rotatechart} and equation (\ref{eq:spherechart})), and the most interesting properties of wormholes are arguably found near the throat region. Therefore, we use a different chart for the study of wormholes which essentially involves tilting the standard spherical chart by $\pi/2$. The wormhole throat is then given via the creation of a surface of revolution of the profile curve, $r=Q(x)$, as shown in Figure~\ref{fig:rotatechart} (also see figure caption). In this new chart, the space-time metric's line element may be written as
\begin{equation}
 ds^{2}=-e^{\Phi(x)}\, dt^{2} + \left\{1+ \left[\partial_{x} Q(x)\right]^{2} \right\}\, dx^{2} + Q^{2}(x)\, d\theta^{2} + Q^{2}(x)\sin^{2}(\theta)\, d\varphi^{2}\,. \label{eq:newchartline}
\end{equation}
Note that in this new chart the metric is analytic at the throat since $\partial_{x} Q(x) \rightarrow 0$, and hence $g_{xx}(x) \rightarrow 1$ as one approaches the throat ($x=0$)\footnote{As an aside, in this coordinate system the exterior Schwarzschild metric is given by $Q(x)=2M+\frac{x^{2}}{8M}$. The horizon is located at $x=0$.}. Also, only one chart is required now to cover the wormhole, as opposed to two charts in the standard spherical coordinates. The function $Q(x)$ must possess the following properties:
\begin{enumerate}[i)]
\item $\qzero:=Q(0) > 0$,\vspace{-0.2cm}
\item $\qzerop:=Q^{\prime}(x)_{|x=0}=0$,\vspace{-0.2cm}
\item $Q^{\prime\prime}(x) > 0$ in some neighborhood of the throat\footnote{More precisely, if $Q$'s first non-zero derivative (higher than first order) at $x=0$ is of even order, the function attains a local minimum if this derivative is positive, and hence we have a wormhole throat. If its first non-zero derivative is of odd order, it is a point of inflection and therefore does not describe a wormhole throat.}.
\end{enumerate}
Aside from the above properties we make the mild assumption that $Q(x)$ is analytic in some nonzero domain about $x=0$.

In the above coordinate chart, the Ricci scalar is given by
\begin{align}
 R=&\frac{1}{2Q^{2}\left[\gxx\right]^{2}}\Bigl[4-4Q(Q^{\prime})^{3}\Phi^{\prime} + 4(Q^{\prime})^{2} -8QQ^{\prime\prime}-4QQ^{\prime}\Phi^{\prime} +2Q^{2}Q^{\prime}Q^{\prime\prime}\Phi^{\prime} \nonumber\\
&\qquad\qquad\qquad\quad\quad -2Q^{2}\Phi^{\prime\prime} -Q^{2}(\Phi^{\prime})^{2}-2Q^{2}(Q^{\prime})^{2}\Phi^{\prime\prime} -Q^{2}(Q^{\prime})^{2}(\Phi^{\prime})^{2} \Bigr]\,, \label{eq:Rscal2}
\end{align}
where here and subsequently the prime denotes differentiation with respect to $x$. We do not expect any serious pathologies (such as infinite tidal forces) in our analysis as, in the domain of validity of the coordinate chart (which is larger than the ``$r$'' chart, and covers the throat) the metric and its derivatives are well behaved and hence we expect the orthonormal Riemann tensor components to also be well behaved. We show later that the function $Q(x)$ describing the spatial geometry, is $C^{\infty}$ and that $\Phi(x)$ is also very smooth.

One issue which is of great interest in wormhole physics is the study of energy condition violation of the matter supporting the wormhole. It is known that in Einstein gravity energy conditions \emph{must} be violated by static wormholes somewhere in the vicinity of the throat \cite{ref:WEC1}, \cite{ref:hochvis}. It is possible that in more complicated gravitational theories, one may circumvent this issue and possess a throat region which respects energy conditions. An analysis in this vein was performed in \cite{ref:furdeb} where the analysis was limited to $n=-1$ and $n=2$ extensions only. A very complete analysis may be found in \cite{ref:lobooliv}, \cite{ref:olivthesis}. In the above cited studies, however, due to the complications involved, the zero tidal force assumption was made. This simplification eliminates some of the complications but still allows one to glean a number of the interesting properties that $f(R)$ wormholes possess. However, constant redshift, and constant weak-field Newtonian potential, from various regions of an inhomogeneous gravitating object may not be particularly realistic and we wish to remove this assumption and allow for a non constant $\Phi(x)$ function. This can alter the physical properties of the throat significantly, as we shall see below. Aside from yielding more complicated equations, a non-constant $\Phi(x)$ now presents us with the problem of how to prescribe this function. For this we appeal to some physical considerations. It is more physical to demand some realistic properties on the matter fields than it is to blindly prescribe $\Phi(x)$, although the former is much more difficult than the latter, as one needs to solve a complicated differential equation for $\Phi(x)$. With this in consideration, what we initially do is prescribe some reasonable energy density profiles, $\rho(x)$, for the matter field (eg. large positive energy density with zero slope at the throat and monotonically decreasing outward towards the surface or infinity, using parameters where this is possible) and numerically solve for $\Phi(x)$ using the numerical code COLSYS \cite{ref:colsys}. This will then give us an idea of what a realistic $\Phi(x)$ function should look like, and we use similar functions for subsequent studies. (The spatial geometry must also be prescribed, and this poses no problem as we apriori know that, spatially, we are dealing with a spherically symmetric wormhole.) We simply use this method to provide an idea of what the functions $\Phi(x)$ should look like in realistic scenarios and then use similar functions for our studies. It should be noted that the function $\Phi(x)$ controls the presence of event horizons. We avoid horizons by ensuring that $\Phi(x)$ does not approach $-\infty$ anywhere in the domain of study.

In the following analysis we wish to discern how the various terms in the action affect the wormhole solution. In the case of $f(R)={\sum} \alpha_{n}\,R^{n}$, each term in the sum contributes independently to the left-hand side of the field equations, and hence also contribute separately in determining the properties of $T^{\mu}_{\;\nu}$. Therefore, it is sufficient to study individual terms in the power series of $f(R)$ separately to see how they each contribute individually to the matter field. Note that in this approach, $\alpha_{n}$ is simply an overall multiplicative constant, and hence we set it equal to one. One may see the effect of changing $\alpha_{n}$ by simply rescaling the results below by whatever value of $\alpha_{n}$ one chooses (with the exception of the anisotropy studies, for which we study the effects of varying $\alpha_{n}$). Negative values of $\alpha_{n}$ also reflect the graphs about the $z=0$ plane.

\subsubsection{{\normalsize $n=1$}}
This case, of course, corresponds to Einstein gravity. It is worthwhile presenting this to summarize the results of Einstein gravity in this coordinate chart, and for comparison with other powers of $n$ later. Regardless of the powers present in the full action, it is expected to have an $n=1$ term. We summarize the $n=1$ results as follows:
\begin{subequations}\romansubs
{\allowdisplaybreaks\begin{align}
\rhot:=&\kappa\,\rho=-\kappa\,T^{0}_{\;\,0}=\frac{1+(Q^{\prime})^{2}-2QQ^{\prime\prime}}{Q^{2}\left[\gxx\right]^{2}}\,, \label{eq:einstrho}\\[0.2cm]
\pr:=&\kappa\,p_{r}= \kappa\,T^{1}_{\;\,1}=\frac{\Phi^{\prime}QQ^{\prime}-1}{Q^{2}\left[\gxx\right]}\,, \label{eq:einstpr}\\[0.2cm]
\pt:=&\kappa\,p_{t}= \kappa\,T^{2}_{\;\,2}=\frac{1}{4Q\left[\gxx\right]^{2}} \left[ 4Q^{\prime\prime} +2 Q^{\prime}\Phi^{\prime} +2(Q^{\prime})^{3}\Phi^{\prime}  +Q(Q^{\prime})^{2}(\Phi^{\prime})^{2} \right. \nonumber \\[0.1cm]
&\qquad\qquad\qquad \left.-2QQ^{\prime}Q^{\prime\prime}\Phi^{\prime} +2Q(Q^{\prime})^{2}\Phi^{\prime\prime} +Q(\Phi^{\prime})^{2} + 2 Q \Phi^{\prime\prime}\right]\,. \label{eq:einstpt}
 \end{align}}
\end{subequations}

There are also the following {combinations} which occur in the energy conditions:

\begin{subequations}\romansubs
{\allowdisplaybreaks\begin{align}
\rhot+\pr=& \frac{Q^{\prime}\Phi^{\prime}+(Q^{\prime})^{3}\Phi^{\prime}-2Q^{\prime\prime}}{Q\left[\gxx\right]^{2}}\,, \label{eq:einstecond1} \\[0.2cm]
\rhot+\pt=&\frac{1}{4Q^{2}\left[\gxx\right]^{2}} \left[4+4(\qp)^{2} -4Q\qpp +2 Q\qp\phip +2 Q (\qp)^{3}\phip \right. \nonumber \\[0.1cm]
&\left.+Q^{2}(\qp)^{2}(\phip)^{2} - 2 Q^{2}\qp \qpp \phip +2Q^{2}(\qp)^{2}\phipp +Q^{2}(\phip)^{2} +2Q^{2}\phipp \right]\,.\label{eq:einstecond2}
 \end{align}}
\end{subequations}

The above expressions, though not overly complicated, do not shed much insight into the behavior of the matter fields near the throat, where we are most interested in. We therefore expand the relevant expressions in Taylor series about the throat ($x=0$):

\begin{subequations}\romansubs
{\allowdisplaybreaks\begin{align}
&\rhot=  \frac{1}{\qzero}\left[1-2 \qzero \qzero^{\prime\prime}\right] -2 \frac{\qzero^{\prime\prime\prime}}{\qzero} x + \frac{1}{\qzero^{3}} \left[4\qzero^{2} (\qzero^{\prime\prime})^{3} - \qzero^{\prime\prime} -\qzero^{2}\qzero^{\prime\prime\prime\prime}\right] x^{2} + \mathcal{O}(x^{3})\,, \label{eq:einstnearthroatrho} \\[0.2cm]
&\rhot+\pr=  -2\frac{\qzero^{\prime\prime}}{\qzero} + \frac{1}{\qzero} \left[\qzero^{\prime\prime} \phizero^{\prime} -2 \qzero^{\prime\prime\prime}\right]x + \frac{1}{\qzero^{2}} \Bigl[4\qzero(\qzero^{\prime\prime})^{3} -\qzero \qzero^{\prime\prime\prime\prime} + \frac{1}{2} \qzero \qzero^{\prime\prime\prime}\phizero^{\prime} \nonumber \\[0.1cm]
&\qquad\qquad\quad  +\qzero\qzero^{\prime\prime}\phizero^{\prime\prime} + (\qzero^{\prime\prime})^{2}\Bigr] x^{2} +\mathcal{O}(x^{3})\,, \label{eq:einstnearthroate1} \\[0.2cm]
&\rhot+\pt= \frac{1}{4\qzero^{2}}\left[4+\qzero^{2}(\phizerop)^{2}-4\qzero\qzeropp+2\qzero^{2}\phizeropp\right] +\frac{1}{\qzero}\left[\qzero\phizerop\phizeropp -\qzero(\qzeropp)^{2}\phizerop\right. \nonumber \\[0.1cm]
&\qquad\qquad\quad + \left.\qzero\phizeroppp -2\qzeroppp +\qzeropp\phizerop \right] x +\mathcal{O}(x^{2})\,.  \label{eq:einstnearthroate2}
 \end{align}}
\end{subequations}
where the zero subscript indicates that the quantity is evaluated at $x=0$.

From the (\ref{eq:einstnearthroatrho}), it can be seen that the energy density in the throat region may be made positive. Regarding (\ref{eq:einstnearthroate1}), if $\qzero^{\prime\prime}$ is non-zero it must be positive (as the throat is a local minimum) and (\ref{eq:einstnearthroate1}) must therefore be negative near the throat. If $\qzero^{\prime\prime}$ is zero, then the condition for a local minimum implies that $\qzero^{\prime\prime\prime}$ is also zero, and hence the lowest order term which would contribute near the throat is the fourth derivative term, $\qzeropppp$. Since this fourth derivative must then be positive under the condition of a minimum, and it appears with a negative sign in (\ref{eq:einstnearthroate1}), this contribution is negative. In such a scenario energy conditions are (barely) met at the throat, but are violated as one moves away from the throat. Similar arguments apply to higher derivatives in case the fourth derivative vanishes at the throat (although one needs to study higher order terms in the expansion, which for brevity we did not write). It is in this way that energy conditions must be violated in the vicinity of a static wormhole throat in Einstein gravity.

For comparison with some of the modified gravity results to follow, which, due to the complicated expressions they yield, must be studied via computational methods, we choose a spatial geometry governed by
\begin{equation}
 Q(x)=A\cosh\left(\frac{x}{x_{o}}\right)\,, \label{eq:Qcosh}
\end{equation}
as this function possesses all the salient properties to describe a throat. The parameter $A$ represents the radius of the throat and $x_{0}$ represents the degree of ``flare-out'' of the wormhole. Both these parameters are strongly related to the degree of energy condition violation, and hence we pay particular attention to these quantities. For all cases studied in this work, when an explicit form of $Q(x)$ is required, we use the form in (\ref{eq:Qcosh}).

For the function $\Phi(x)$ we appeal to physical considerations (with the aid of the numerical code COLSYS, as mentioned previously). For positive $n$ we expect $\Phi(x)$ to slowly asymptote to a constant value far away from the throat, where the geometry is expected to approach Minkowski space-time. For negative $n$ (and for generality, for all $n$) the asymptotic value should approach the deSitter or anti-deSitter value.  As we are interested in the near-throat region only it is not crucial that the asymptote is manifest in the domain of consideration, but an indication of such an asymptote is desirable. We consider both scenarios where $g_{tt}=-e^{\Phi(x)}$ smoothly increases towards the asymptote, as well as scenarios where $g_{tt}$ smoothly decreases towards the asymptote. (One cannot necessarily rule out one scenario over the other in extended gravity theories.) Specifically, we choose fitting functions of the form
\begin{subequations}
\begin{align}
g_{tt}(x)=&\mathsf{A}_{0}\frac{\mathsf{B_{0}}+x^2}{1+x^2}-\mathsf{C}_{0} & \mbox{for concave-down $g_{tt}$\,,}  \label{eq:gttfuna}\\
g_{tt}(x)=& -\frac{\mathsf{A}_{0} + \mathsf{B}_{0}x^{2}}{\mathsf{C}_{0}+x^{2}} & \mbox{for concave-up $g_{tt}$\,,}\label{eq:gttfunb}
\end{align}
\end{subequations}
Where $\mathsf{A}_{0}$, $\mathsf{B}_{0}$ and $\mathsf{C}_{0}$ are fitting constants to fit generally to the COLSYS numerical results.
Since the wormhole profile chosen in (\ref{eq:Qcosh}) is symmetric about the throat, we also make similar symmetry demands on $g_{tt}$. There is no great loss of generality in doing this since if one wishes to model a non-symmetric throat, the solution presented can be viewed as being valid only on one side of the throat, and a different solution can be patched to the other side. The forms of (\ref{eq:gttfuna},b) are used throughout this manuscript in the analysis of non-zero tidal force models. 

\begin{figure}[!ht]
\begin{framed}
\begin{center}
\vspace{-0.5cm}
\begin{tabular}{cc}
\subfloat[$g_{tt}(x)$]{\includegraphics[width=45mm,height=45mm,clip]{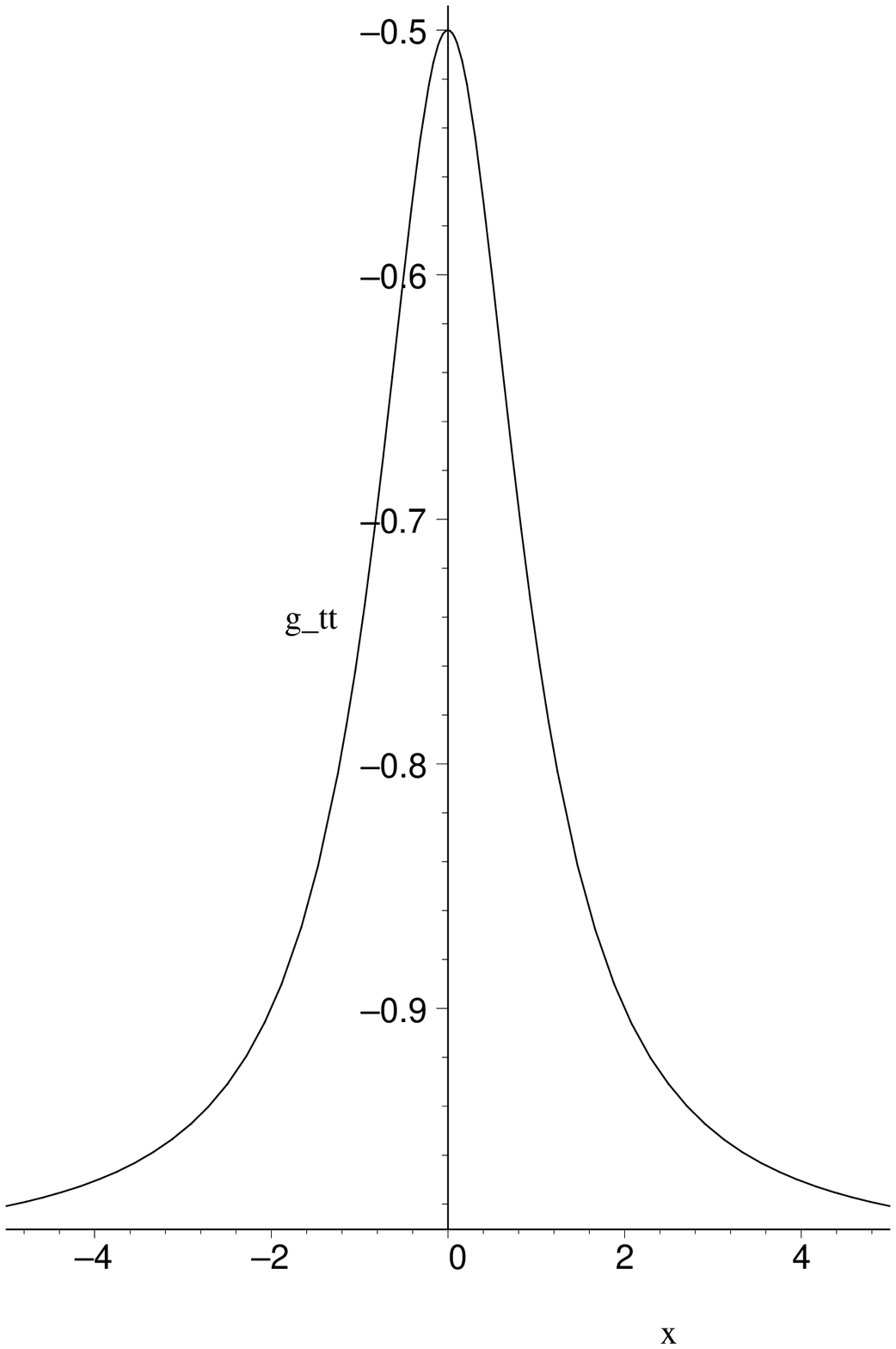}}&\hspace{0.5cm}
\subfloat[$\tilde{\rho}(x)$]{\includegraphics[width=45mm,height=45mm,clip]{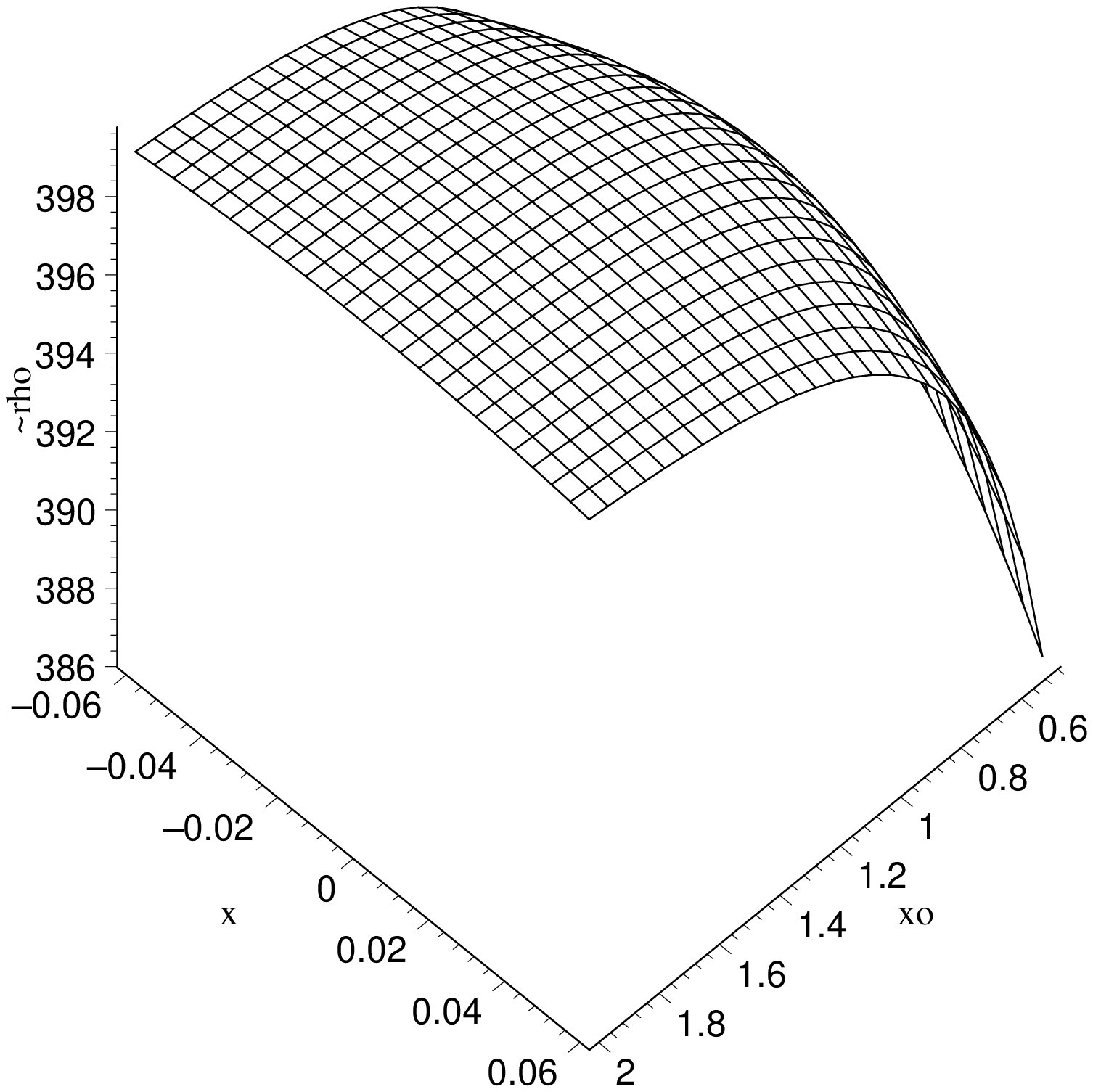}} \\
\subfloat[$\tilde{\rho}+\tilde{p}_{r}$]{\includegraphics[width=45mm,height=45mm,clip]{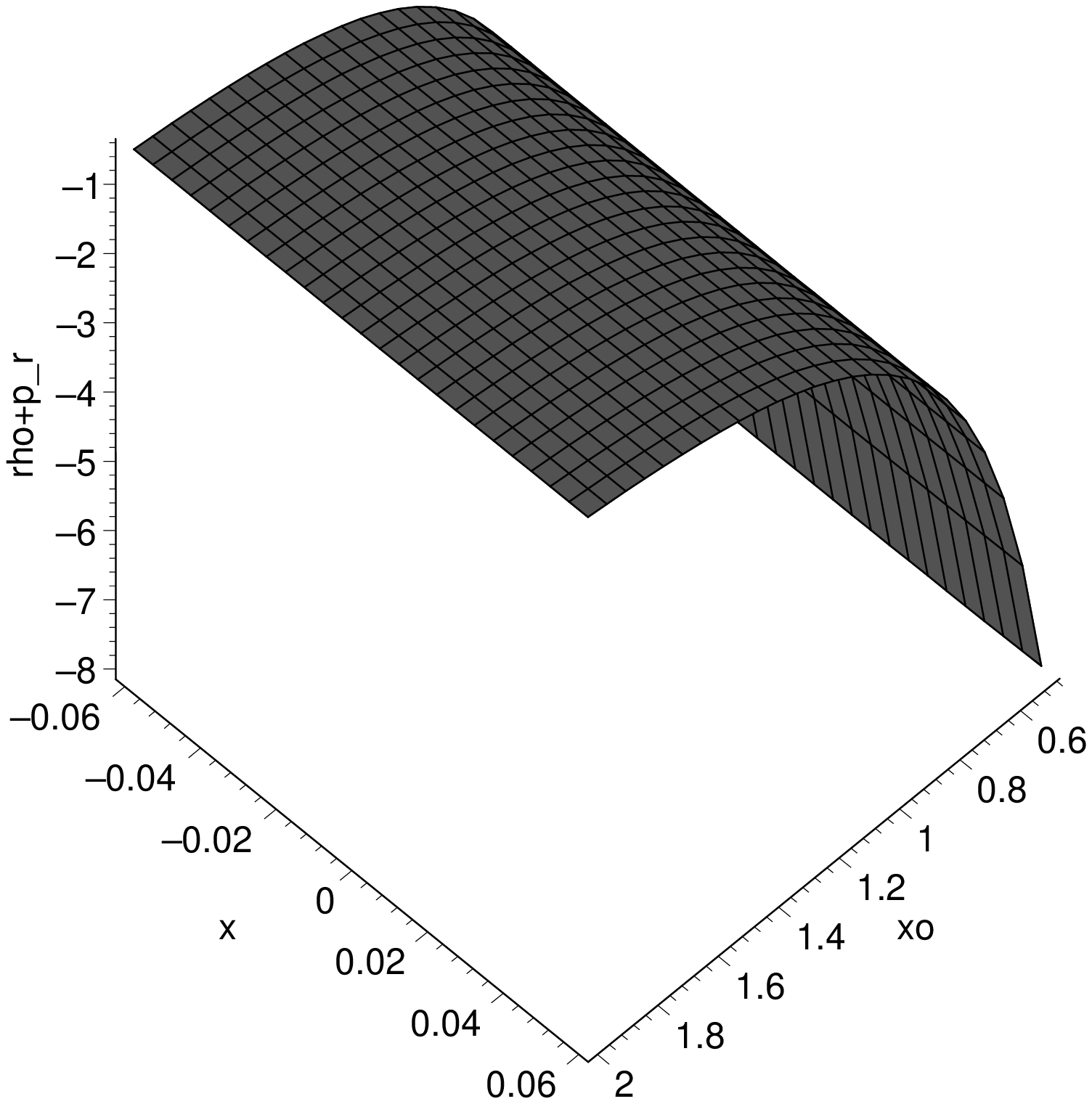}}&\hspace{0.5cm}
\subfloat[$\tilde{\rho}+\tilde{p}_{t}$]{\includegraphics[width=45mm,height=45mm,clip]{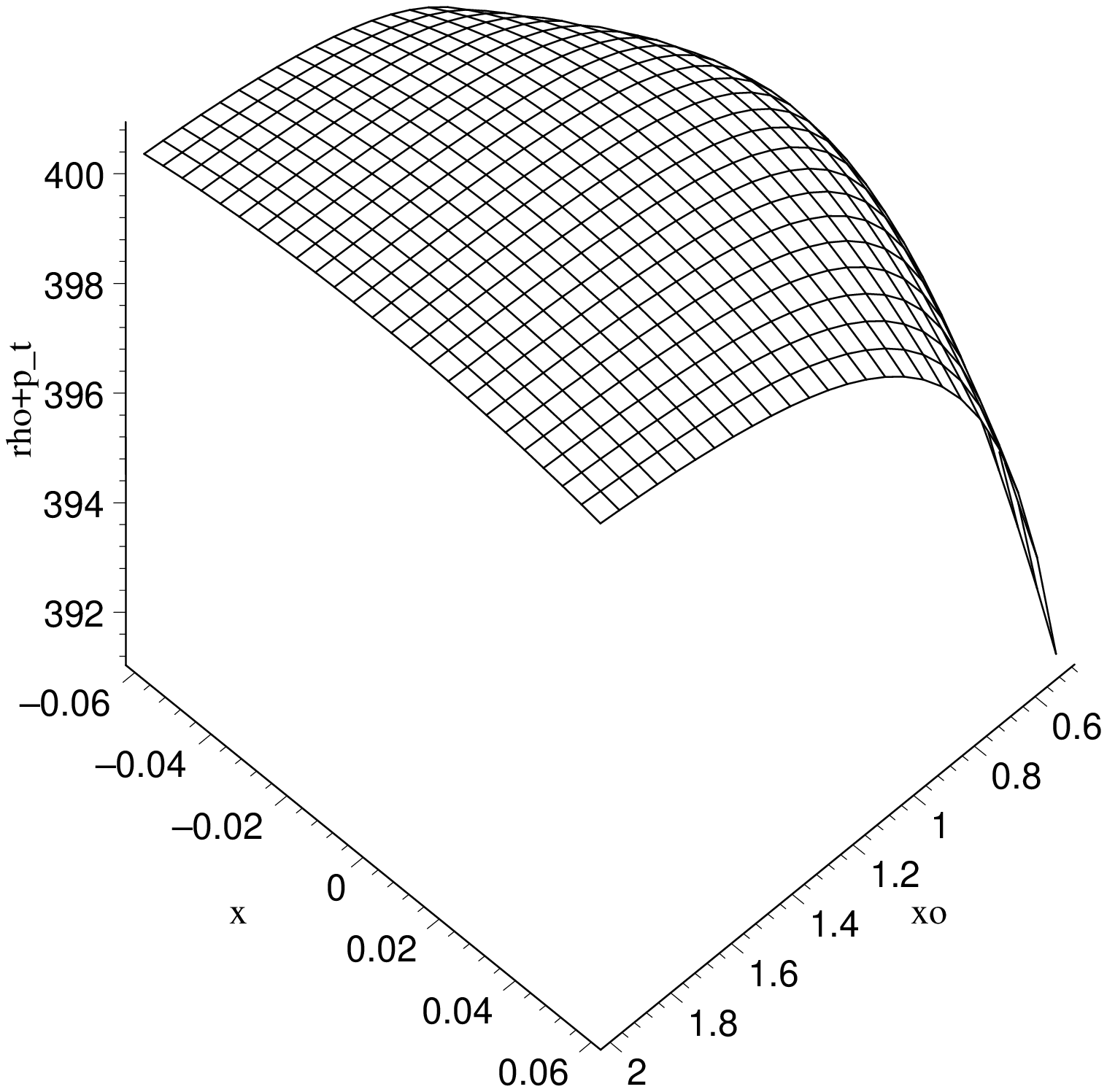}}
\end{tabular}
\end{center}
\caption{\small{Einstein gravity, $g_{tt}$ \emph{decreasing}, throat radius=0.05, varying flare-out $x_{0}$.}}
\label{fig:einstxoconcdown}
\end{framed}
\end{figure}

\begin{figure}[!ht]
\begin{framed}
\begin{center}
\vspace{-0.5cm}
\begin{tabular}{cc}
\subfloat[$g_{tt}(x)$]{\includegraphics[width=45mm,height=45mm,clip]{squared_gtt_c_down.eps}}&\hspace{0.5cm}
\subfloat[$\tilde{\rho}(x)$]{\includegraphics[width=45mm,height=45mm,clip]{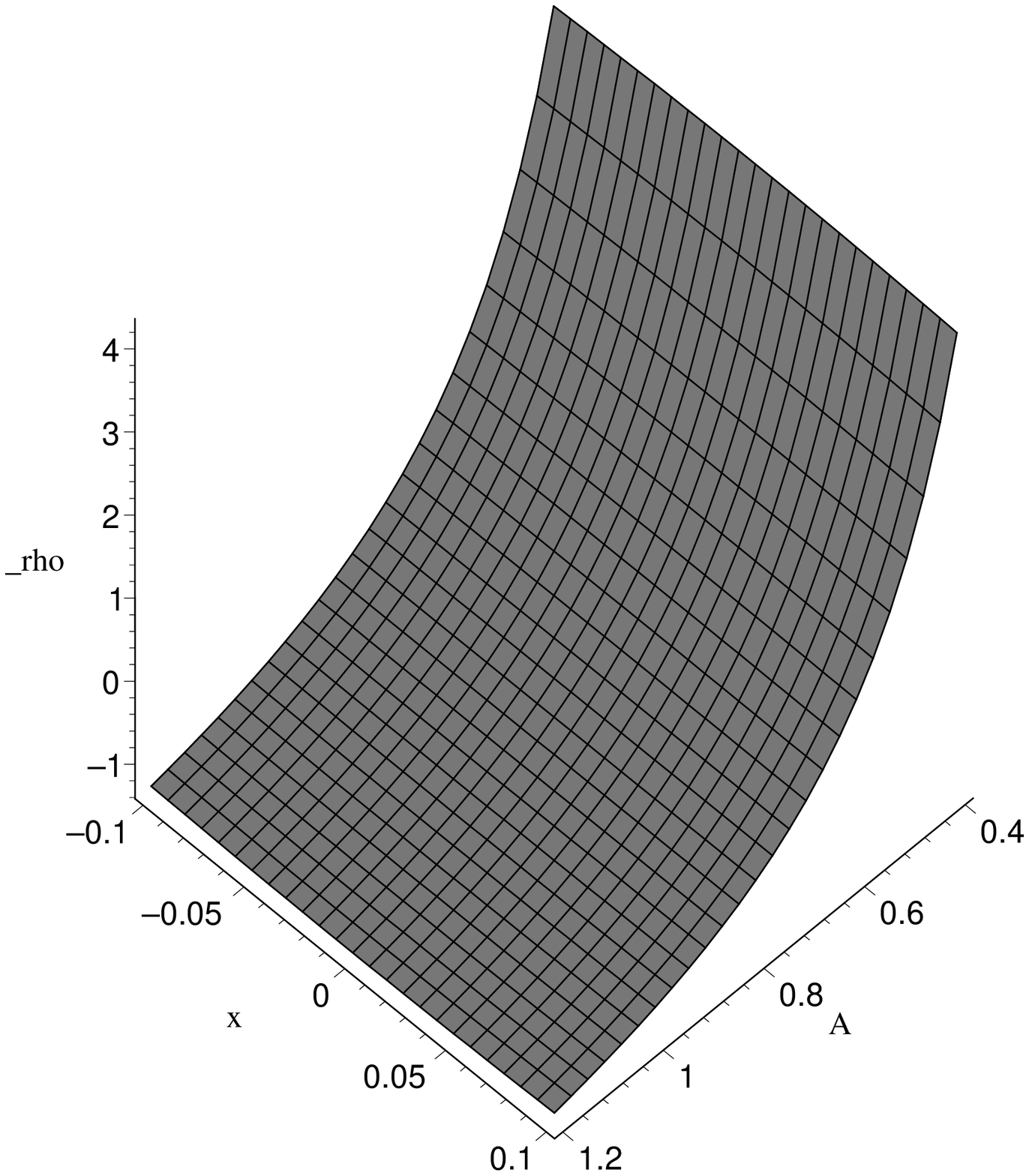}} \\
\subfloat[$\tilde{\rho}+\tilde{p}_{r}$]{\includegraphics[width=45mm,height=45mm,clip]{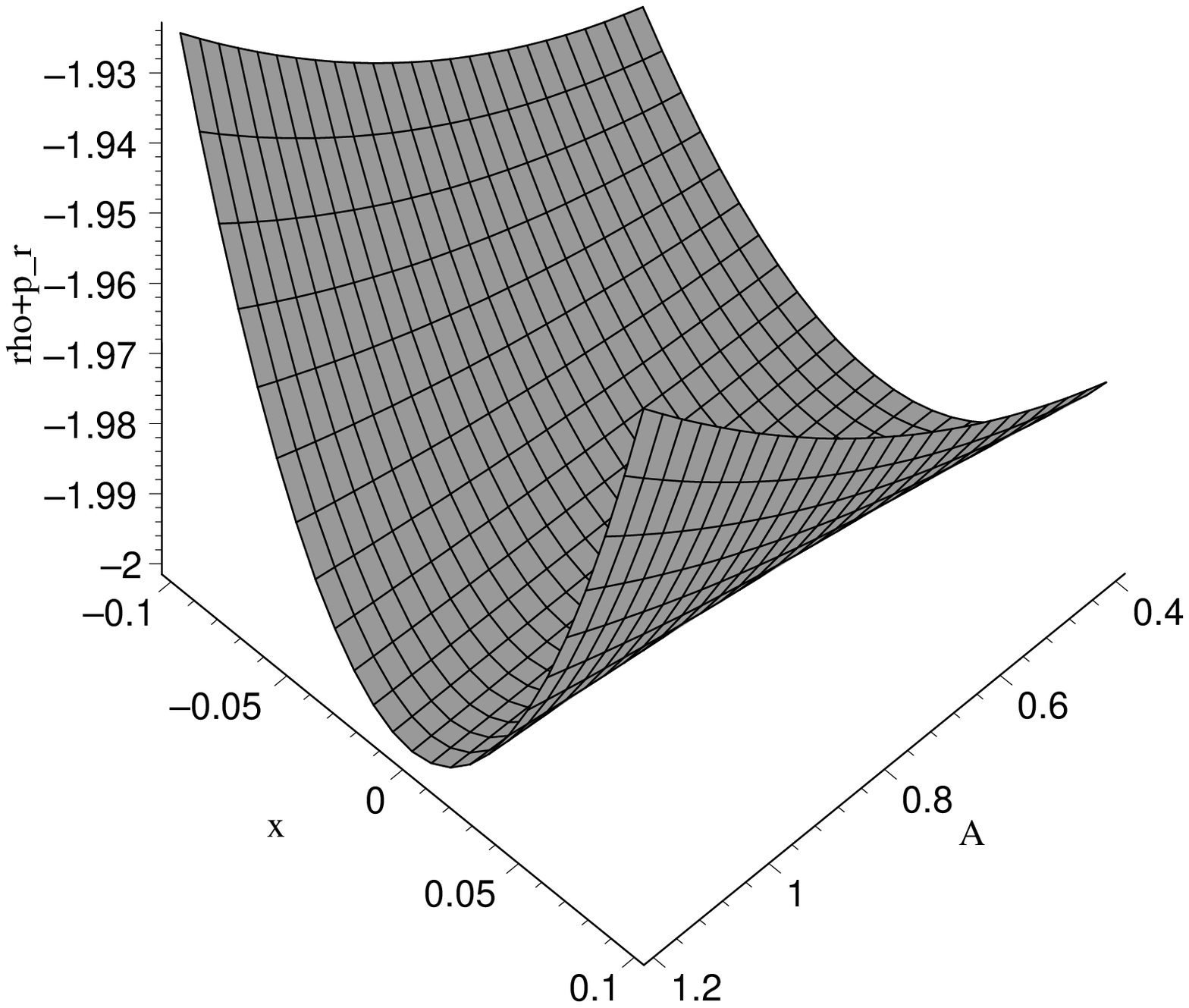}}&\hspace{0.5cm}
\subfloat[$\tilde{\rho}+\tilde{p}_{t}$]{\includegraphics[width=45mm,height=45mm,clip]{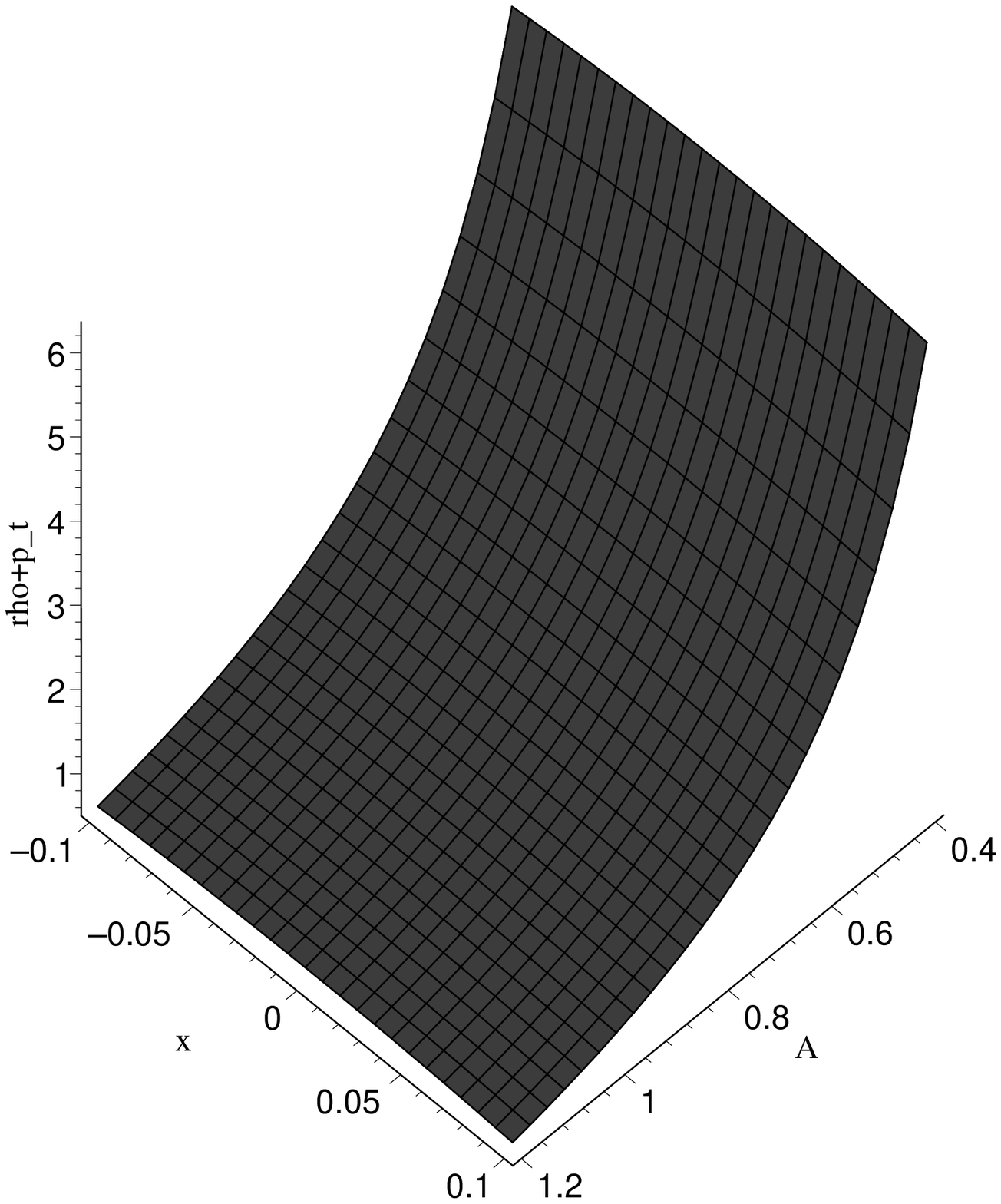}}
\end{tabular}
\end{center}
\caption{\small{Einstein gravity, $g_{tt}$ \emph{decreasing}, $x_{0}=1$, varying throat radius.}}
\label{fig:einstradiusconcdown}
\end{framed}
\end{figure}

\begin{figure}[!ht]
%%\begin{figure}[H]
\begin{framed}
\begin{center}
\vspace{0.0cm}
\includegraphics[width=60mm,height=45mm, clip]
{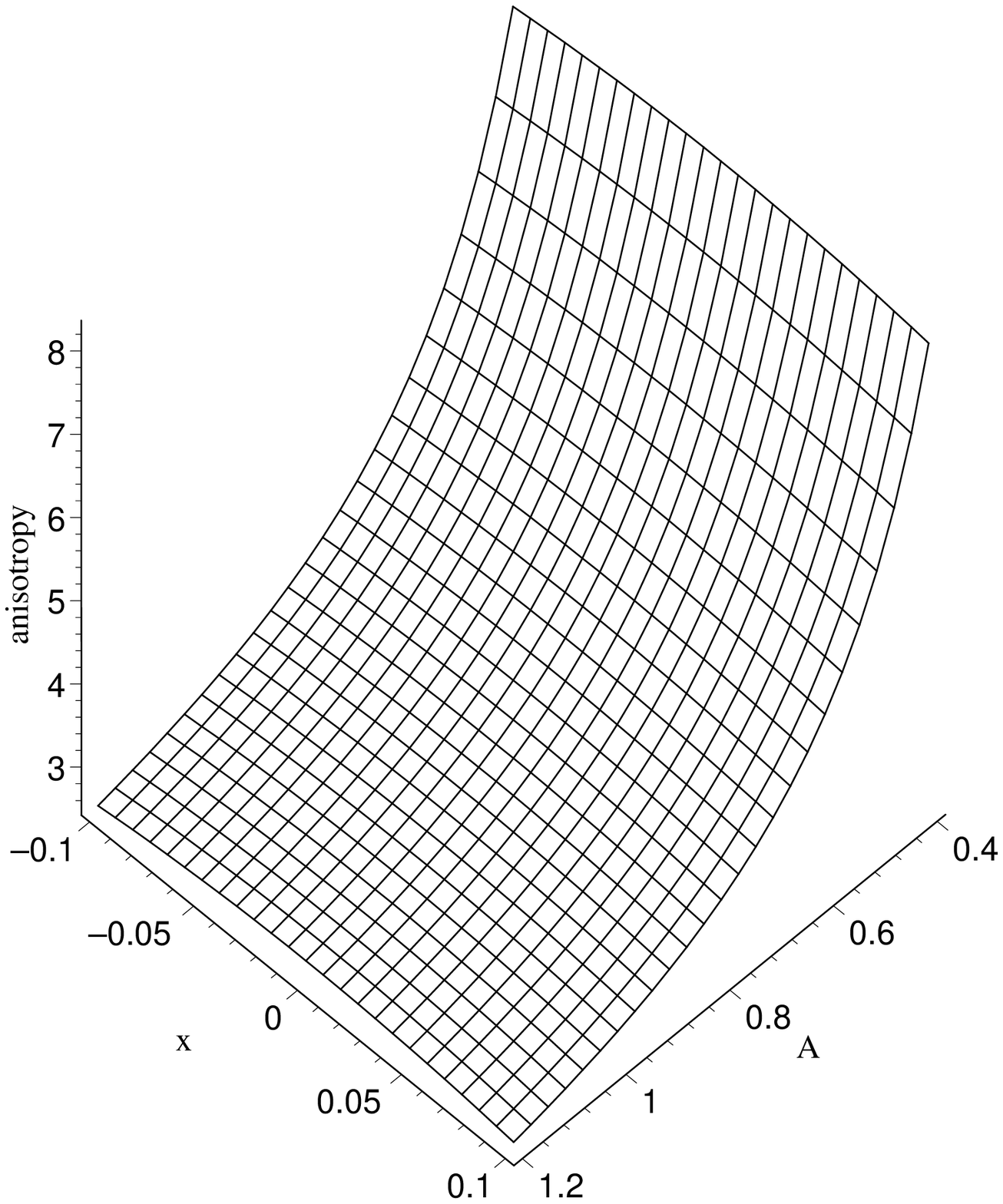}
\end{center}\vspace{-0.1cm}
\caption{\small{Anisotropy of throat region in Einstein case, $g_{tt}$ \emph{decreasing}, $x_{0}=1$, varying throat radius.}}
\label{fig:einstradiusconcdownaniso}
\end{framed}
\end{figure}

\begin{figure}[!ht]
\begin{framed}
\begin{center}
\vspace{-0.5cm}
\begin{tabular}{cc}
\subfloat[$g_{tt}(x)$]{\includegraphics[width=45mm,height=45mm,clip]{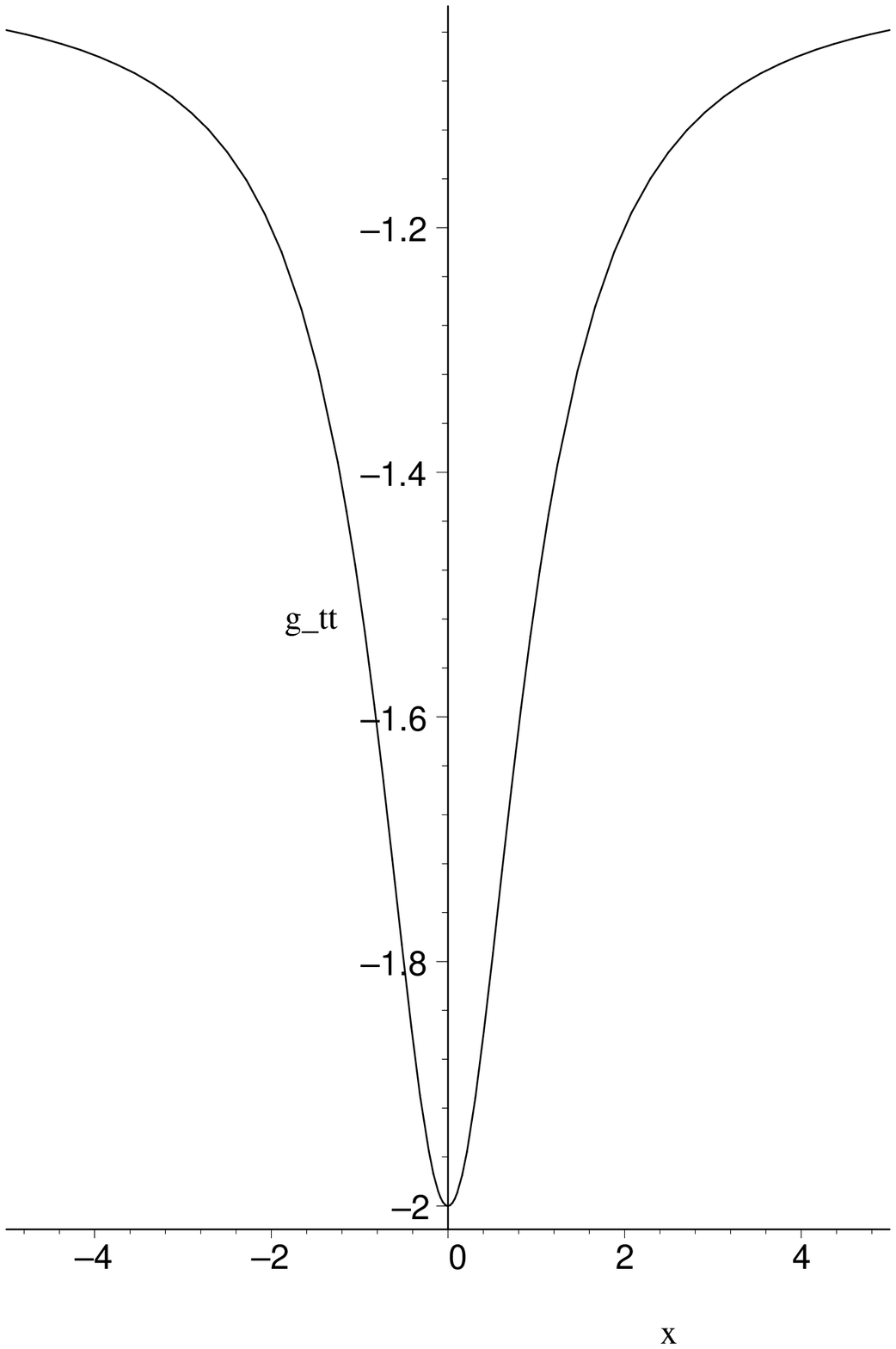}}&\hspace{0.5cm}
\subfloat[$\tilde{\rho}(x)$]{\includegraphics[width=45mm,height=45mm,clip]{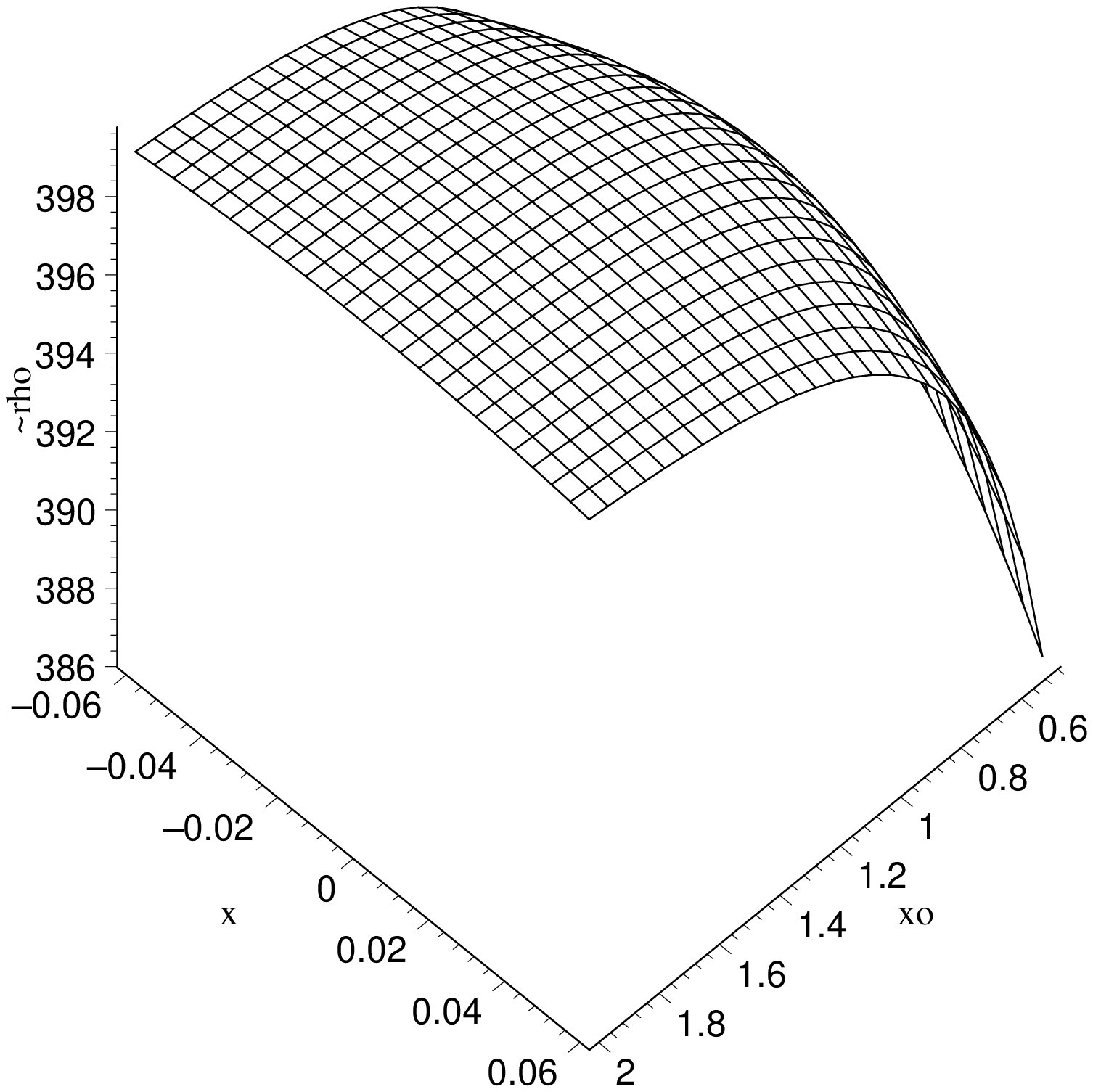}} \\
\subfloat[$\tilde{\rho}+\tilde{p}_{r}$]{\includegraphics[width=45mm,height=45mm,clip]{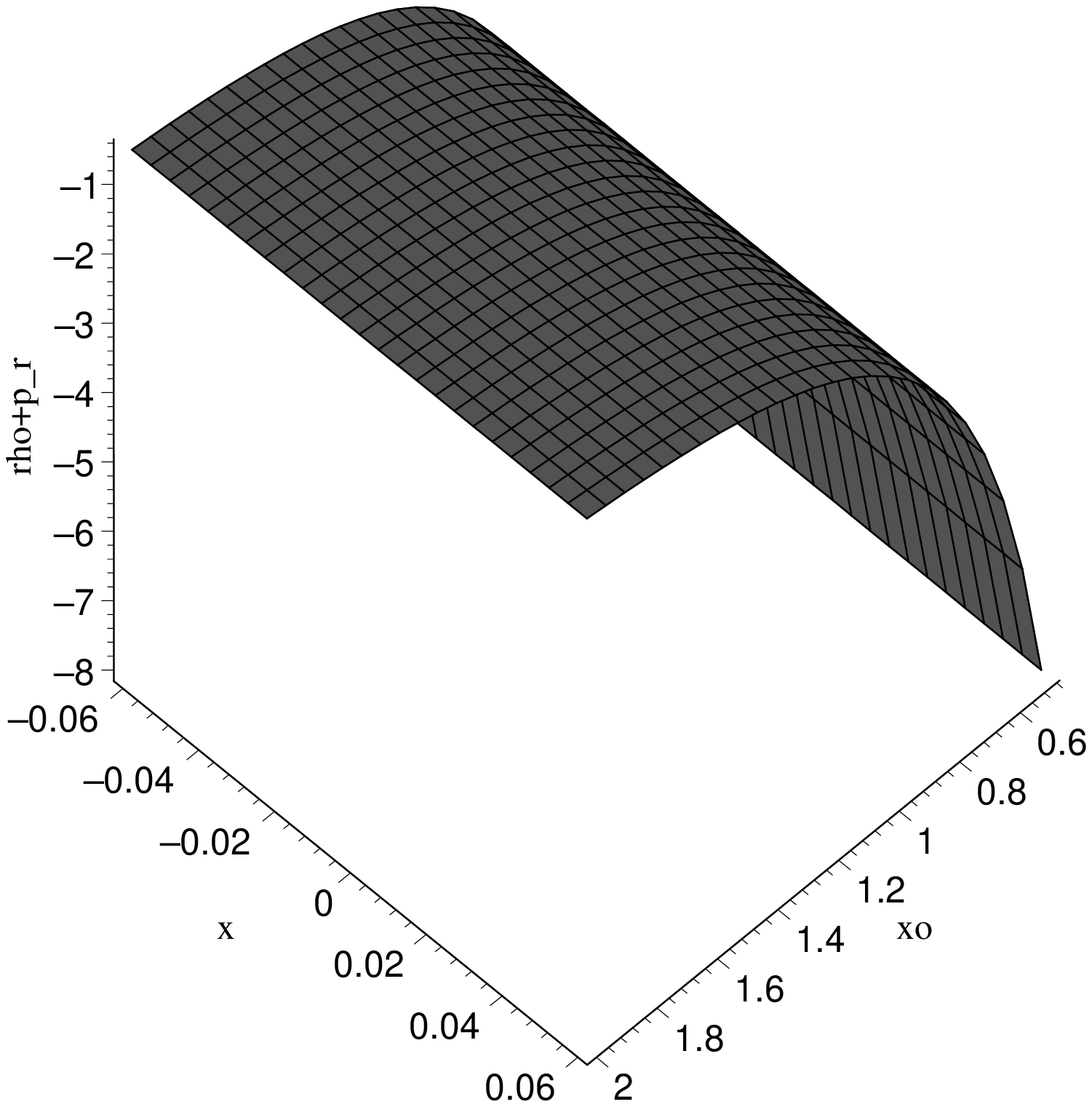}}&\hspace{0.5cm}
\subfloat[$\tilde{\rho}+\tilde{p}_{t}$]{\includegraphics[width=45mm,height=45mm,clip]{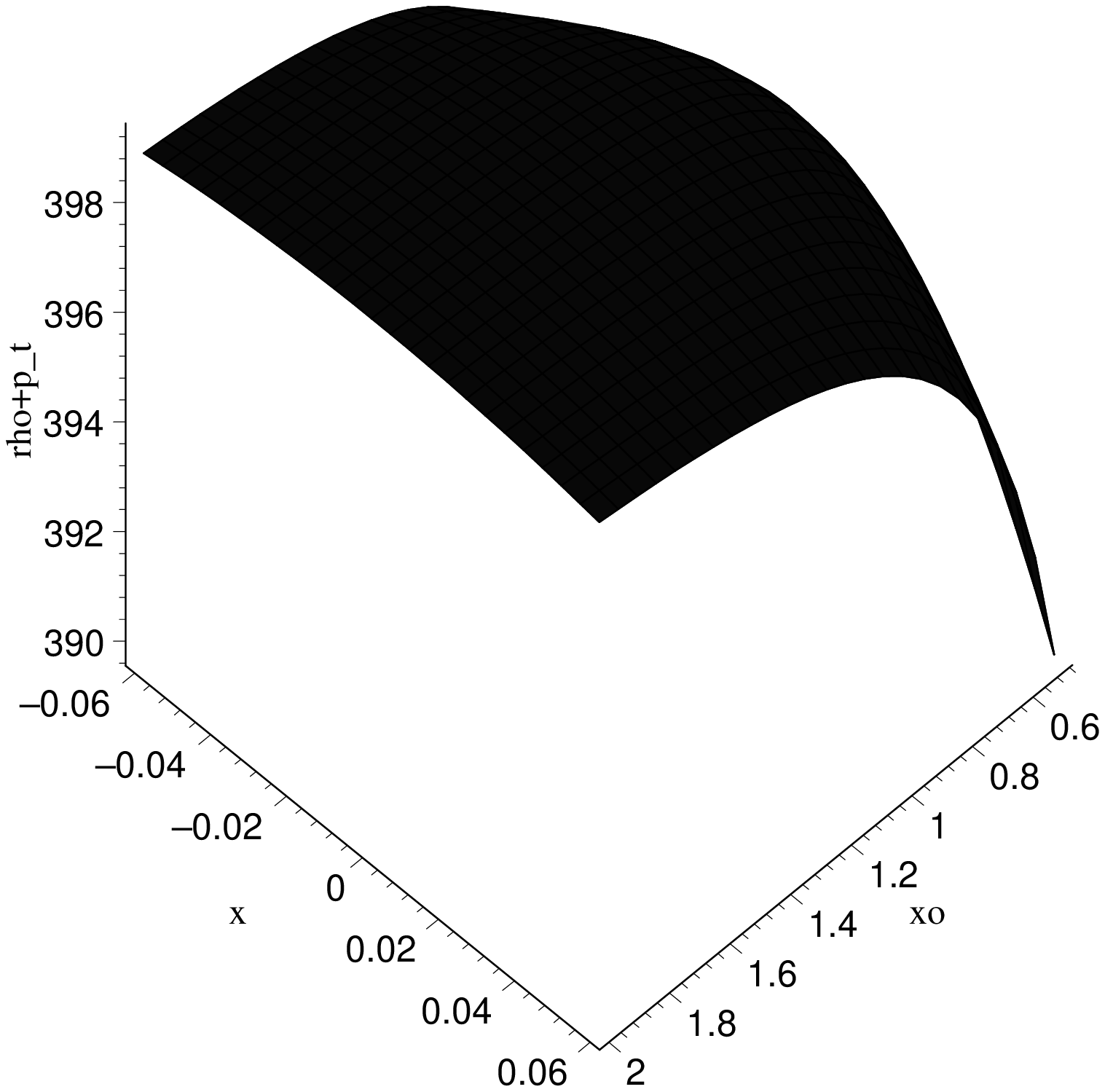}}
\end{tabular}
\end{center}
\caption{\small{Einstein gravity, $g_{tt}$ \emph{increasing}, throat radius=0.05, varying flare-out $x_{0}$.}}
\label{fig:einstxoconcup}
\end{framed}
\end{figure}

\begin{figure}[!ht]
\begin{framed}
\begin{center}
\vspace{-0.5cm}
\begin{tabular}{cc}
\subfloat[$g_{tt}(x)$]{\includegraphics[width=45mm,height=45mm,clip]{squared_gtt_c_up.eps}}&\hspace{0.5cm}
\subfloat[$\tilde{\rho}(x)$]{\includegraphics[width=45mm,height=45mm,clip]{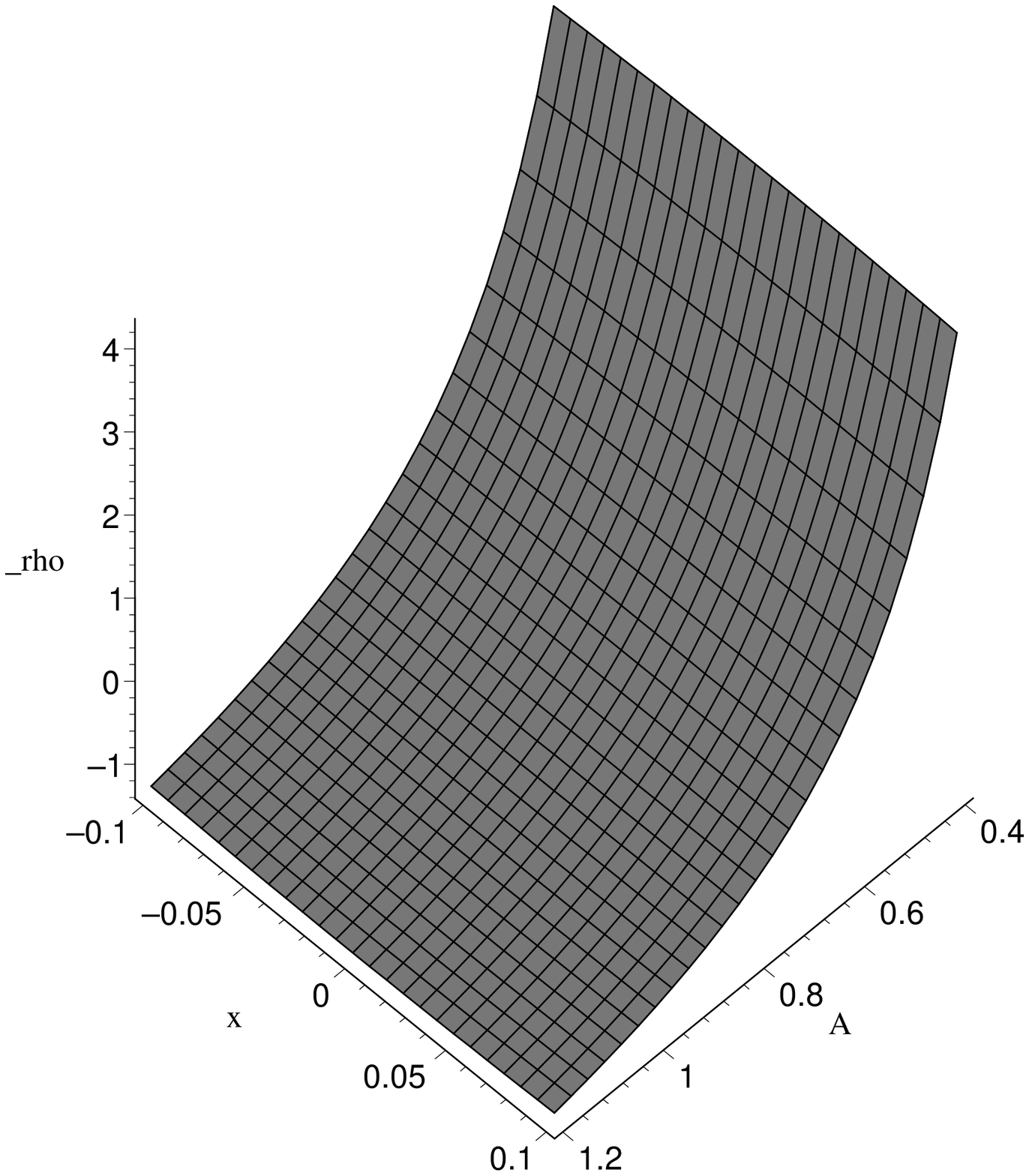}} \\
\subfloat[$\tilde{\rho}+\tilde{p}_{r}$]{\includegraphics[width=45mm,height=45mm,clip]{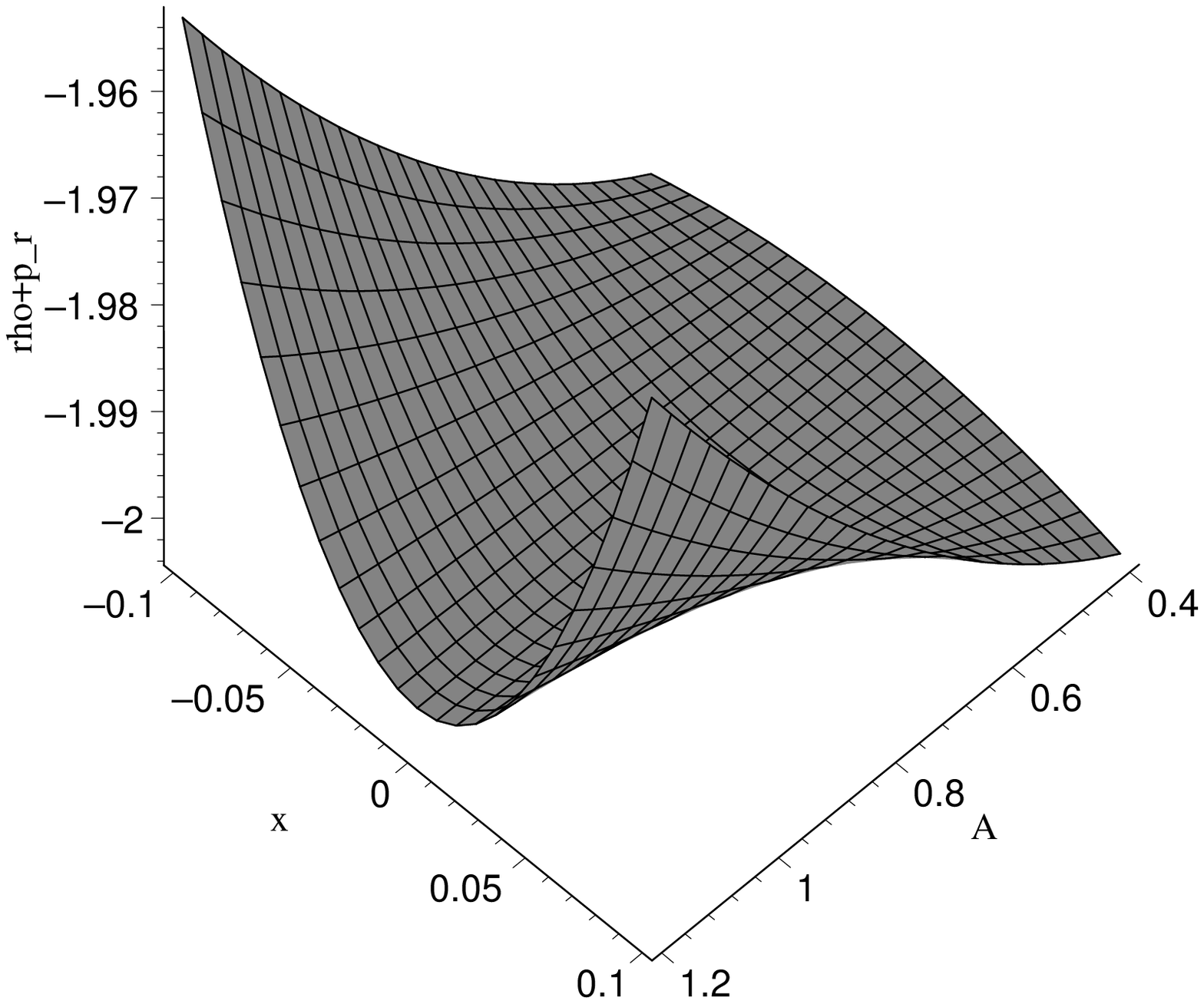}}&\hspace{0.5cm}
\subfloat[$\tilde{\rho}+\tilde{p}_{t}$]{\includegraphics[width=45mm,height=45mm,clip]{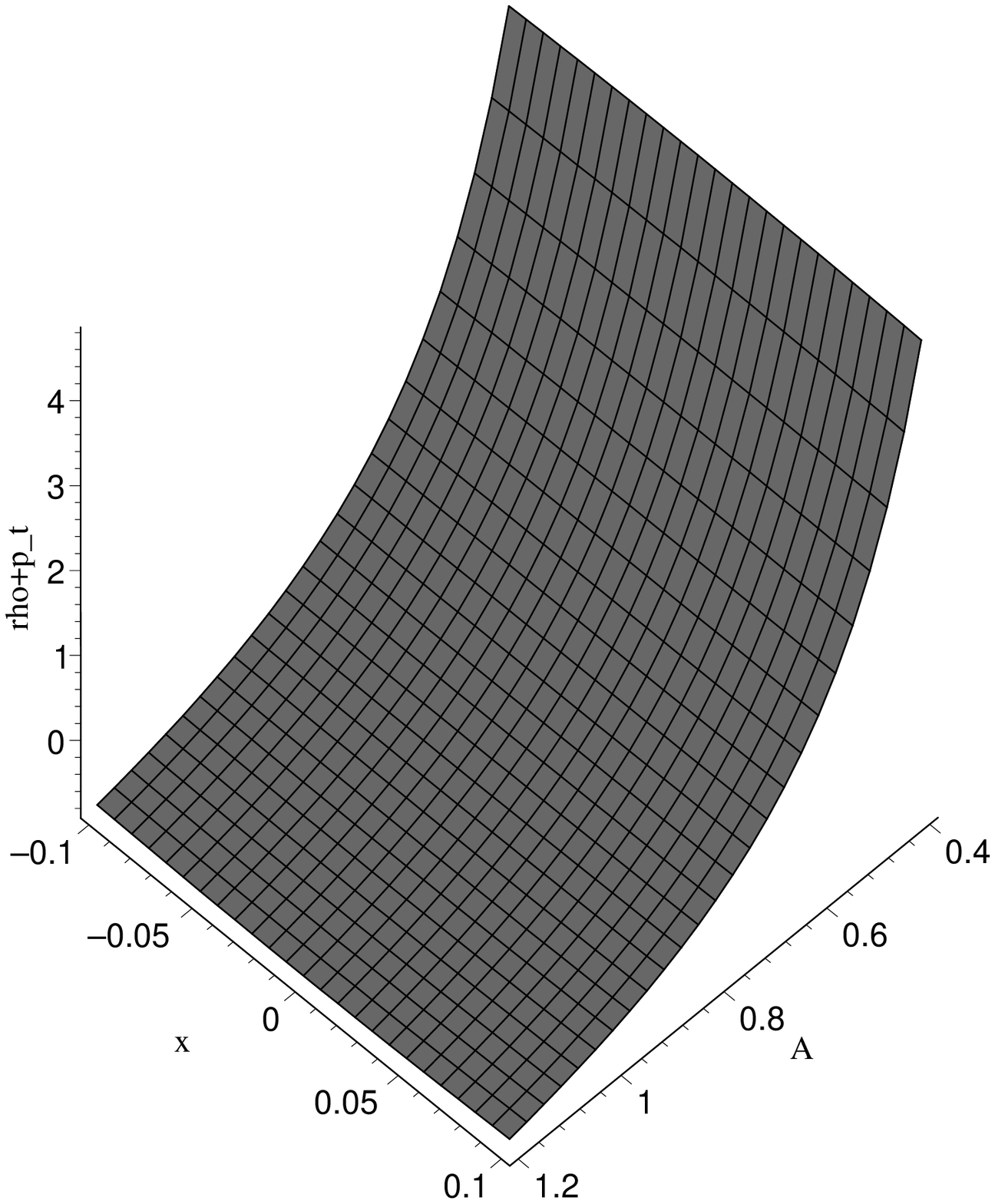}}
\end{tabular}
\end{center}
\vspace{-0.1cm}
\caption{\small{Einstein gravity, $g_{tt}$ \emph{increasing}, $x_{0}=1$, varying throat radius.}}
\label{fig:einstradiusconcup}
\end{framed}
\end{figure}

\begin{figure}[!ht]
%%\begin{figure}[H]
\begin{framed}
\begin{center}
\vspace{0.5cm}
%\fbox{
\includegraphics[width=60mm,height=45mm, clip]
{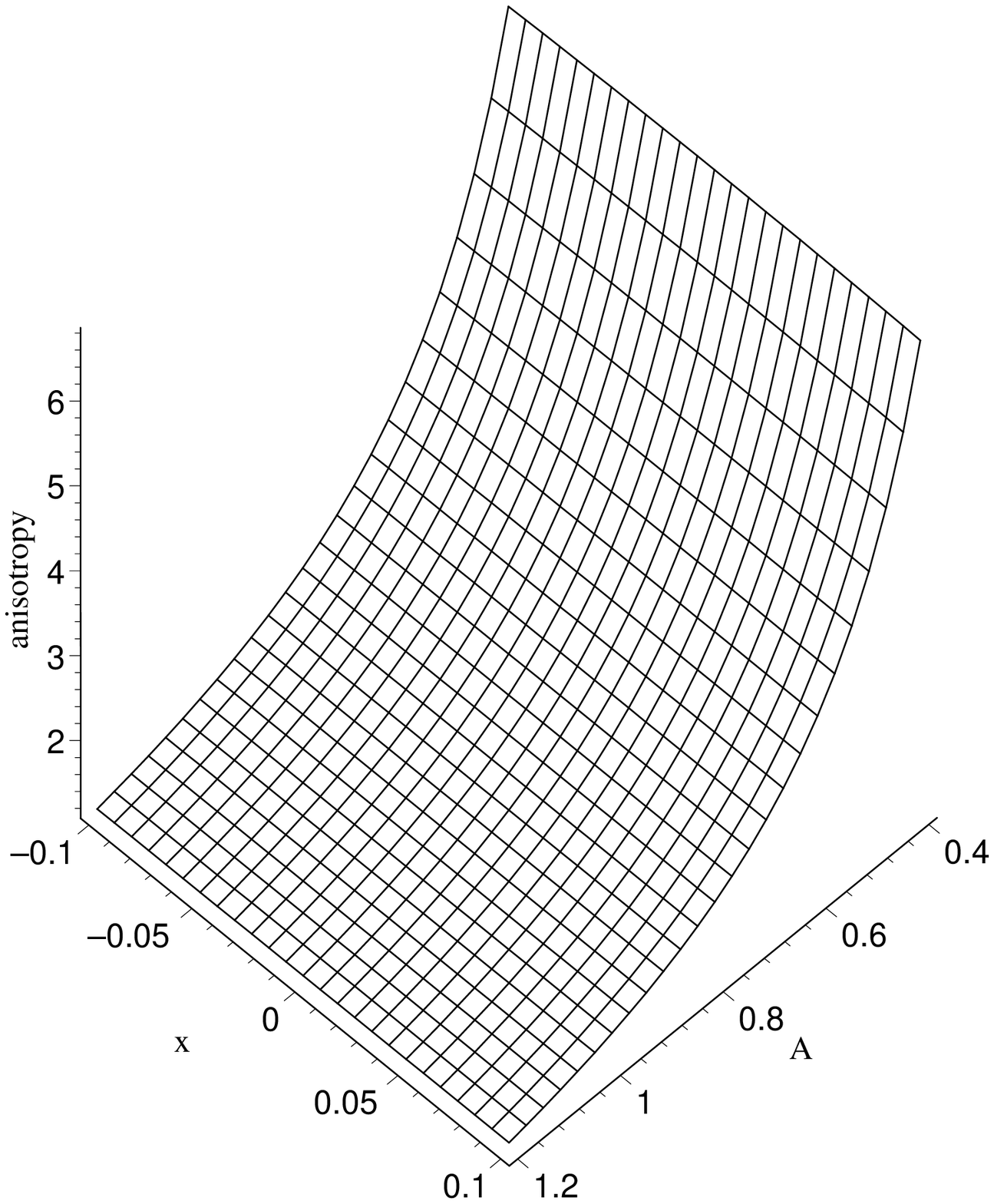}
%}
\end{center}\vspace{-0.1cm}
\caption{\small{Anisotropy of throat region in Einstein case, $g_{tt}$ \emph{increasing}, $x_{0}=1$, varying throat radius.}}
\label{fig:einstradiusconcupaniso}
\end{framed}
\end{figure}

In the figures below (Figures \ref{fig:einstxoconcdown} - \ref{fig:einstradiusconcupaniso}) we display the results of Einstein theory ($n=1$) for the spatial geometry governed by (\ref{eq:Qcosh}). The first set of figures (figs.~\ref{fig:einstxoconcdown}a-d) show the behavior of $g_{tt}$, along with the energy conditions (\ref{eq:einstrho}) and (\ref{eq:einstecond1},ii) as a function of the \emph{flare-out parameter}, $x_{0}$. Although the energy density is positive (as is $\rhot+\pt$), the other energy condition is not positive. We can also see here the well-known situation that the greater the degree of flare-out (equivalent to small $x_{0}$) the more severe the energy condition violation in Einstein gravity.

Figure \ref{fig:einstradiusconcdown} displays the behavior of $g_{tt}$, along with the energy conditions (\ref{eq:einstrho}) and (\ref{eq:einstecond1},ii) as a function of the \emph{throat radius}, $A$. Note that although the energy density can be made positive near the throat if the throat is not too large, the quantity (\ref{eq:einstecond1}) is negative in the throat vicinity, confirming the analytic analysis above.

In Figure \ref{fig:einstradiusconcdownaniso} we display the anisotropy, $\pt-\pr$, as a function of \emph{throat radius}. Note that larger throats require less anisotropy (i.e. are ``more isotropic'').

\clearpage
In Figures \ref{fig:einstxoconcup}, \ref{fig:einstradiusconcup} and \ref{fig:einstradiusconcupaniso} we present a similar analysis as above, except that in these figures, $g_{tt}$ is \emph{concave-up}. Note the similarity of these results to the previous analysis. 
%% \clearpage
%%\vspace{-2cm}

\vspace{-2.85cm}
%%\clearpage
\FloatBarrier

\subsubsection{{\normalsize $n=2$}}
We now study the $n=2$ contribution. Adding an $\alpha_{2}R^{2}$ term to the gravitational action is arguably the most popular
supplement to the Einstein-Hilbert action (see \cite{ref:rsq1}-\cite{ref:rsqlast}). For example, this modification has been utilized to drive inflation purely from the gravitational sector (the Starobinsky inflationary theory \cite{ref:starob}) or to capture some low-order quantum corrections \cite{ref:birrellanddavies}. 

As mentioned earlier, since in the theory given by the action (\ref{eq:lag}) each term in the series contributes to the energy conditions separately, it is more instructive to look at this contribution on its own, to see whether it helps or hinders with respect to the energy conditions. For the anisotropy analysis, we add the Einstein term to it as well to see if the modification can make the system more or less isotropic compared to Einstein gravity alone. We consider zero and non-zero tidal force cases.

\paragraph{{\small Zero tidal force:}}
In the case of zero tidal force ($\Phi(x)=\mbox{const.}$) we can get a handle on the behavior of the energy conditions at the throat by performing a series expansion as was done in the Einstein case. The results are summarized as:
\begin{subequations}\romansubs
{\allowdisplaybreaks\begin{align}
&\tilde{\rho}=\frac{2\alpha_{2}}{\qzero^{4}} \left[1+4\qzero^{3}\qzeropppp-16\qzero^{3}(\qzeropp)^{3}+4\qzero^{2}(\qzeropp)^{2}\right] \nonumber \\
&\qquad\quad\quad +\frac{8\alpha_{2}}{\qzero^{2}} \left[\qzero\qzeroppppp -25\qzero(\qzeropp)^{2}\qzeroppp + 3\qzeropp\qzeroppp\right]\,x + \mathcal{O}(x^{2})\,, \label{eq:squaredzerorho}\\[0.2cm]
&\rhot+\pr=\frac{8\alpha_{2}}{\qzero^{2}}\left[\qzero\qzeropppp-4\qzero(\qzeropp)^{3}+2(\qzeropp)^{2}\right] \nonumber \\
&\qquad\quad\quad +\frac{8\alpha_{2}}{\qzero^{2}} \left[\qzero\qzeroppppp -25 \qzero(\qzeropp)^{2}\qzeroppp + 3\qzeropp\qzeroppp\right] \,x + \mathcal{O}(x^{2})\,, \label{eq:squaredzeroecond1}\\[0.2cm]
&\rhot+\pt=\frac{4\alpha_{2}}{\qzero^{4}} \left[1+2\qzero^{2}(\qzeropp)^{2} -3\qzero\qzeropp\right] +\frac{12\alpha_{2}}{\qzero^{3}}\qzeroppp\left(2\qzero\qzeropp-1\right)\,x + \mathcal{O}(x^{2})\,.\label{eq:squaredzeroecond2}
\end{align}}
\end{subequations}
For $\alpha_{2} > 0$, which is the more physical sector \cite{ref:faraonibook}, it can be seen that the above can all be made positive at the throat (for example, by considering models where $\qzeropp=0$ and $\qzeropppp>0$). By analyticity in a non-zero neighborhood, in such scenarios there therefore must be a non-zero domain about the throat for which the functions are non-negative. Therefore, in this class of models, throats (and we stress again here that we are not considering asymptotics at infinity) \emph{are allowed} which respect energy conditions. This will remain true even when adding the Einstein term to the action, as in the Einstein case the near-throat violation of energy conditions may be made arbitrarily small \cite{ref:VKD}, \cite{ref:nandi} independently of the $\alpha_{2}$ parameter.

\paragraph{{\small Non-zero tidal force:}}
Unfortunately, for non-zero tidal force the analytic expressions with a general $Q(x)$ and $\Phi(x)$ are very long and complicated, even as a near throat expansion, and not very revealing. We must therefore specify these functions in order to perform numerical studies. For this purpose we choose the same function as in the Einstein case so that comparisons may be easily made. That is, we choose the profile $Q(x)$ as given in (\ref{eq:Qcosh}) since, to reiterate, this function possesses all the required properties to describe a throat. Recall that the parameter $A$ represents the radius of the throat and $x_{0}$ represents the degree of ``flare-out'' of the wormhole. We also use similar $\Phi(x)$ functions as presented in the Einstein case.

\begin{figure}[!ht]
\begin{framed}
\begin{center}
\vspace{-0.5cm}
\begin{tabular}{cc}
\subfloat[$g_{tt}(x)$]{\includegraphics[width=45mm,height=45mm,clip]{squared_gtt_c_down.eps}}&\hspace{0.5cm}
\subfloat[$\tilde{\rho}(x)$]{\includegraphics[width=45mm,height=45mm,clip]{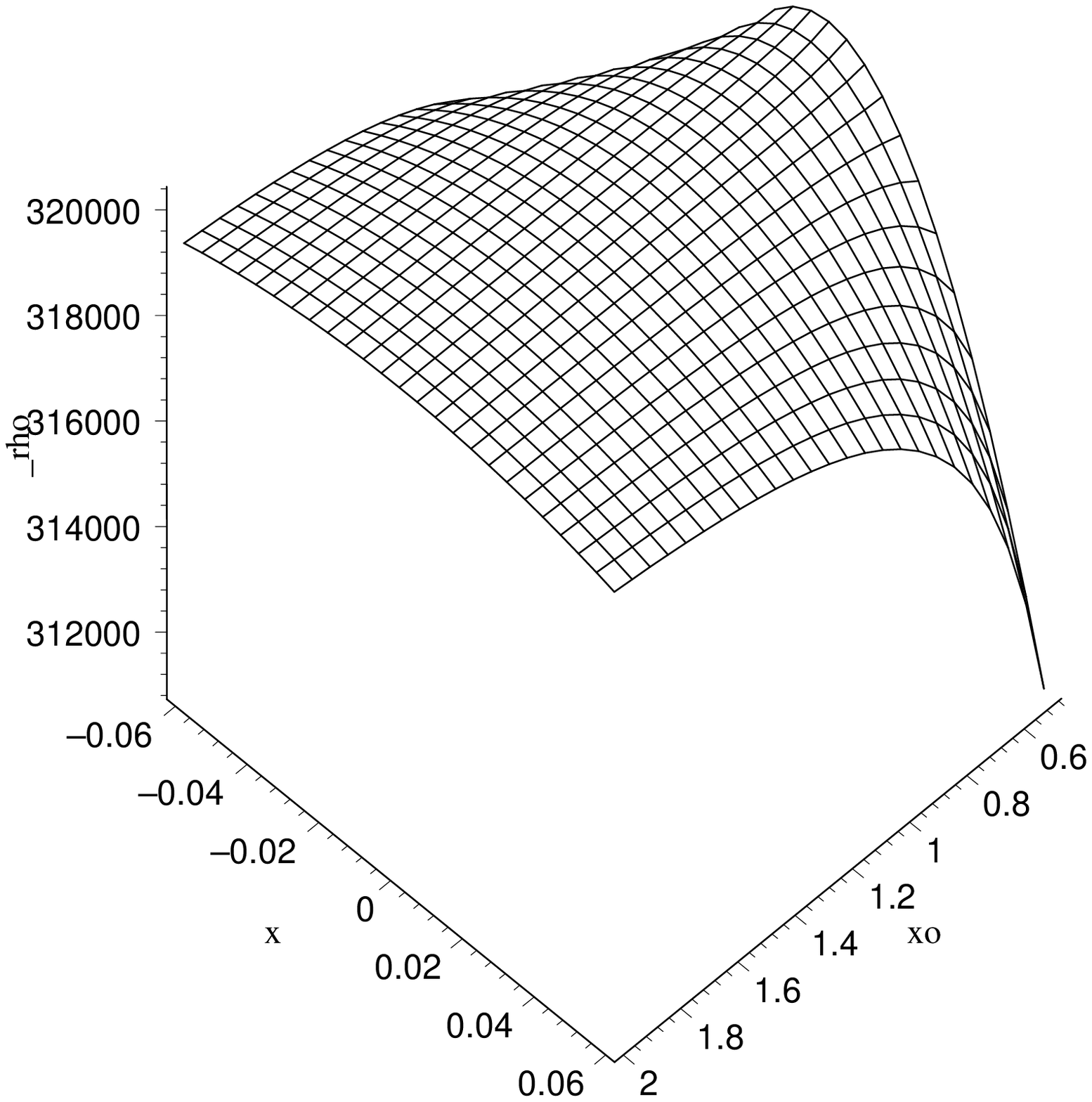}} \\
\subfloat[$\tilde{\rho}+\tilde{p}_{r}$]{\includegraphics[width=45mm,height=45mm,clip]{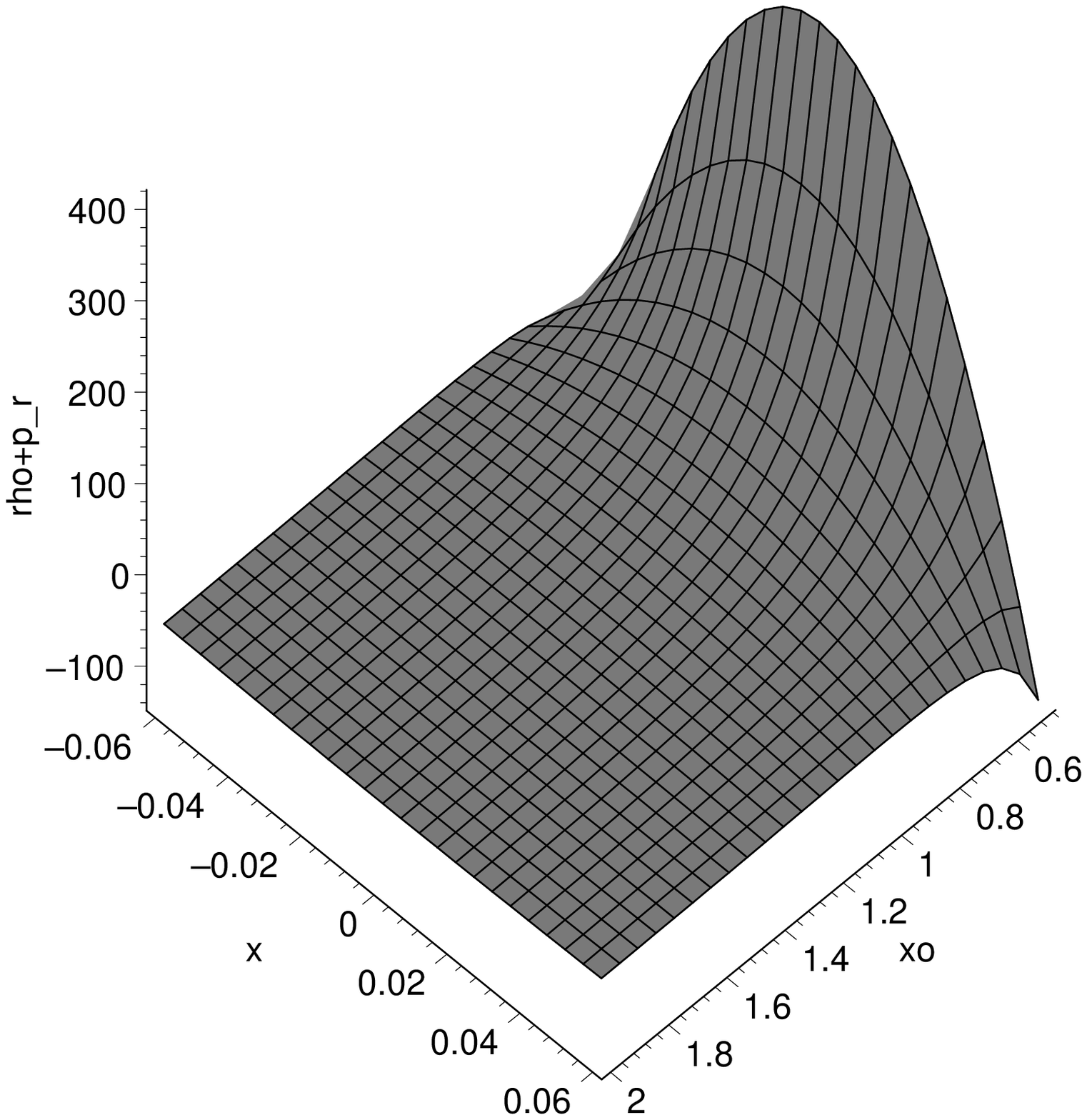}}&\hspace{0.5cm}
\subfloat[$\tilde{\rho}+\tilde{p}_{t}$]{\includegraphics[width=45mm,height=45mm,clip]{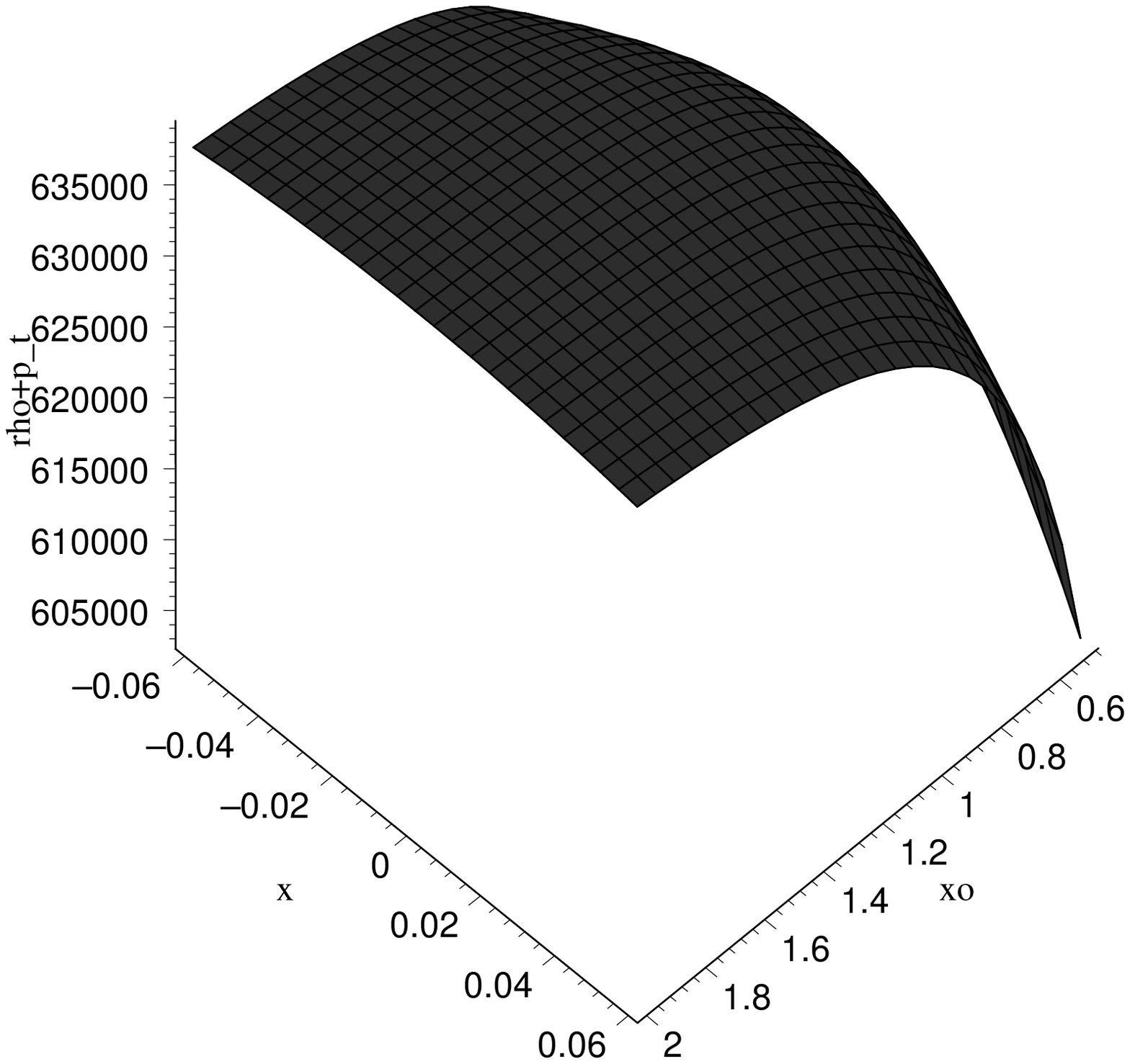}}
\end{tabular}
\end{center}
\vspace{-0.1cm}
\caption{\small{$R^{2}$ contribution, $g_{tt}$ \emph{decreasing}, throat radius=0.05, and varying $x_{0}$.}}
\label{fig:rsquaredalphaconcdown}
\end{framed}
\end{figure}

As mentioned previously, we set $\alpha_{2}=1$, as all results here can be rescaled by whatever value of $\alpha_{2}$ one wishes to study, including negative values, which result in a reflection about the horizontal planes in the graphs. As mentioned above, it should be noted that $\alpha_{2}>0$ is the preferred model. The first set of results are summarized in Figure~\ref{fig:rsquaredalphaconcdown} (also please refer to figure captions for details).

Note that like the zero tidal-force case, there is a large region of the parameter space where all energy conditions can be respected. In fact, for positive $\alpha_{2}$, a large flare-out (small $x_{0}$) actually helps the condition $\rhot+\pr>0$ near the throat.

\begin{figure}[!ht]
\begin{framed}
\begin{center}
\vspace{0.0cm}
%\fbox{
\includegraphics[width=60mm,height=45mm, clip]
{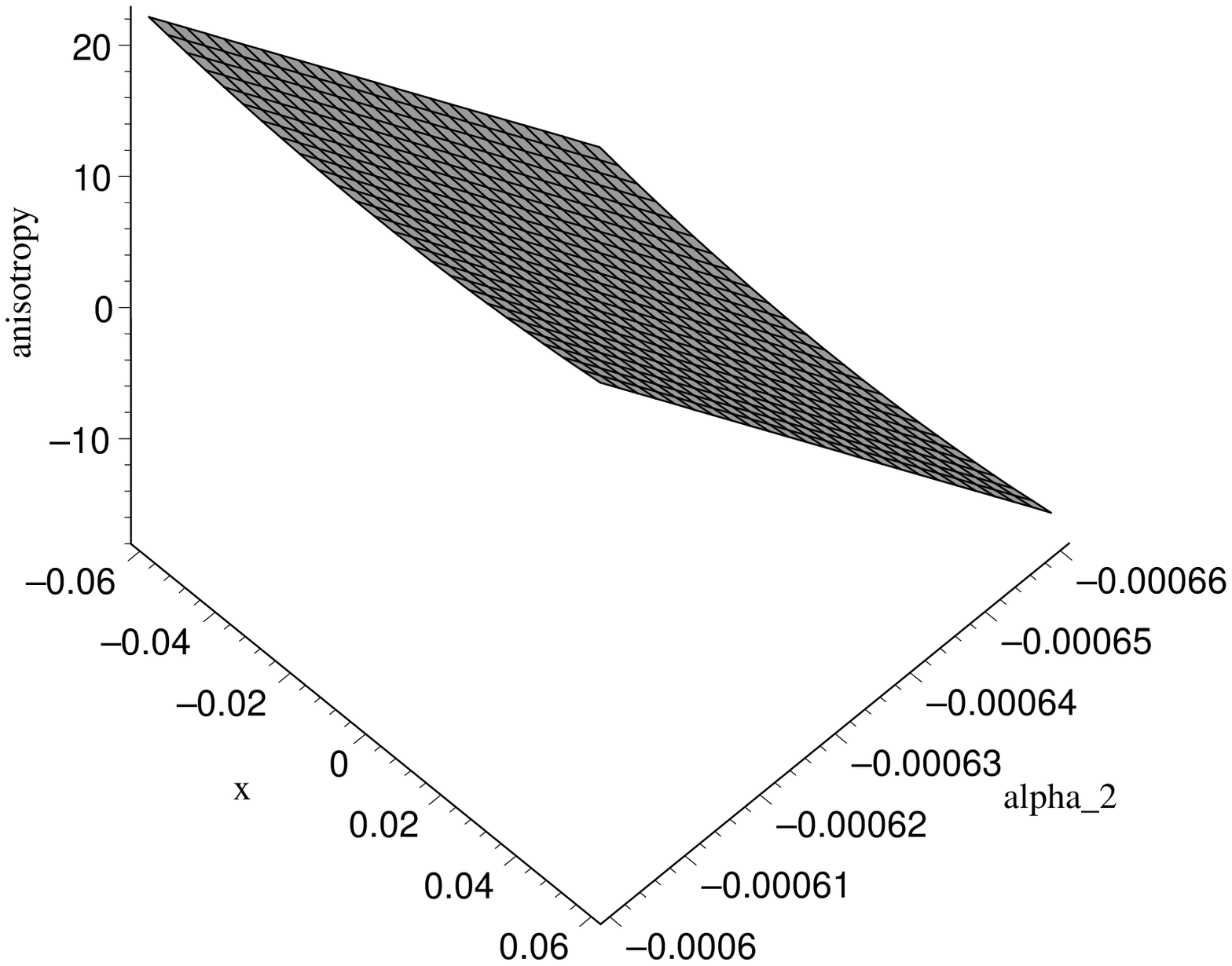}
%}
\end{center}\vspace{-0.1cm}
\caption{\small{Minimum anisotropy region: $R+\alpha_{2}R^{2}$ contribution, $g_{tt}$ \emph{decreasing}, throat radius=0.05, $x_{0}=1$, and varying $\alpha_{2}$.}}
\label{fig:rsquaredalphaconcdownaniso}
\end{framed}
\end{figure}

Also of interest is the anisotropy. Namely, can the presence of the extra terms in $f(R)$ gravity lessen the amount of anisotropy required to support a throat, compared to Einstein gravity alone. To study this we present Figure~\ref{fig:rsquaredalphaconcdownaniso} where the anisotropy, defined here as $\pt-\pr$, is presented for the Lagrangian $R+\alpha_{2}R^{2}$ and is plotted in the vicinity of the minimum anisotropy curve. Note from this figure that anisotropy is a minimum for $\alpha_{2}<0$, and not for the Einstein case ($\alpha_{2}=0$), although the minimum curve is very close to $\alpha_{2}=0$.

\begin{figure}[!ht]
\begin{framed}
\begin{center}
\vspace{-0.0cm}
\begin{tabular}{ccc}
%%\subfloat[$g_{tt}(x)$]{\includegraphics[width=45mm,height=45mm,clip]{squared_gtt_c_down.eps}}&
\subfloat[$\tilde{\rho}(x)$]{\includegraphics[width=42mm,height=42mm,clip]{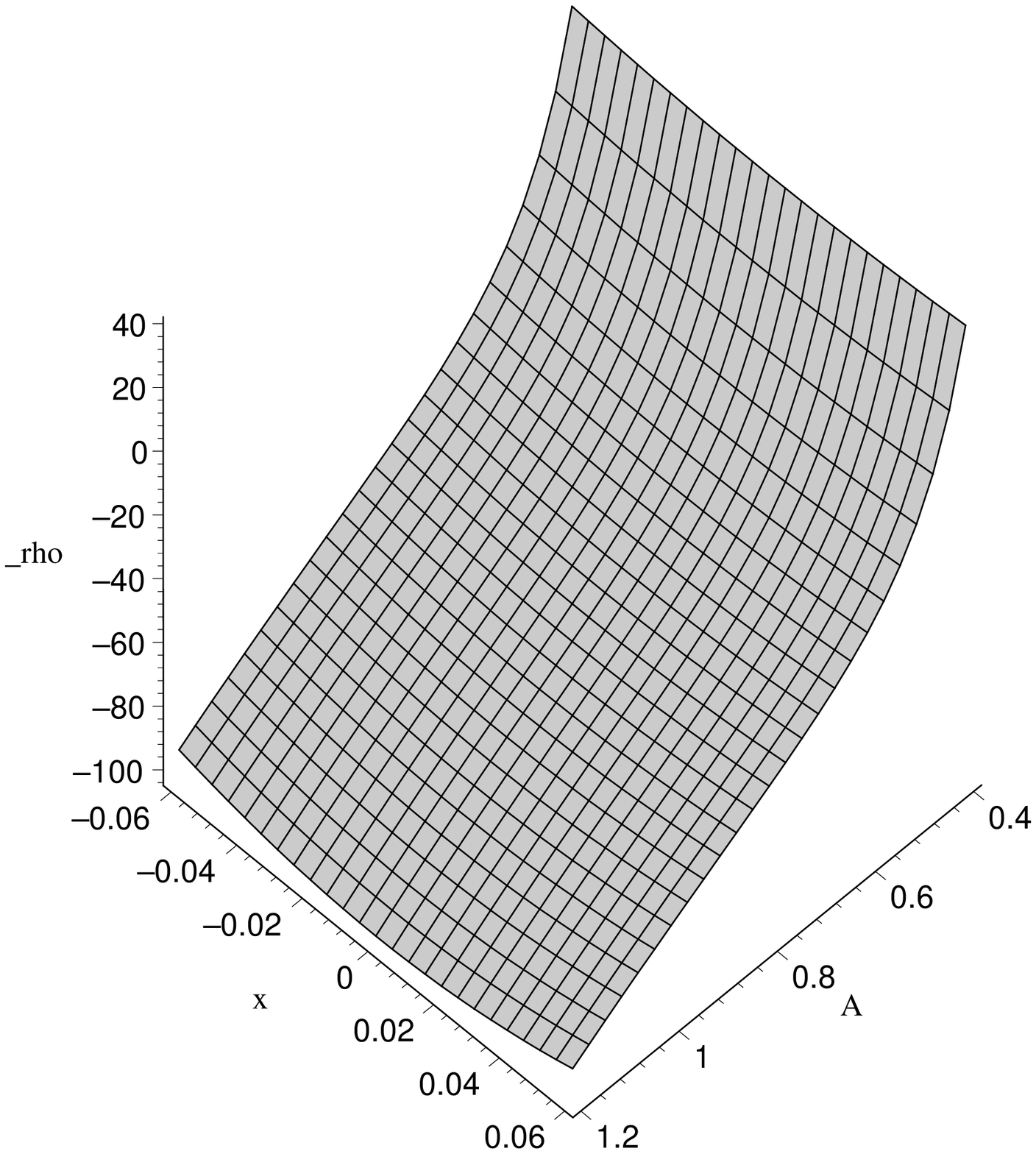}} &
\subfloat[$\tilde{\rho}+\tilde{p}_{r}$]{\includegraphics[width=42mm,height=42mm,clip]{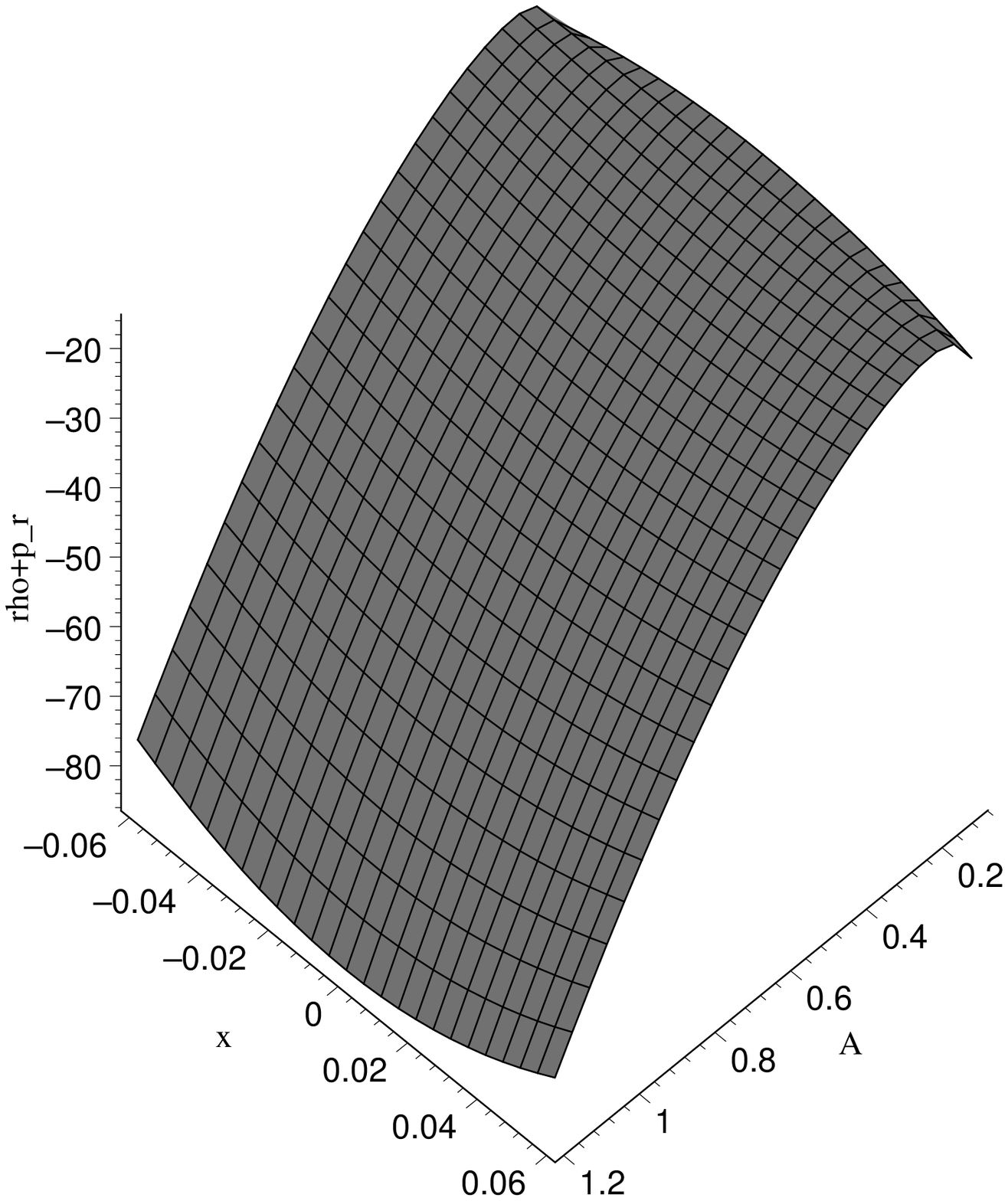}}&
\subfloat[$\tilde{\rho}+\tilde{p}_{t}$]{\includegraphics[width=42mm,height=42mm,clip]{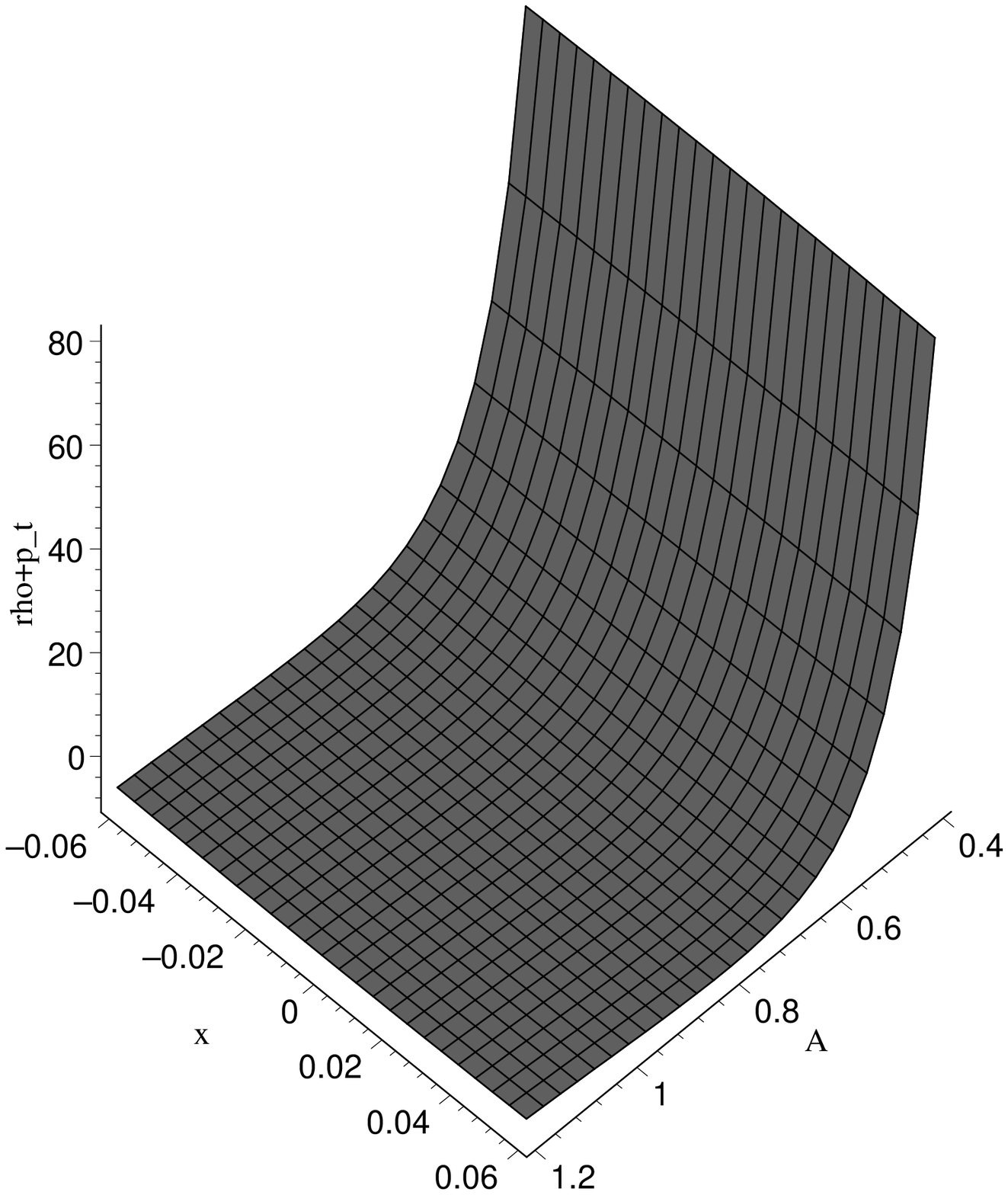}}
\end{tabular}
\end{center}
\vspace{-0.1cm}
\caption{\small{$R^{2}$ contribution for $x_{0}=1$, $g_{tt}$ \emph{decreasing} and varying throat radius. For negative $\alpha_{2}$ the graphs should be flipped around the plane $z=0$.}}
\label{fig:rsquaredAconcdown}
\end{framed}
\end{figure}

In Figure~\ref{fig:rsquaredAconcdown} we present another non-zero tidal force analysis (with the same $g_{tt}(x)$ as in the previous case) where, instead of allowing $x_{0}$ to vary, we vary the throat radius. This study is interesting as even in Einstein gravity the size of the throat affects the amount of energy condition violation. We find that, generally, smaller throat radius is more favorable for respecting energy conditions. (This result is reversed for $\alpha_{2}<0$.)

\begin{figure}[!ht]
\begin{framed}
\begin{center}
\vspace{-0.5cm}
\begin{tabular}{cc}
\subfloat[$g_{tt}(x)$]{\includegraphics[width=45mm,height=45mm,clip]{squared_gtt_c_up.eps}}&\hspace{0.5cm}
\subfloat[$\tilde{\rho}(x)$]{\includegraphics[width=45mm,height=45mm,clip]{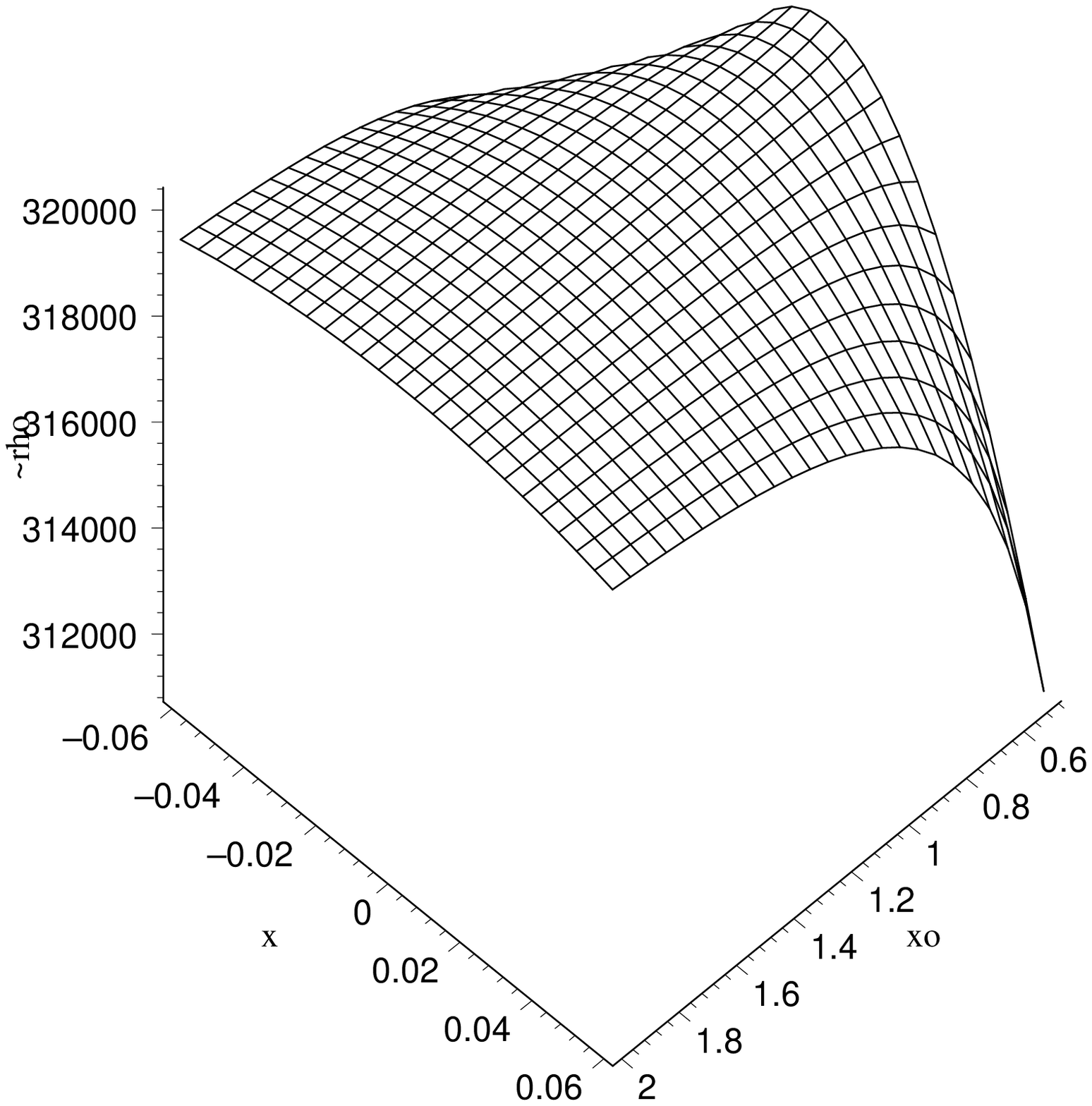}} \\
\subfloat[$\tilde{\rho}+\tilde{p}_{r}$]{\includegraphics[width=45mm,height=45mm,clip]{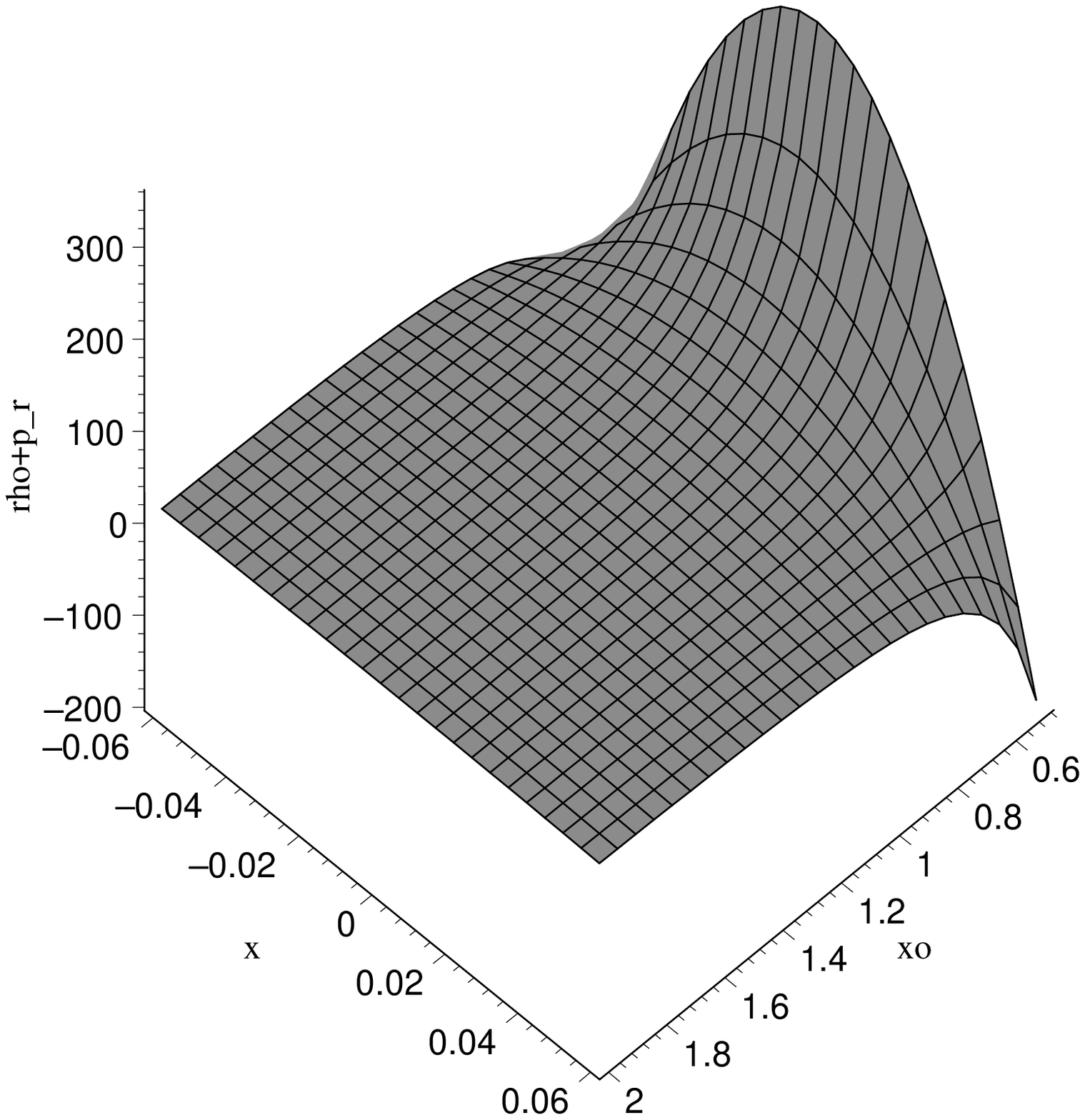}}&\hspace{0.5cm}
\subfloat[$\tilde{\rho}+\tilde{p}_{t}$]{\includegraphics[width=45mm,height=45mm,clip]{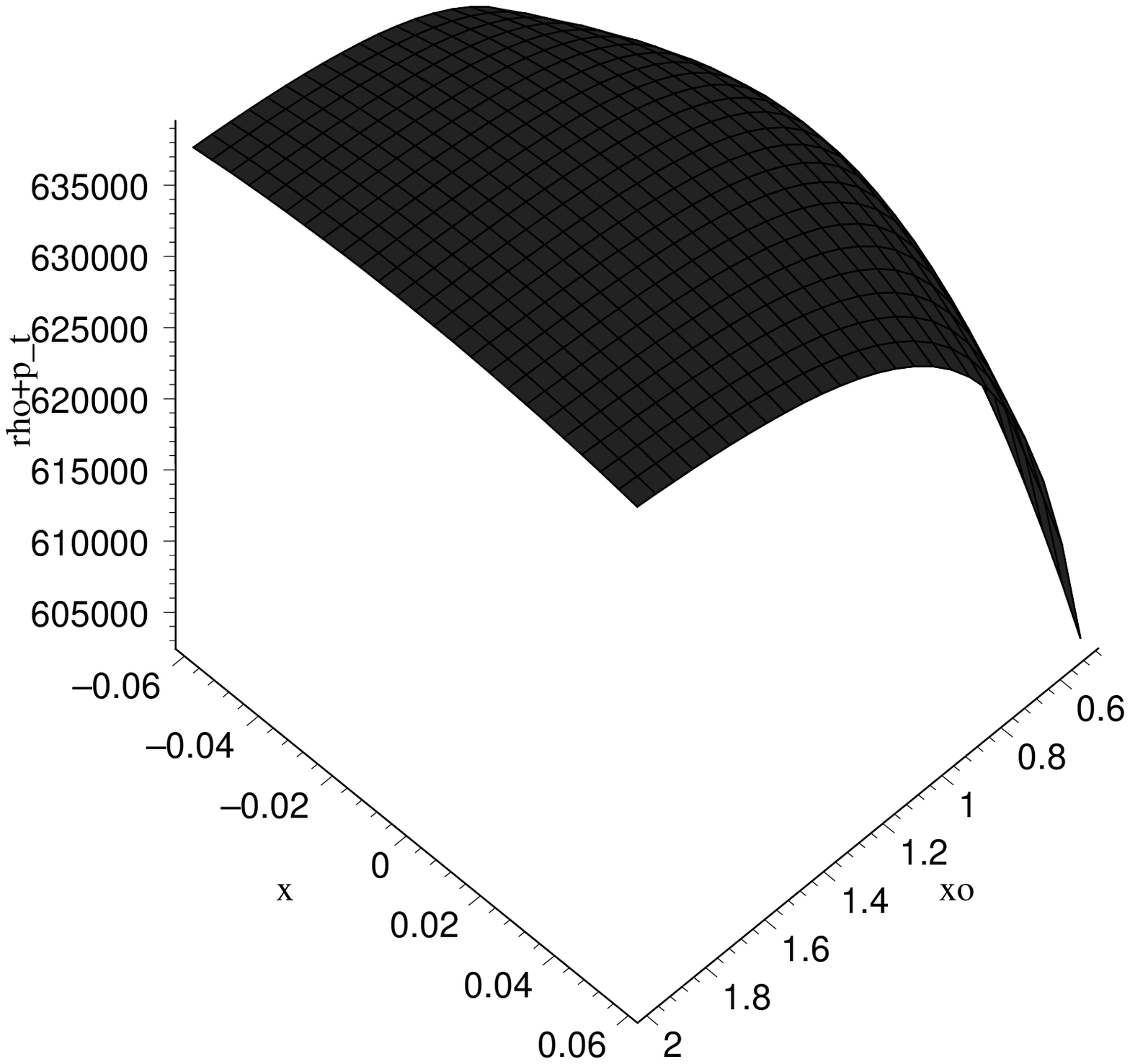}}
\end{tabular}
\end{center}
\vspace{-0.1cm}
\caption{\small{ $R^{2}$ contribution, $g_{tt}$ \emph{increasing}, throat radius=0.05, and varying $x_{0}$.}}
\label{fig:rsquaredalphaconcup}
\end{framed}
\end{figure}

\begin{figure}[!ht]
\begin{framed}
\begin{center}
\vspace{0.0cm}
%\fbox{
\includegraphics[width=60mm,height=45mm, clip]{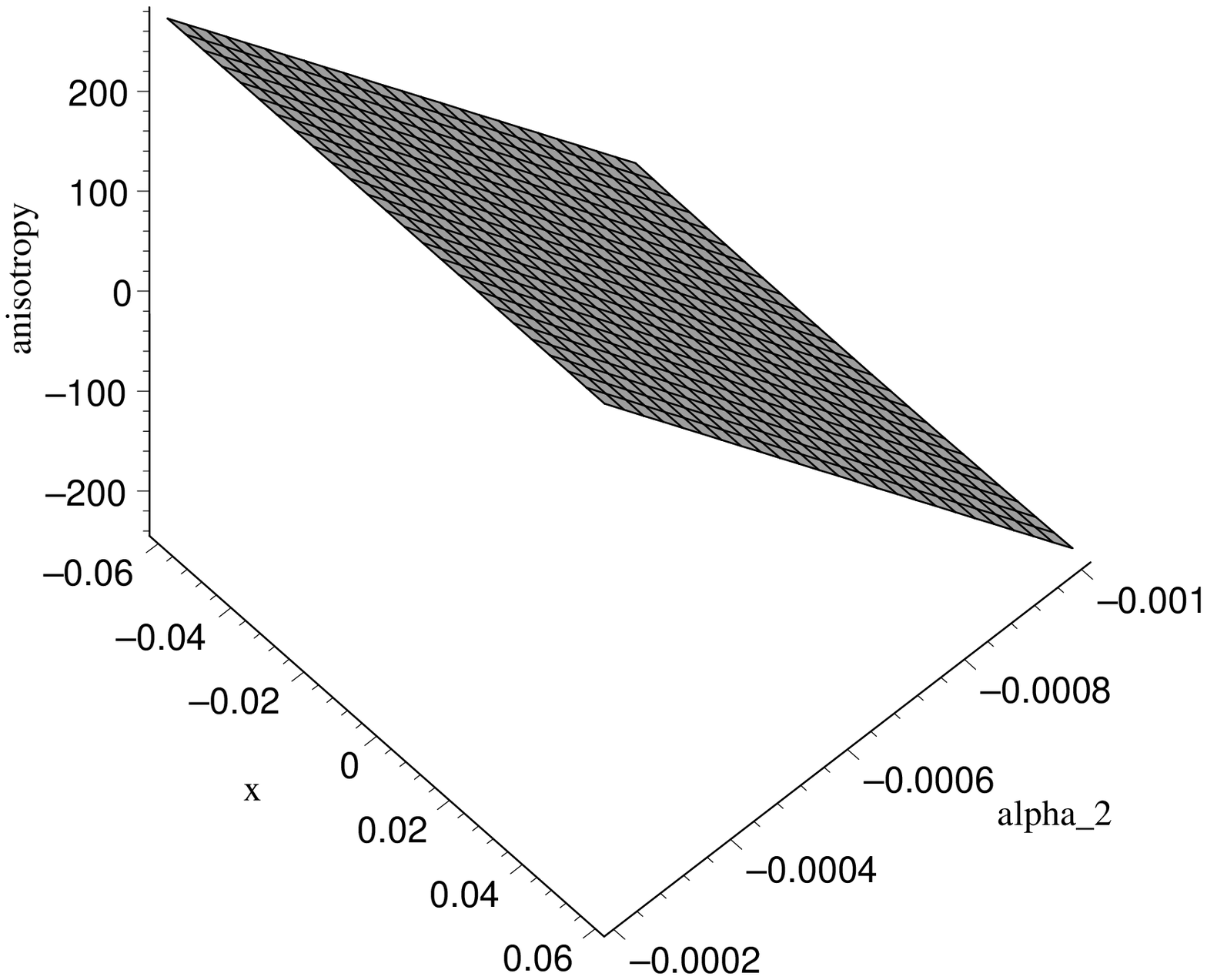}
%}
\end{center}\vspace{-0.1cm}
\caption{\small{Minimum anisotropy region: $R+\alpha_{2}R^{2}$ contribution, $g_{tt}$ \emph{increasing}, throat radius=0.05, $x_{0}=1$, and varying $\alpha_{2}$.}}
\label{fig:rsquaredalphaconcupaniso}
\end{framed}
\end{figure}

Next we consider non-zero tidal force models where $g_{tt}(x)$ is increasing instead of decreasing. The figures below (figs.~\ref{fig:rsquaredalphaconcup} and \ref{fig:rsquaredalphaconcupaniso}) summarize the results for varying $x_{0}$ and $\alpha_{2}$.
%%\vspace{-0.5cm}
Note that in Figure~\ref{fig:rsquaredalphaconcup} the graphs mimic the previous scenario with decreasing $\gtt$, indicating insensitivity to the form of the lapse function. Therefore, in this scenario, again energy conditions can be respected for the more physical sector of positive $\alpha_{2}$. We also present the anisotropy in Figure~\ref{fig:rsquaredalphaconcupaniso} for the full Lagrangian of $R+\alpha_{2}R^{2}$. Anisotropy is again minimized for $\alpha_{2}\neq 0$ and in the negative sector.
%%\vspace{-0.5cm}

\begin{figure}[H]
\begin{framed}
\begin{center}
\vspace{-0.5cm}
\begin{tabular}{ccc}
%%\subfloat[$g_{tt}(x)$]{\includegraphics[width=45mm,height=45mm,clip]{squared_gtt_c_up.eps}}&
\subfloat[$\tilde{\rho}(x)$]{\includegraphics[width=42mm,height=42mm,clip]{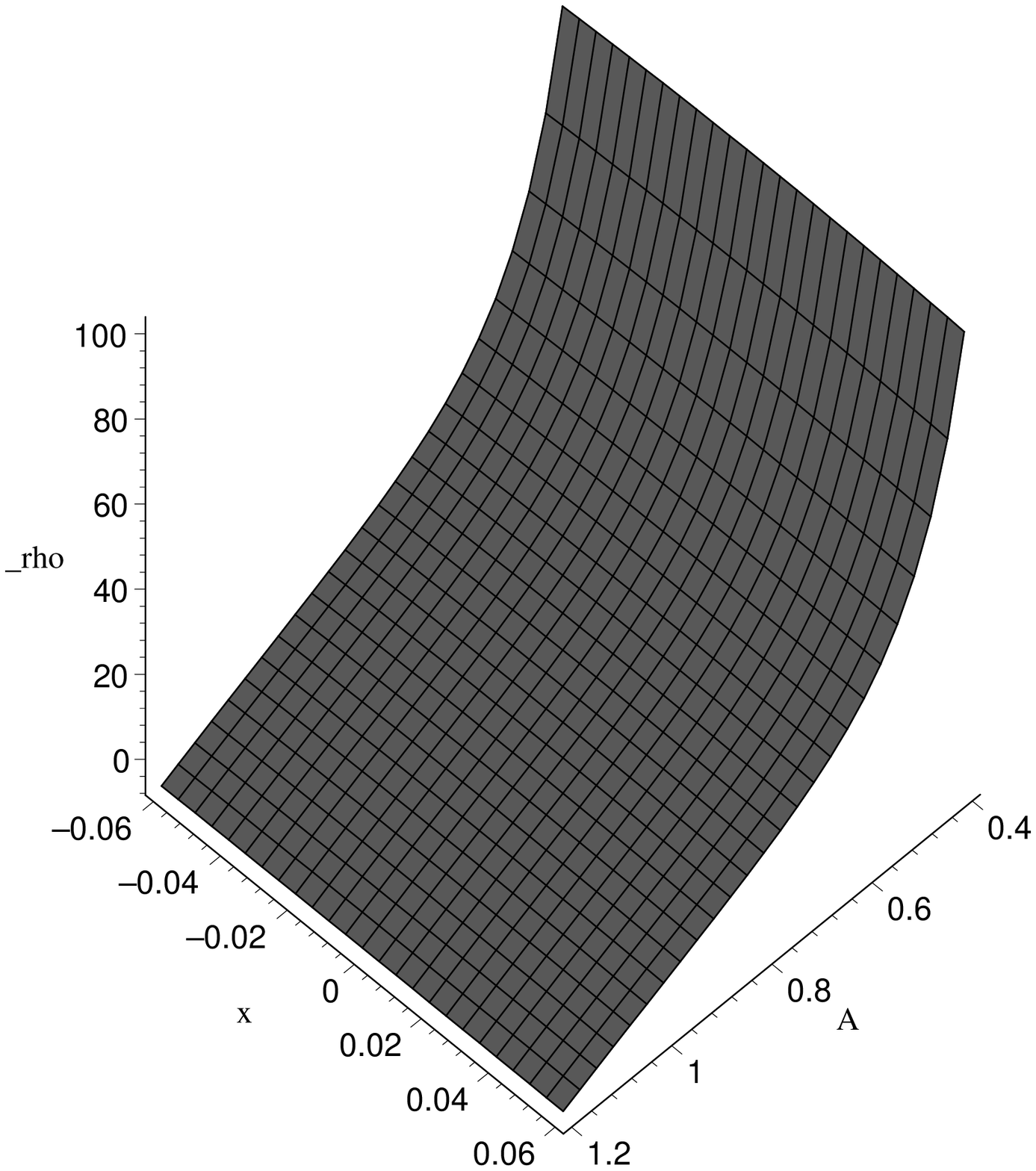}} &
\subfloat[$\tilde{\rho}+\tilde{p}_{r}$]{\includegraphics[width=42mm,height=42mm,clip]{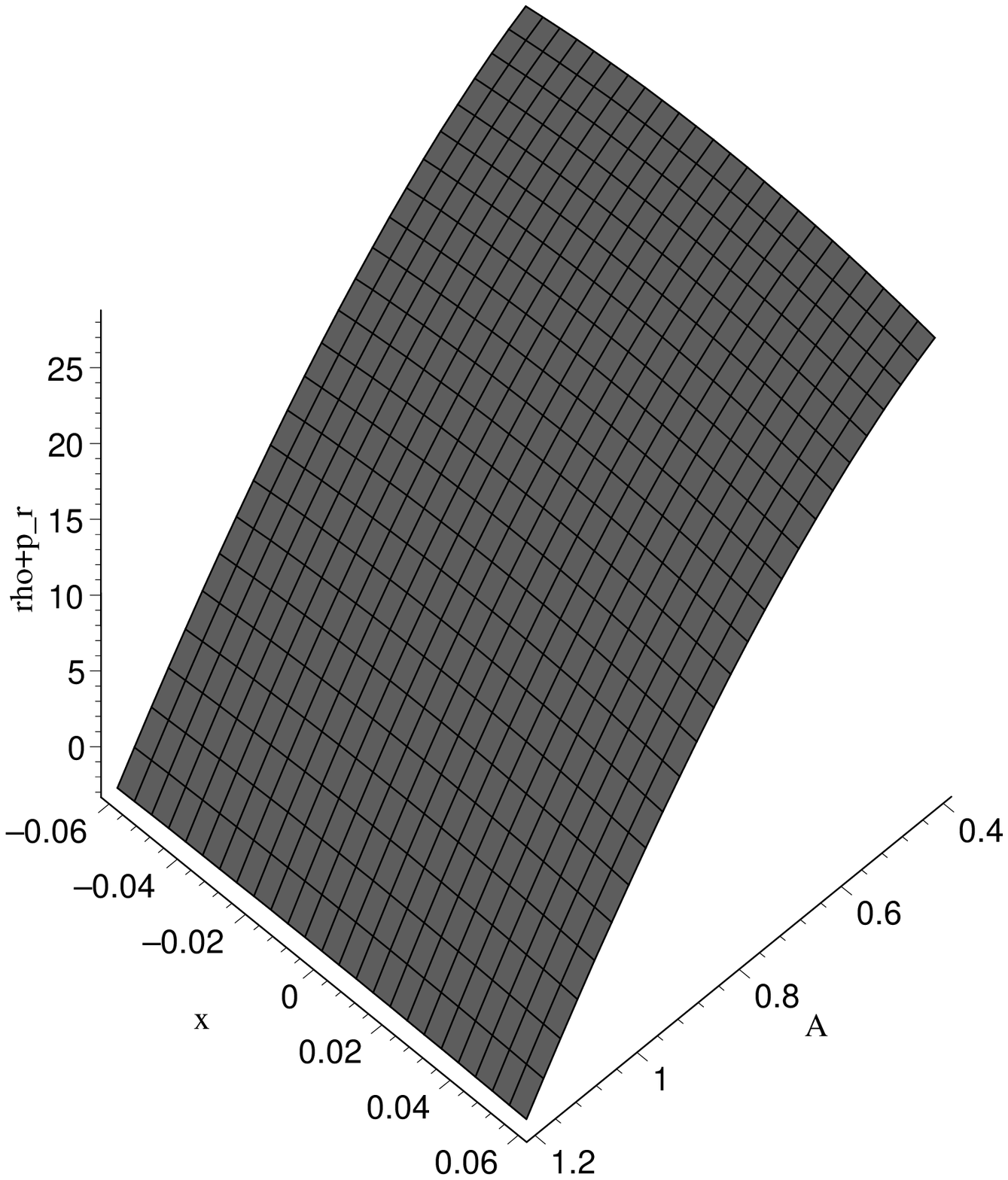}}&
\subfloat[$\tilde{\rho}+\tilde{p}_{t}$]{\includegraphics[width=42mm,height=42mm,clip]{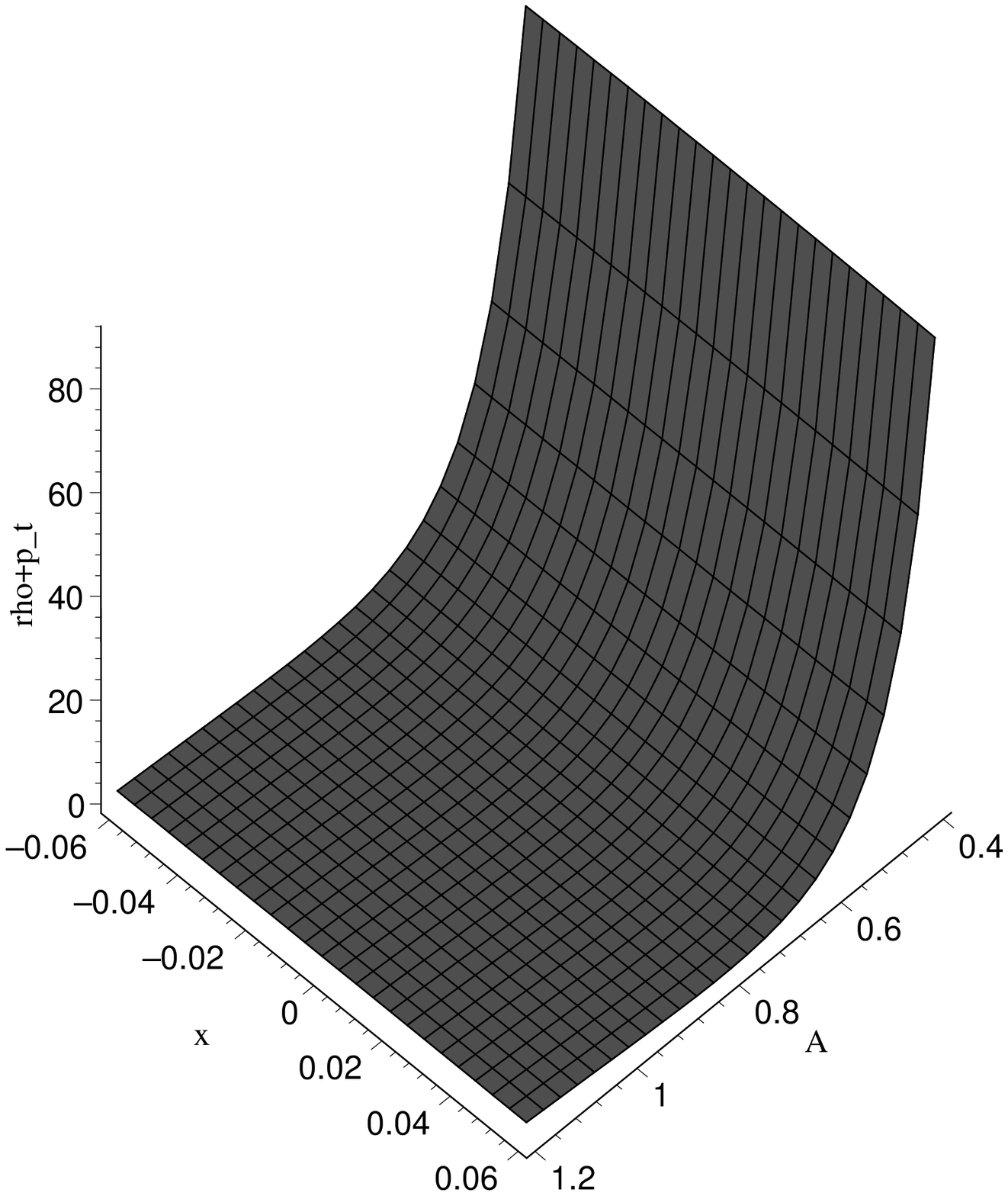}}
\end{tabular}
\end{center}
\vspace{-0.1cm}
\caption{\small{ $R^{2}$ contribution for $x_{0}=1$, $g_{tt}$ \emph{increasing}, and varying throat radius. For negative $\alpha_{2}$ the graphs should be flipped around the plane $z=0$.}}
\label{fig:rsquaredAconcup}
\end{framed}
\end{figure}

Finally for this sub-section, we present the results for the increasing $g_{tt}(x)$ but where the throat radius is allowed to vary. These results are summarized in Figure~\ref{fig:rsquaredAconcup}.
\clearpage
%%\begin{figure}[!ht]
%%\vspace{-0.5cm}
\noindent Note that again smaller throat radius is more favorable for energy conditions than larger throat radius if $\alpha_{2}$ is positive. (The result is reversed for negative $\alpha_{2}$.) Parameters exist where energy conditions may again be satisfied, even when adding these results to the energy condition violating Einstein gravity terms.
\FloatBarrier
%%\clearpage

%%\vspace{-16cm}
\subsubsection{{\normalsize $n=3$}}
The $n=3$ scenarios are also of interest and it has been shown in a cosmological setting how this case can be transformed to a system consisting of Einstein gravity and a scalar field with a self-coupling proportional to $\alpha_{3}\phi^{4}$ \cite{ref:n3a}. 
\paragraph{{\small Zero tidal force:}}
Again for the $\Phi(x)=\mbox{const.}$ scenario we present an analytic expansion about the throat, although, due to the complexity of the coefficients, we produce only the lowest order term. The results are:
\begin{subequations}\romansubs
{\allowdisplaybreaks\begin{align}
&\tilde{\rho}= \frac{4\alpha_{3}}{\qzero^{6}} \left[1- 24\qzero^{4}\qzeropp\qzeropppp + 6\qzero\qzeropp -24 \qzero^{4}(\qzeroppp)^{2} +12 \qzero^{3}\qzeropppp -12\qzero^{2}(\qzeropp)^{2}\right. \nonumber \\
&\qquad\left. -56\qzero^{3}(\qzeropp)^{3} +96\qzero^{4}(\qzeropp)^{4}\right] + \mathcal{O}(x)\,, \label{eq:cubedzerorho}\\[0.2cm]
&\rhot+\pr=\frac{24\alpha_{3}}{\qzero^{5}} \left[\qzeropp-4\qzero^{3}\qzeropp\qzeropppp -4\qzero^{3}(\qzeroppp)^{2} +2\qzero^{2}\qzeropppp -12\qzero^{2}(\qzeropp)^{3} \right. \nonumber \\
&\qquad\quad\quad \left. +16 \qzero^{3}(\qzeropp)^{4} \right] + \mathcal{O}(x)\,,  \label{eq:cubedzeroecond1}\\[0.2cm]
&\rhot+\pt= \frac{12\alpha_{3}}{\qzero^{6}} \left[1-5\qzero\qzeropp +8\qzero^{2}(\qzeropp)^{2} -4 \qzero^{3} (\qzeropp)^{3}\right] + \mathcal{O}(x)\,.\label{eq:cubedzeroecond2}
\end{align}}
\end{subequations}
Here, in the case of positive $\alpha_{3}$,  (\ref{eq:cubedzerorho}) - (\ref{eq:cubedzeroecond2}) can in principle be made positive. The simplest models obeying energy conditions would again be those where $\qzeropp=0$ (which, due to the condition for a minimum, also requires $\qzeroppp=0$) and $\qzeropppp>0$, as was the case for the corresponding situation in $n=2$. Note that if $\qzeropp=0$, the contributions from Einstein gravity, from (\ref{eq:einstnearthroatrho}) - (\ref{eq:einstnearthroate2}), exactly at the throat are non-negative (although (\ref{eq:einstnearthroate1}) becomes negative in some neighborhood away from the throat), and hence with the addition of these $\alpha_{3}R^{3}$ contributions, $f(R)=R+\alpha_{3}R^{3}$ gravity can be made to obey energy conditions (and therefore, including the previous analysis, $f(R)=R+\alpha_{2}R^{2}+\alpha_{3}R^{3}$ gravity can also be made to obey energy conditions).

\begin{figure}[!ht]
\begin{framed}
\begin{center}
\vspace{-0.5cm}
\begin{tabular}{cc}
\subfloat[$g_{tt}(x)$]{\includegraphics[width=45mm,height=45mm,clip]{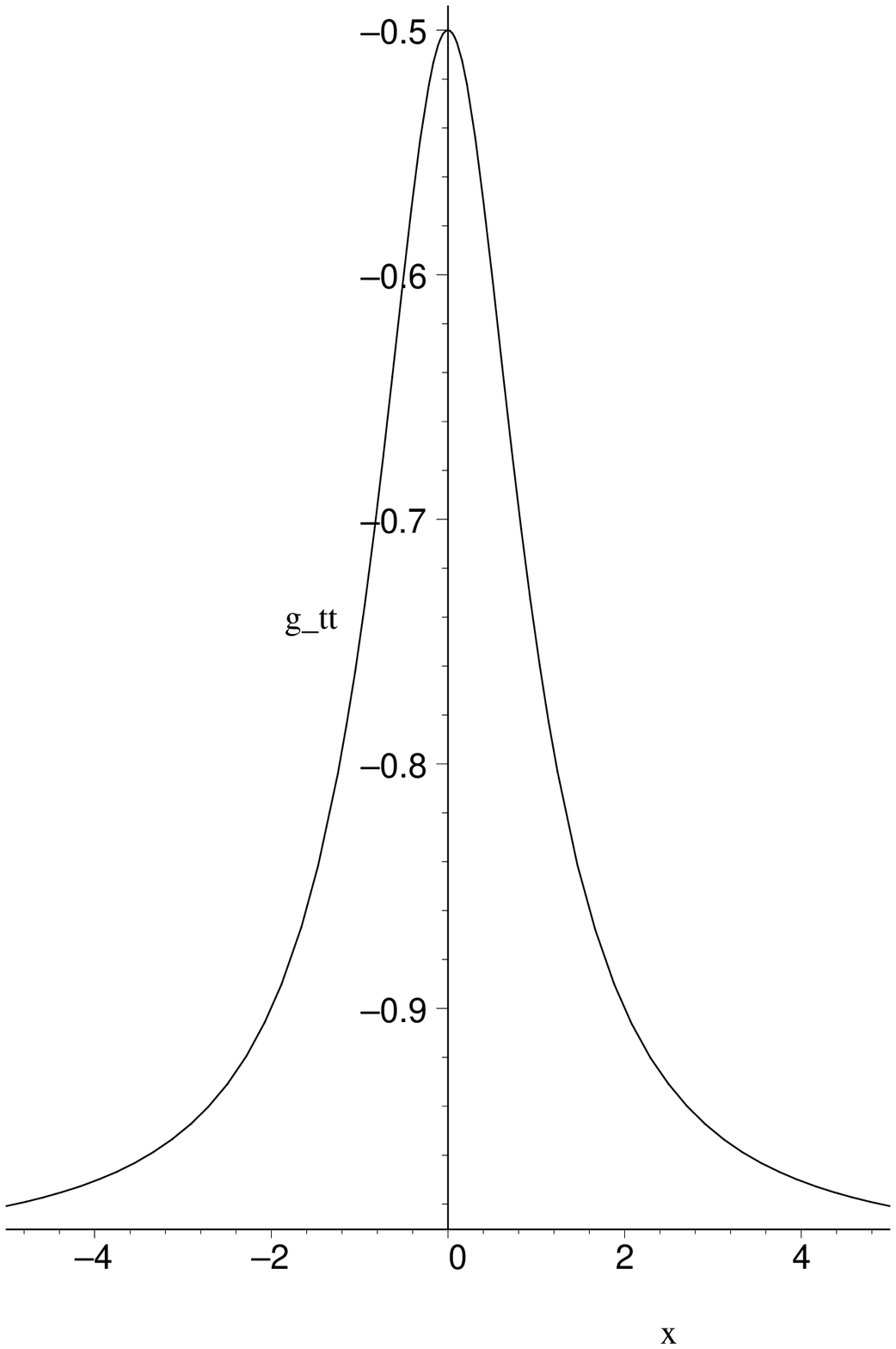}}&\hspace{0.5cm}
\subfloat[$\tilde{\rho}(x)$]{\includegraphics[width=45mm,height=45mm,clip]{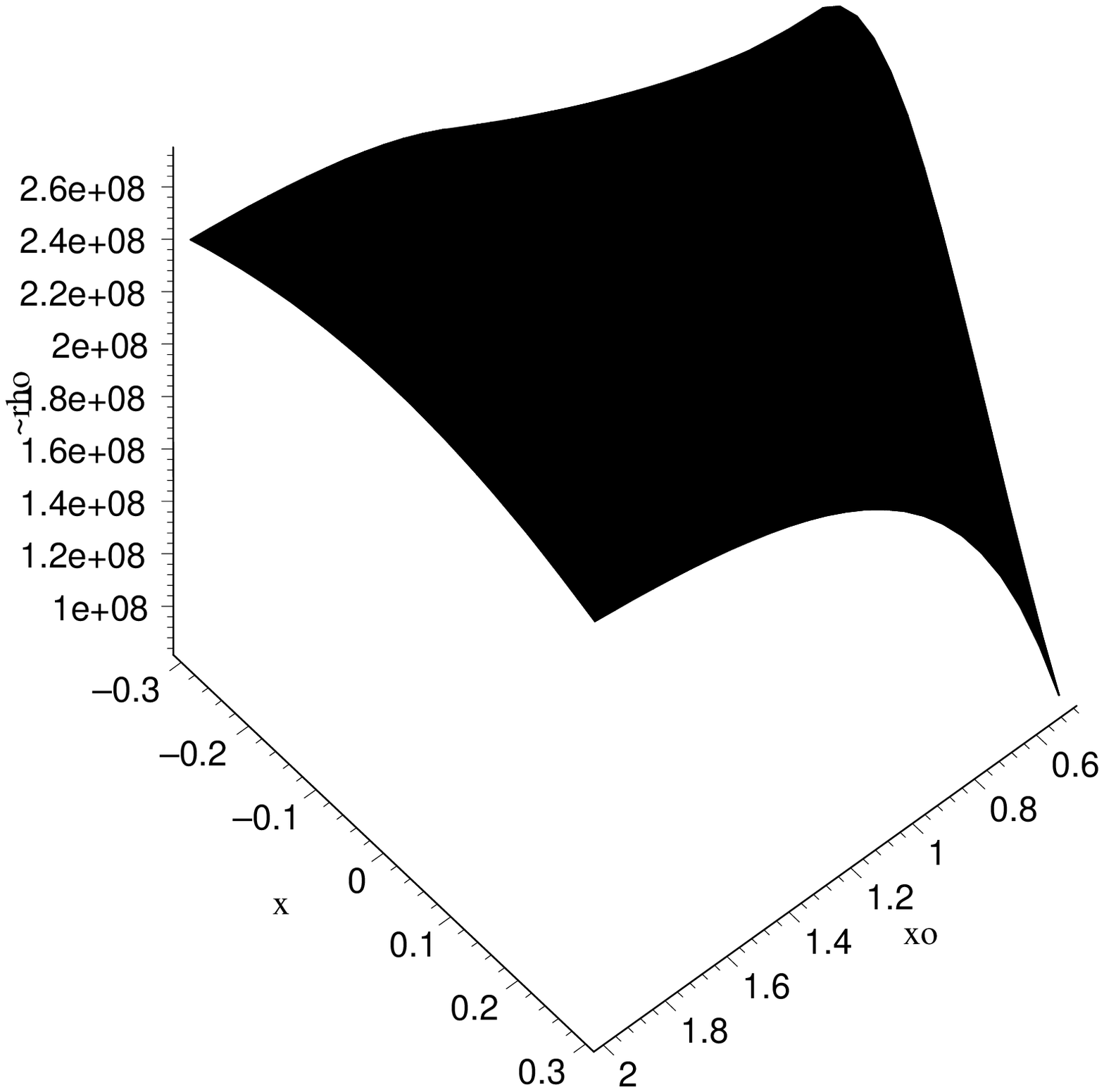}} \\
\subfloat[$\tilde{\rho}+\tilde{p}_{r}$]{\includegraphics[width=45mm,height=45mm,clip]{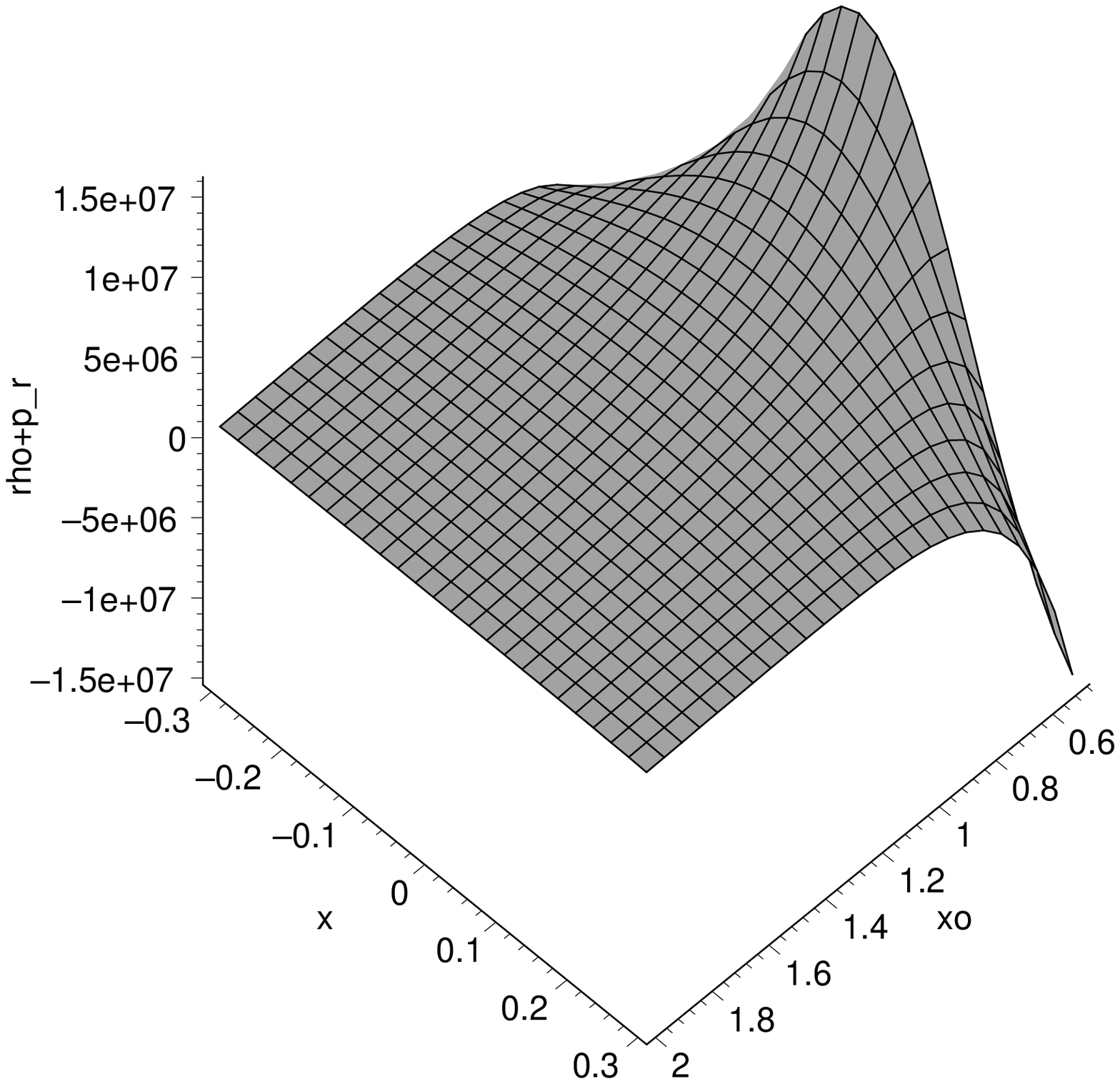}}&\hspace{0.5cm}
\subfloat[$\tilde{\rho}+\tilde{p}_{t}$]{\includegraphics[width=45mm,height=45mm,clip]{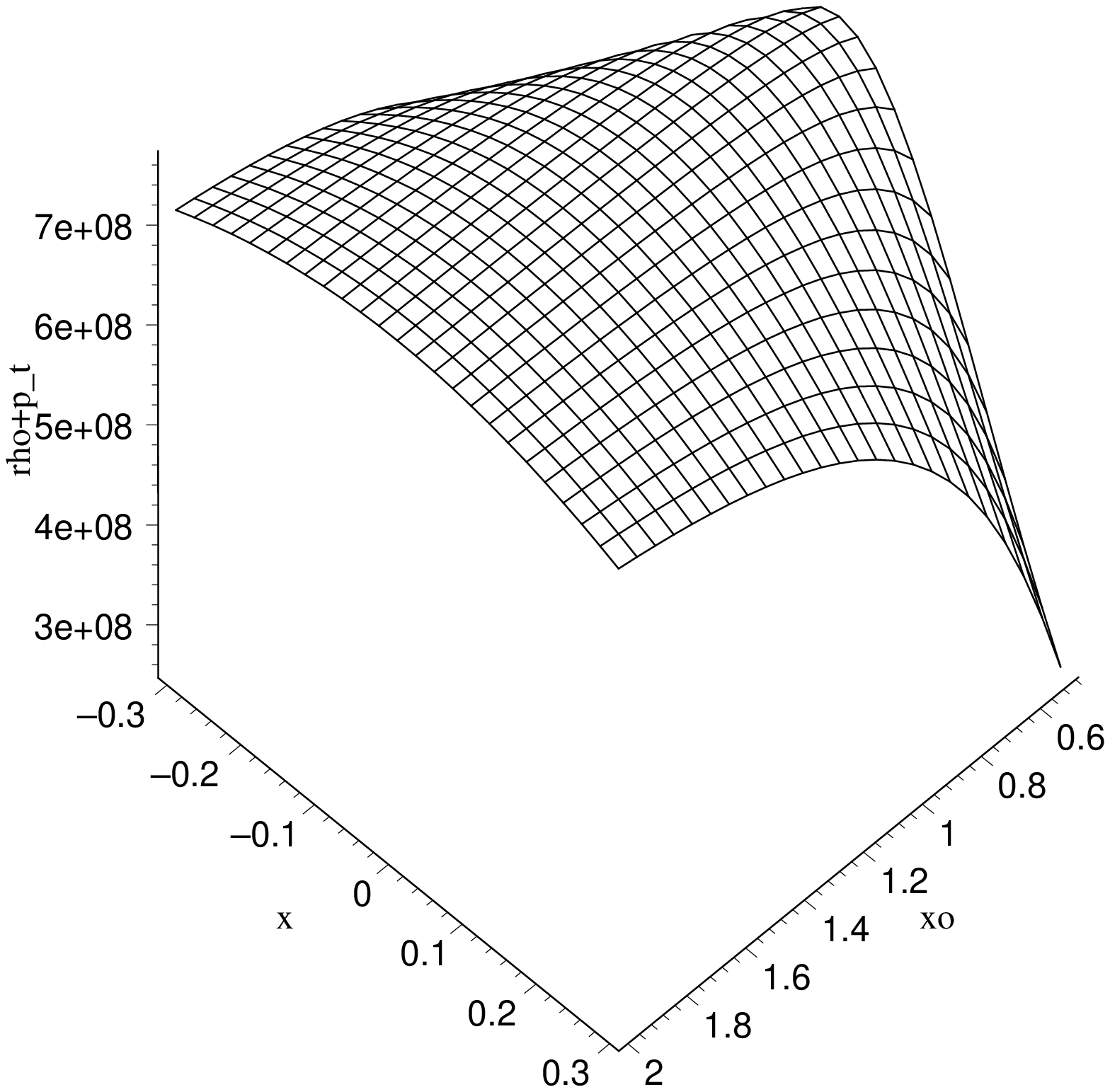}}
\end{tabular}
\end{center}
\vspace{-0.1cm}
\caption{\small{ $R^{3}$ contribution, $g_{tt}$ \emph{decreasing}, throat radius=0.05 and varying $x_{0}$.}}
\label{fig:rcubedalphaconcdown}
\end{framed}
\end{figure}

\begin{figure}[!ht]
\begin{framed}
\begin{center}
\vspace{0.0cm}
%\fbox{
\includegraphics[width=60mm,height=45mm, clip]{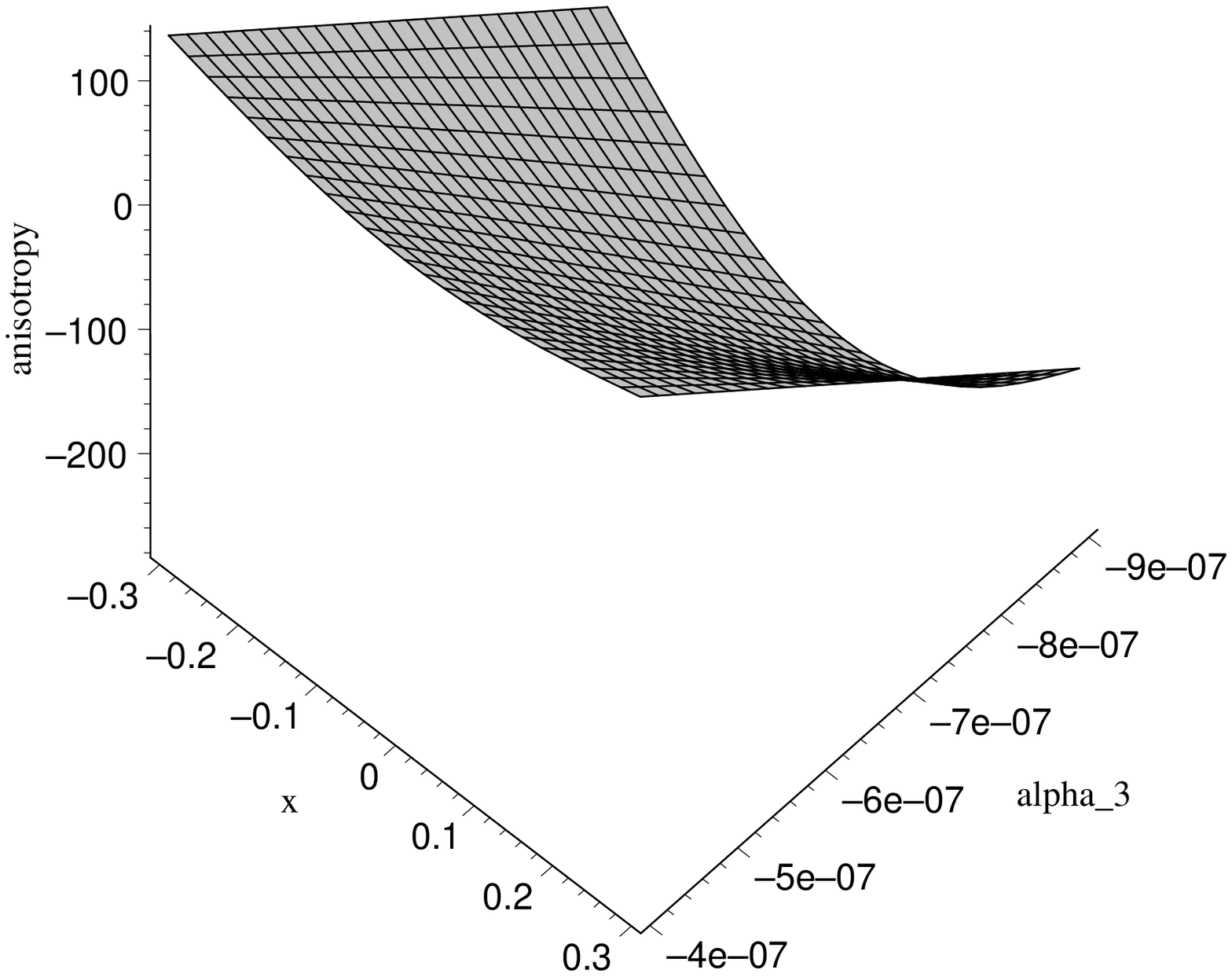}
%}
\end{center}\vspace{-0.1cm}
\caption{\small{Minimum anisotropy region: $R+\alpha_{3}R^{3}$ contribution, $g_{tt}$ \emph{decreasing}, throat radius=0.05, $x_{0}=1$, and varying $\alpha_{3}$.}}
\label{fig:rcubedalphaconcdownaniso}
\end{framed}
\end{figure}

\begin{figure}[!ht]
\begin{framed}
\begin{center}
\vspace{-0.5cm}
\begin{tabular}{ccc}
%%\subfloat[$g_{tt}(x)$]{\includegraphics[width=45mm,height=45mm,clip]{squared_gtt_c_down.eps}}&
\subfloat[$\tilde{\rho}(x)$]{\includegraphics[width=42mm,height=42mm,clip]{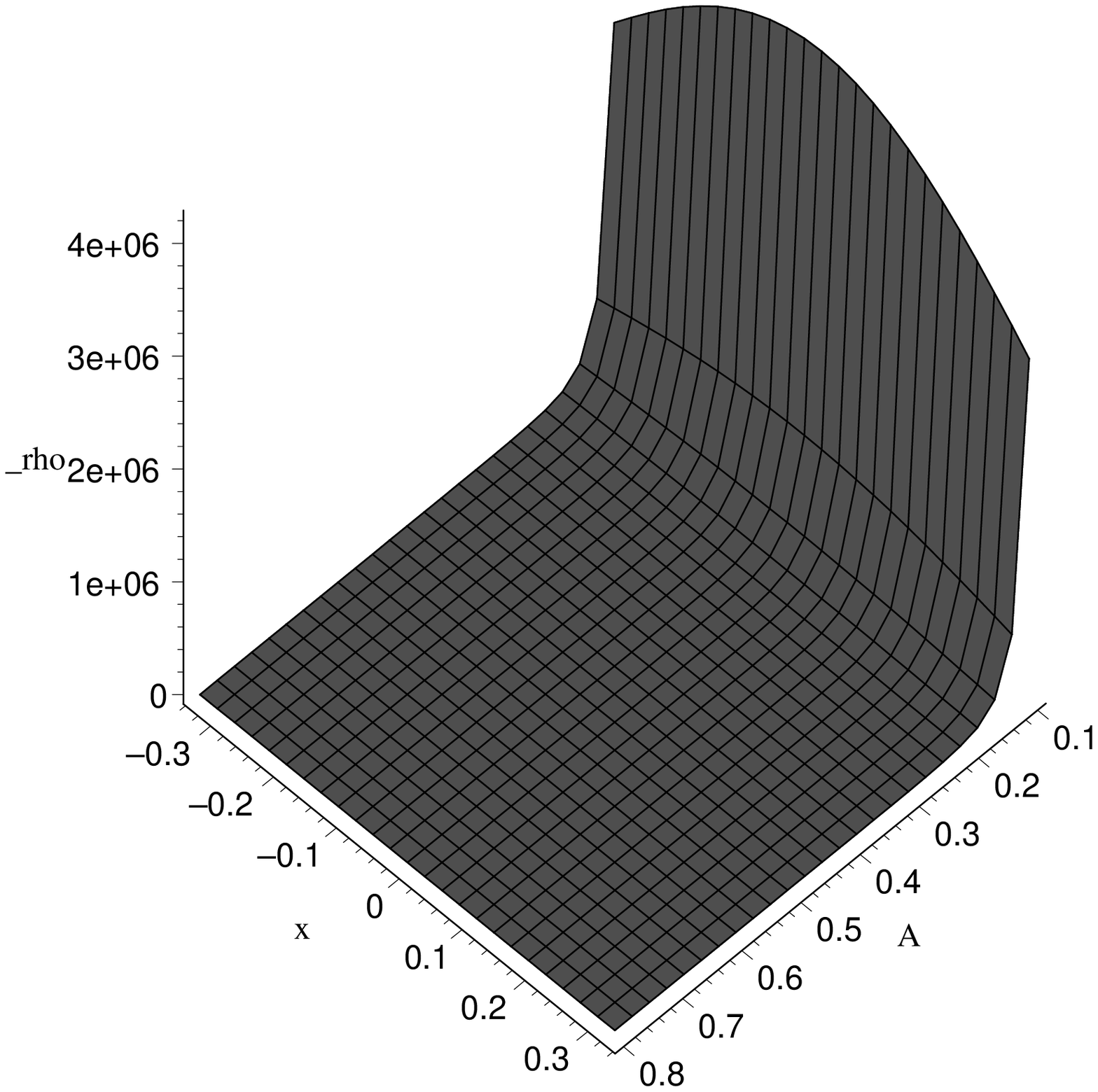}} &
\subfloat[$\tilde{\rho}+\tilde{p}_{r}$]{\includegraphics[width=42mm,height=42mm,clip]{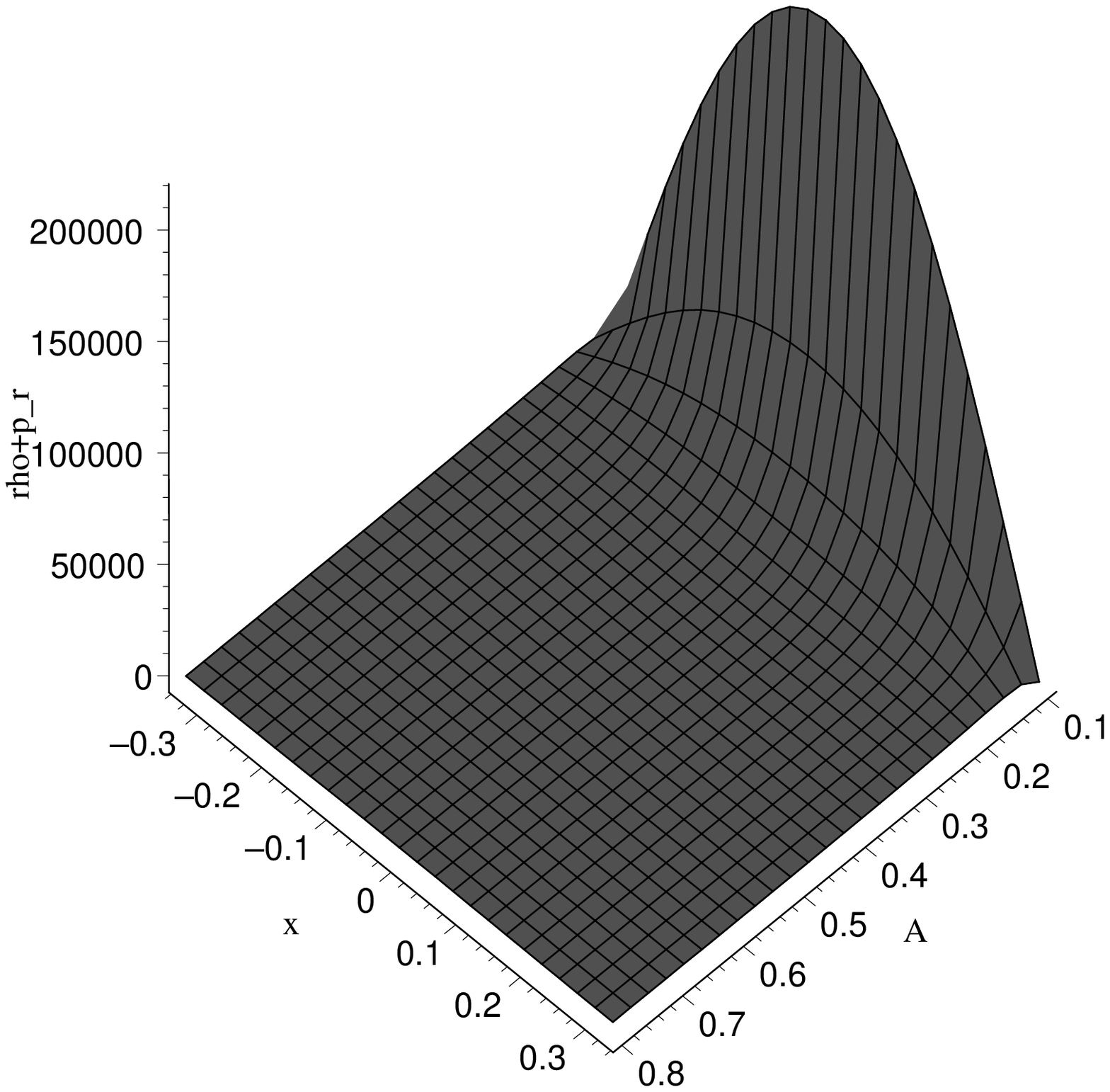}}&
\subfloat[$\tilde{\rho}+\tilde{p}_{t}$]{\includegraphics[width=42mm,height=42mm,clip]{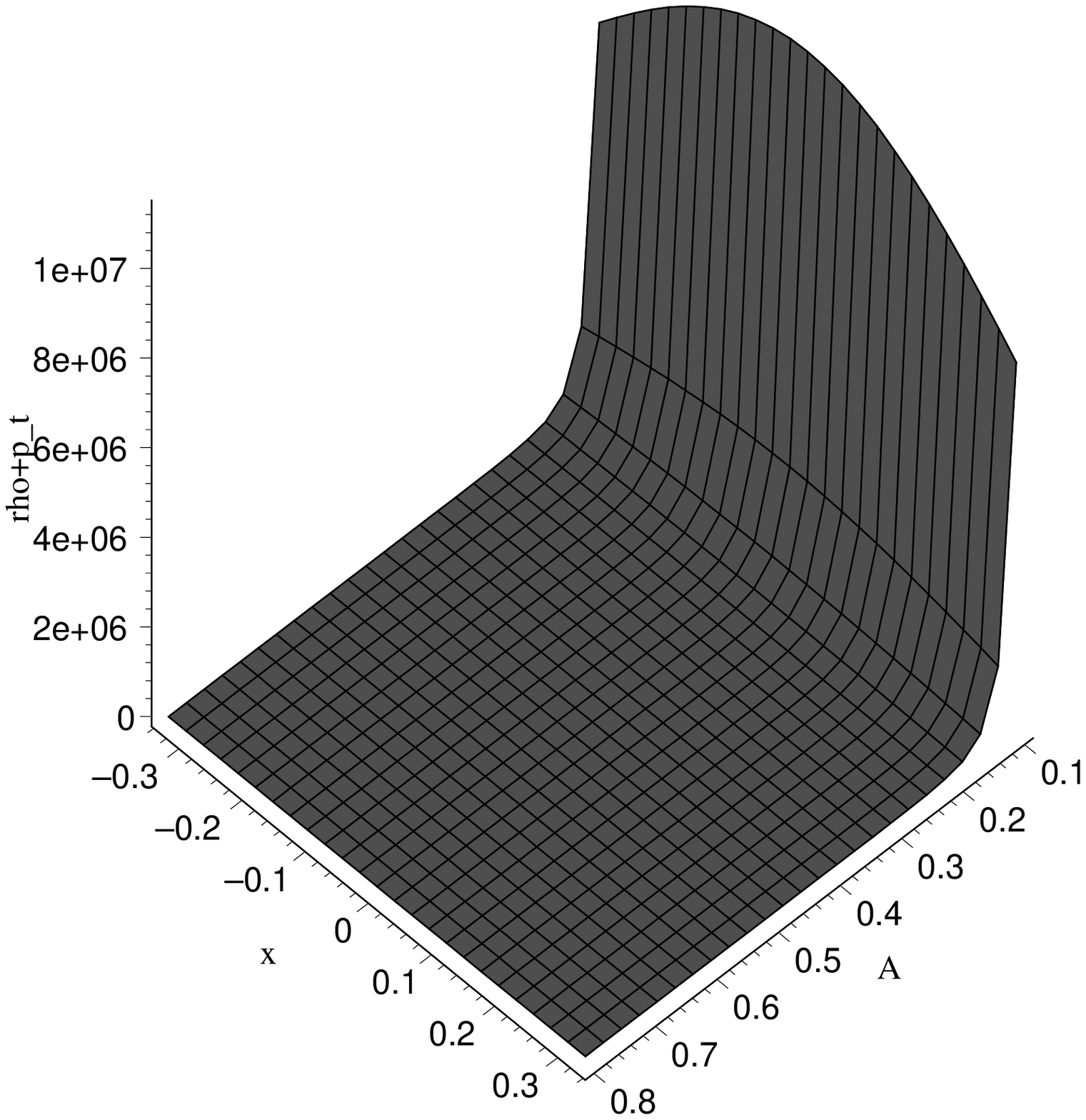}}
\end{tabular}
\end{center}
\vspace{-0.1cm}
\caption{\small{ $R^{3}$ contribution for $x_{0}=1$, $g_{tt}$ \emph{decreasing}, and varying throat radius. For negative $\alpha_{3}$ the graphs should be flipped around the plane $z=0$.}}
\label{fig:rcubedAconcdown}
\end{framed}
\end{figure}

\begin{figure}[!ht]
\begin{framed}
\begin{center}
\vspace{-0.5cm}
\begin{tabular}{cc}
\subfloat[$g_{tt}(x)$]{\includegraphics[width=45mm,height=45mm,clip]{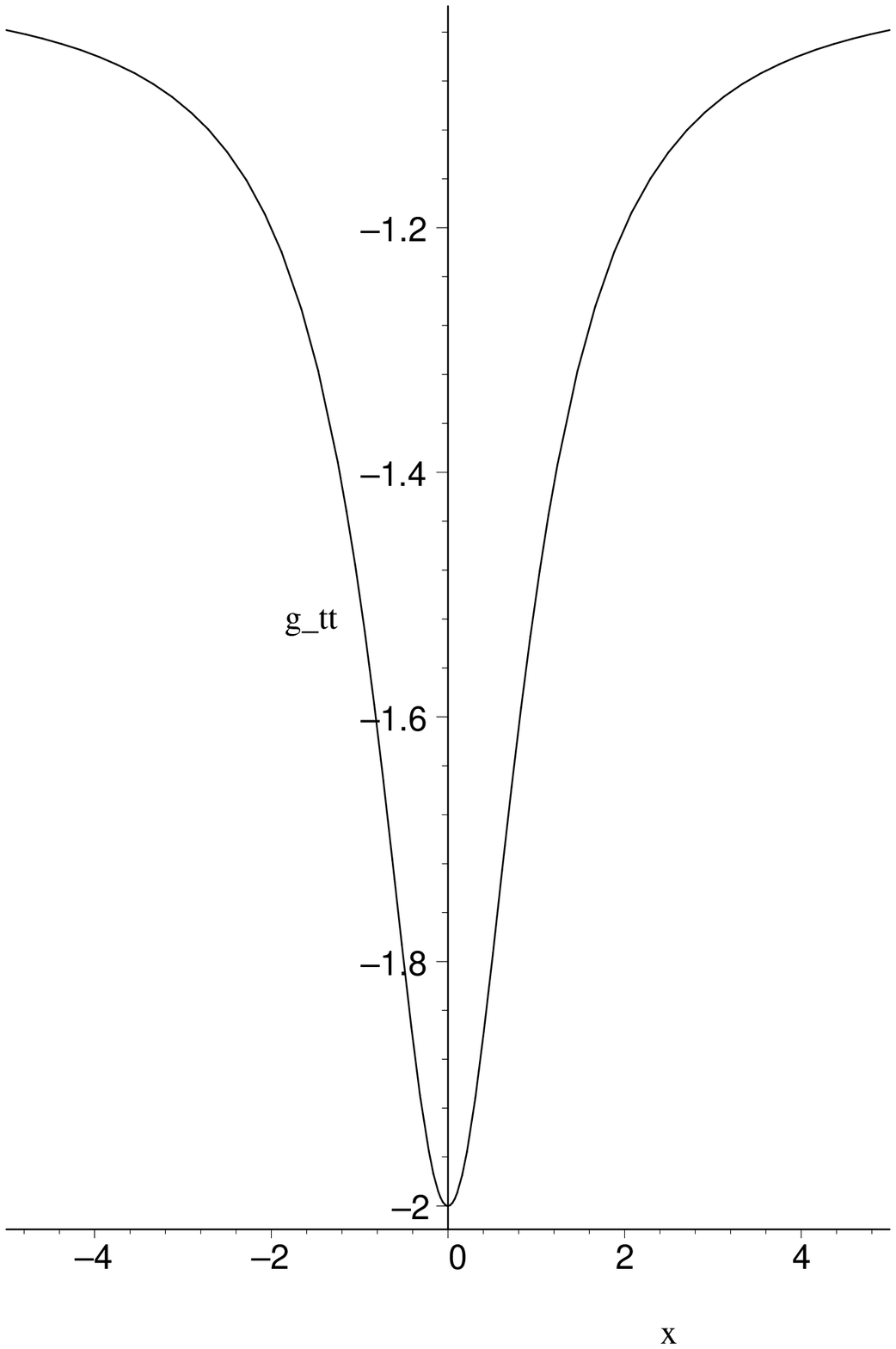}}&\hspace{0.5cm}
\subfloat[$\tilde{\rho}(x)$]{\includegraphics[width=45mm,height=45mm,clip]{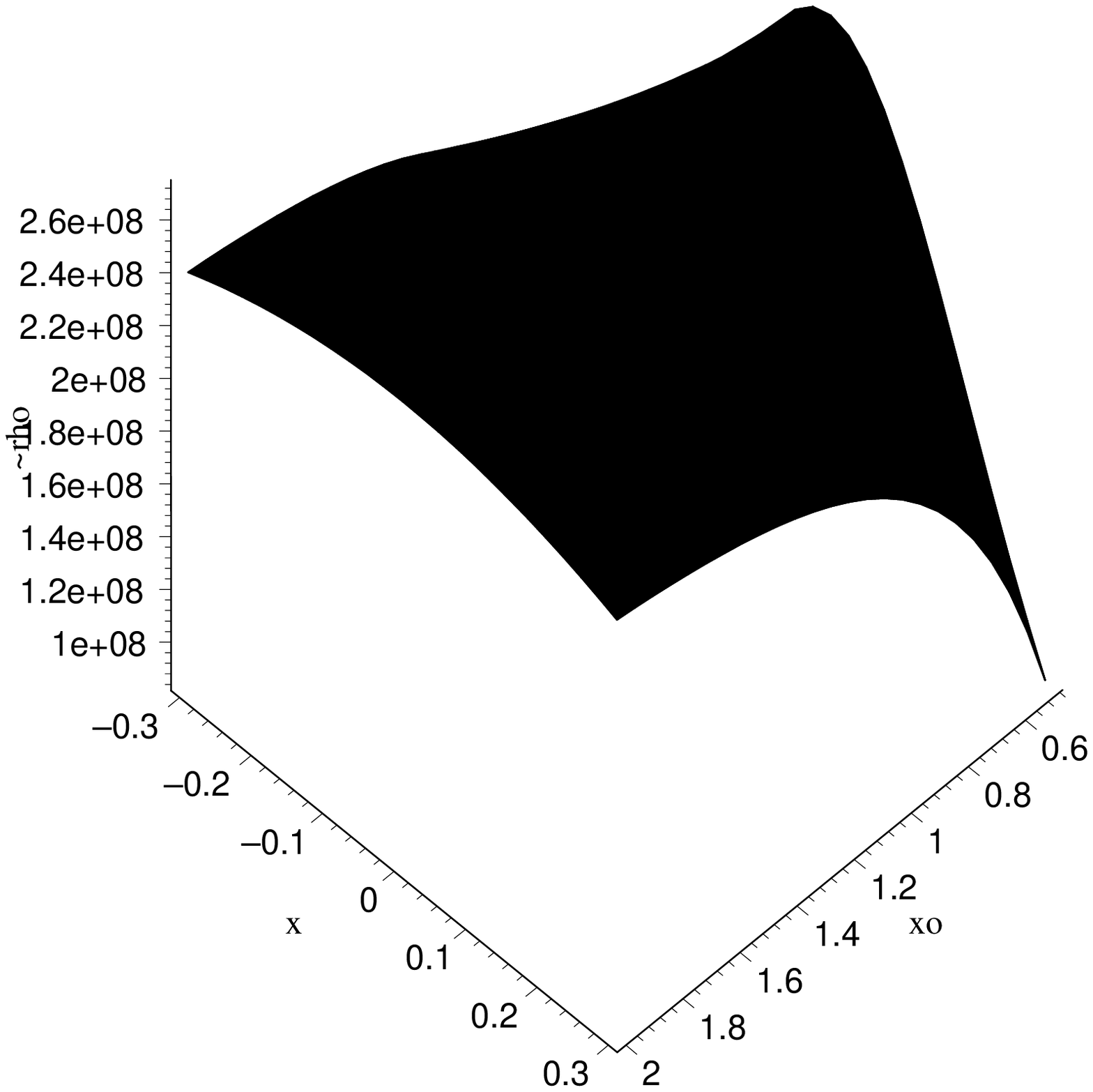}} \\
\subfloat[$\tilde{\rho}+\tilde{p}_{r}$]{\includegraphics[width=45mm,height=45mm,clip]{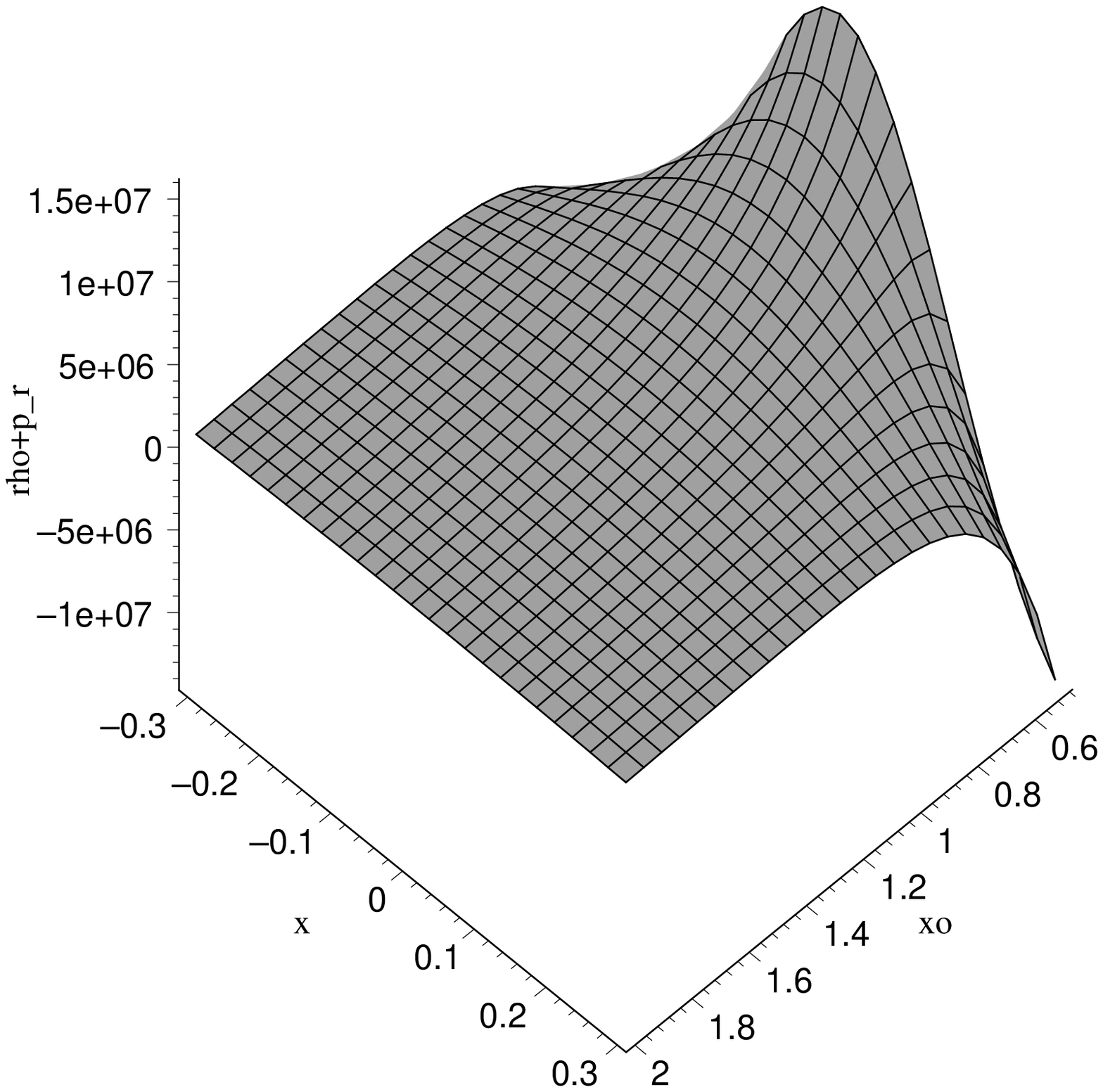}}&\hspace{0.5cm}
\subfloat[$\tilde{\rho}+\tilde{p}_{t}$]{\includegraphics[width=45mm,height=45mm,clip]{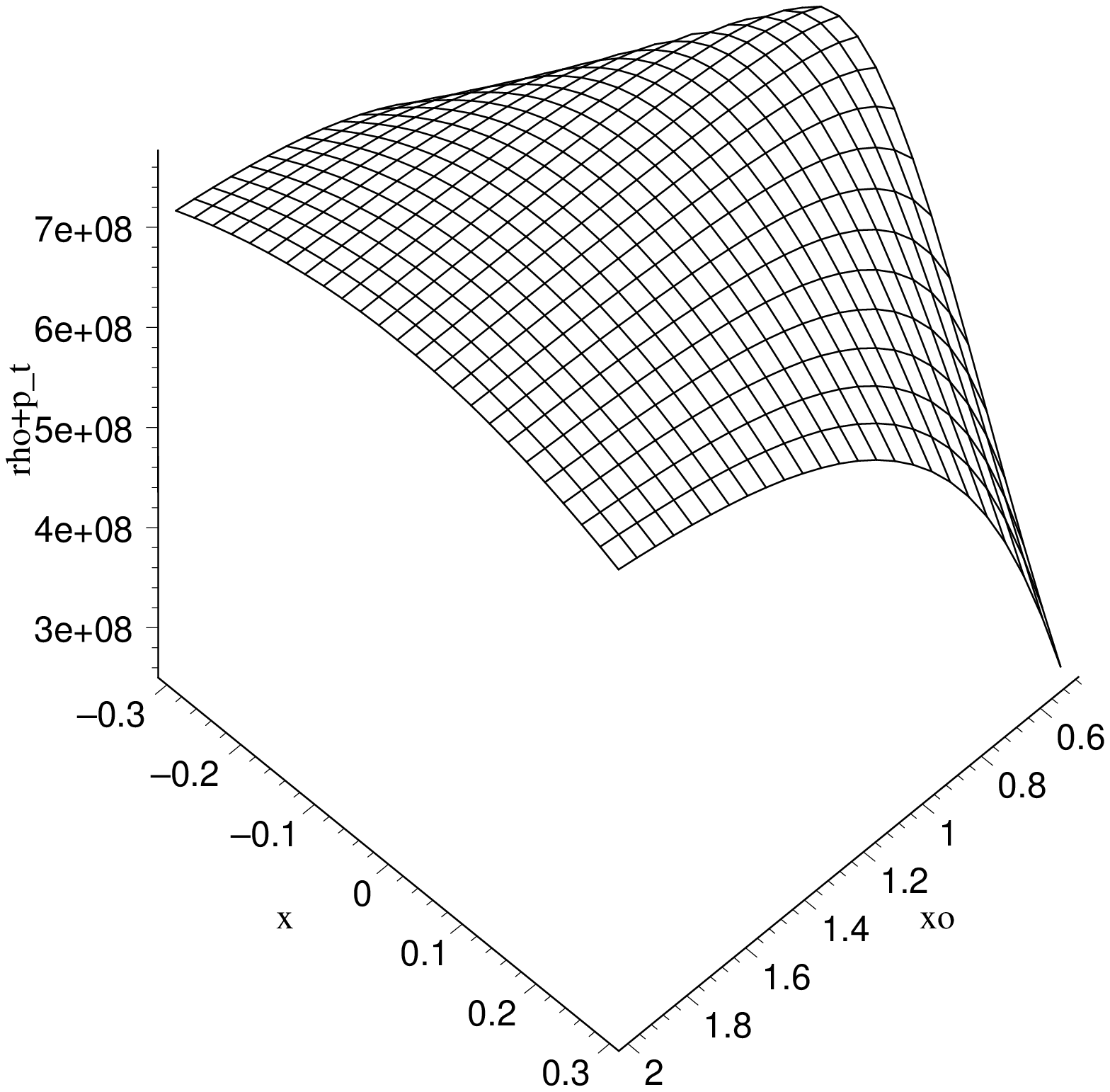}}
\end{tabular}
\end{center}
\vspace{-0.1cm}
\caption{\small{ $R^{3}$ contribution, $g_{tt}$ \emph{increasing}, throat radius=0.05, and varying $x_{0}$.}}
\label{fig:rcubedalphaconcup}
\end{framed}
\end{figure}

\begin{figure}[!ht]
\begin{framed}
\begin{center}
\vspace{0.0cm}
%\fbox{
\includegraphics[width=60mm,height=45mm, clip]{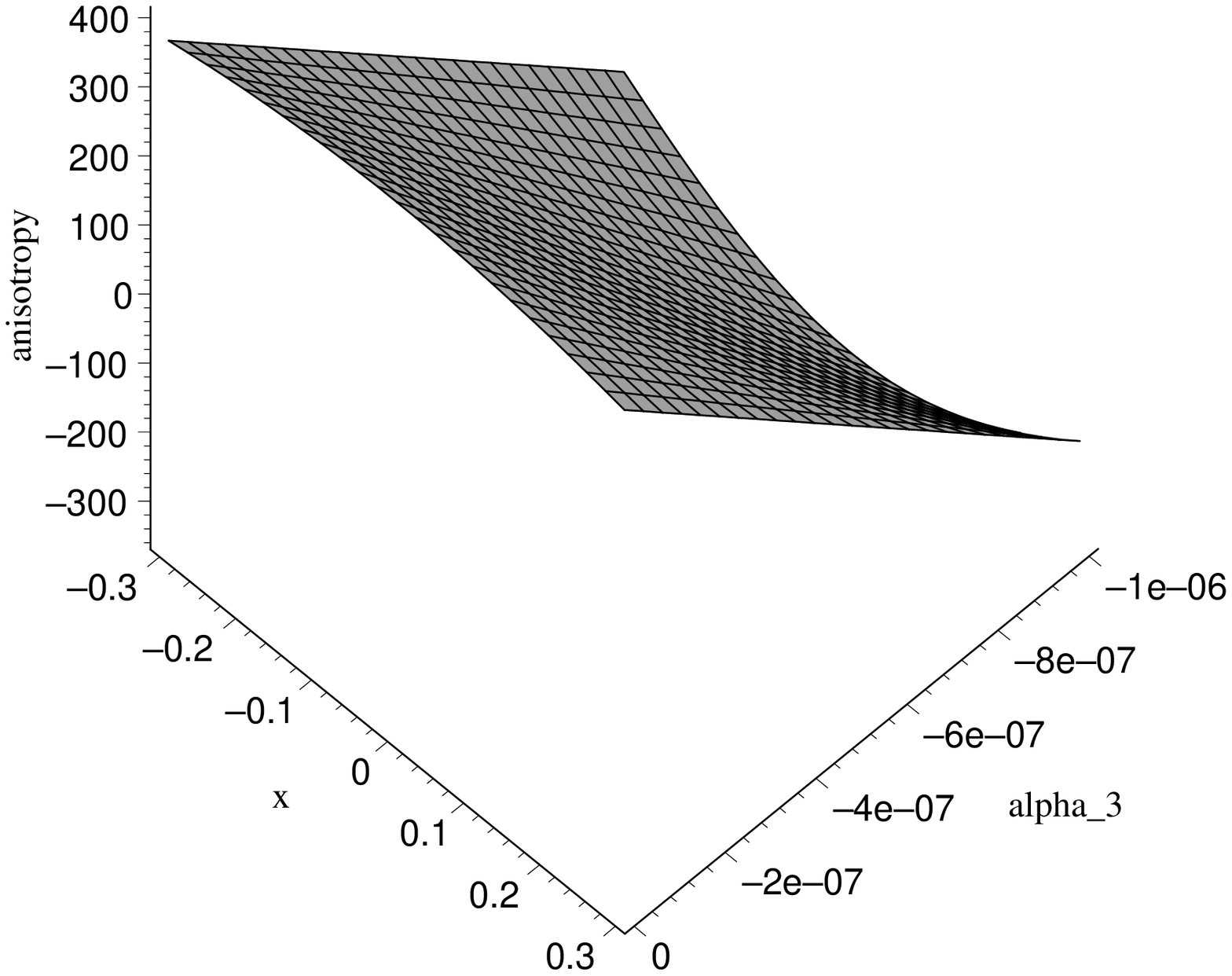}
%}
\end{center}\vspace{-0.1cm}
\caption{\small{Minimum anisotropy region: $R+\alpha_{3}R^{3}$ contribution, $g_{tt}$ \emph{increasing}, throat radius=0.05, $x_{0}=1$, and varying $\alpha_{3}$.}}
\label{fig:rcubedalphaconcupaniso}
\end{framed}
\end{figure}

\begin{figure}[!ht]
\begin{framed}
\begin{center}
\vspace{-0.0cm}
\begin{tabular}{ccc}
%%\subfloat[$g_{tt}(x)$]{\includegraphics[width=45mm,height=45mm,clip]{squared_gtt_c_down.eps}}&
\subfloat[$\tilde{\rho}(x)$]{\includegraphics[width=42mm,height=42mm,clip]{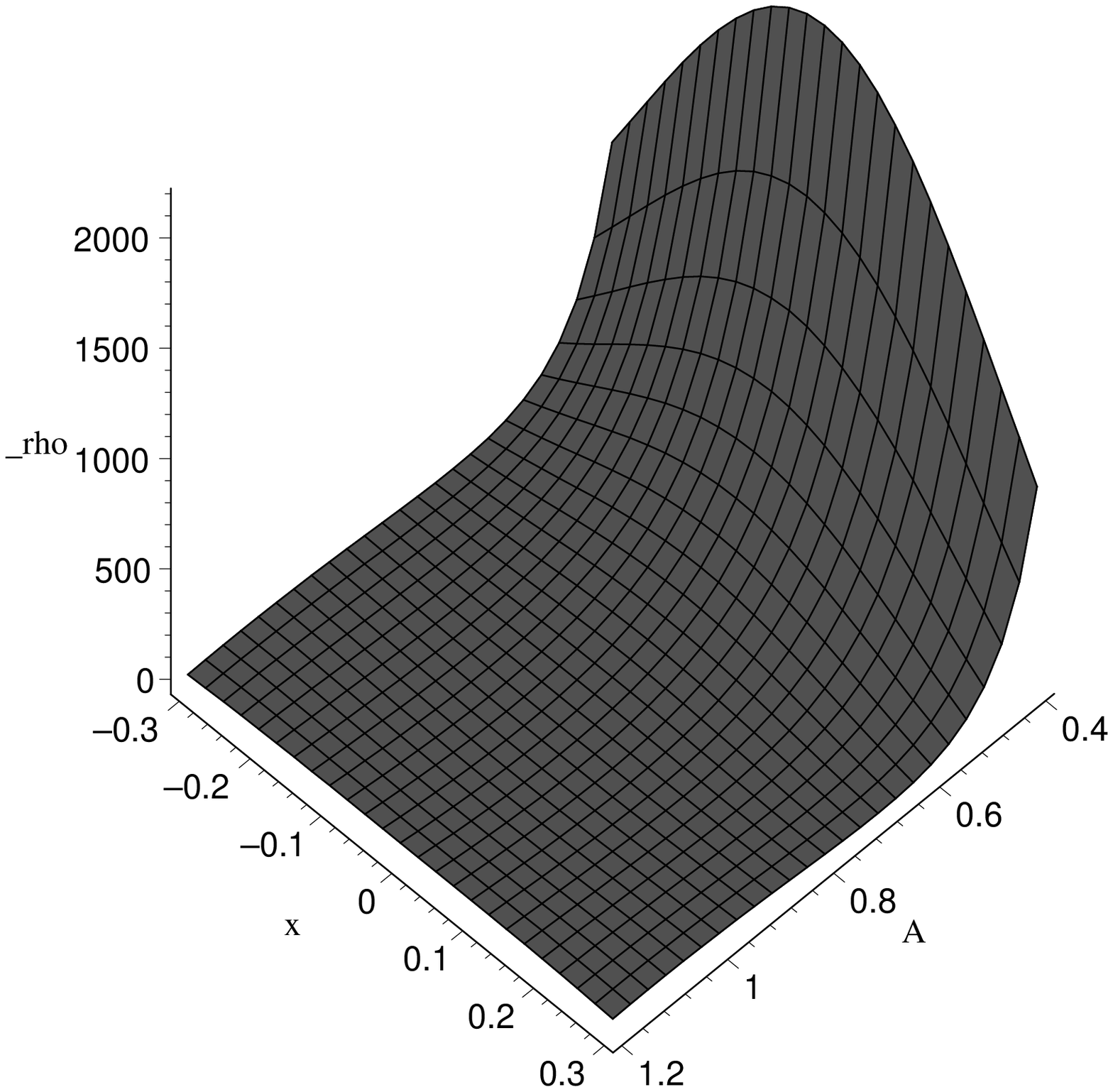}} &
\subfloat[$\tilde{\rho}+\tilde{p}_{r}$]{\includegraphics[width=42mm,height=42mm,clip]{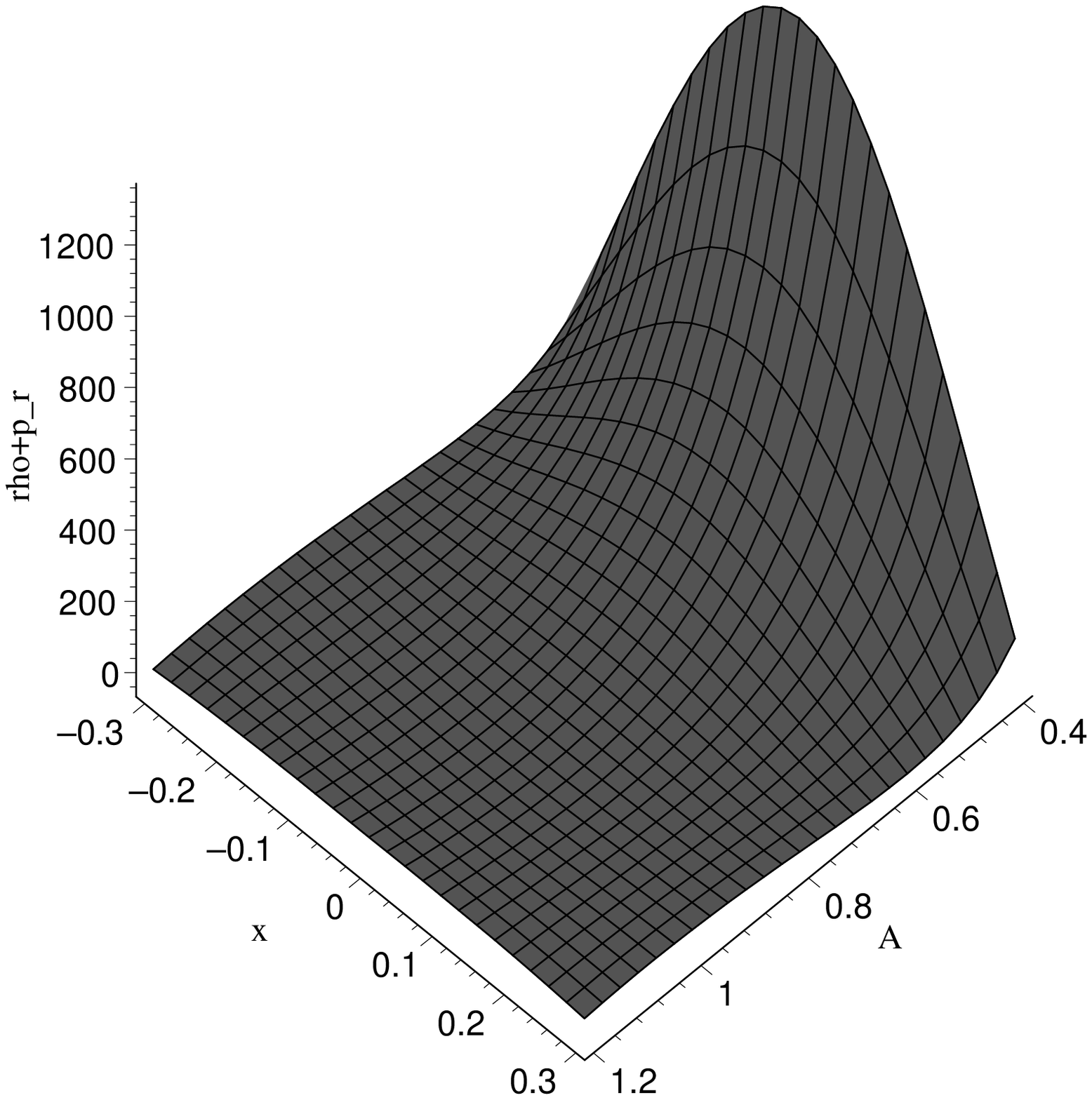}}&
\subfloat[$\tilde{\rho}+\tilde{p}_{t}$]{\includegraphics[width=42mm,height=42mm,clip]{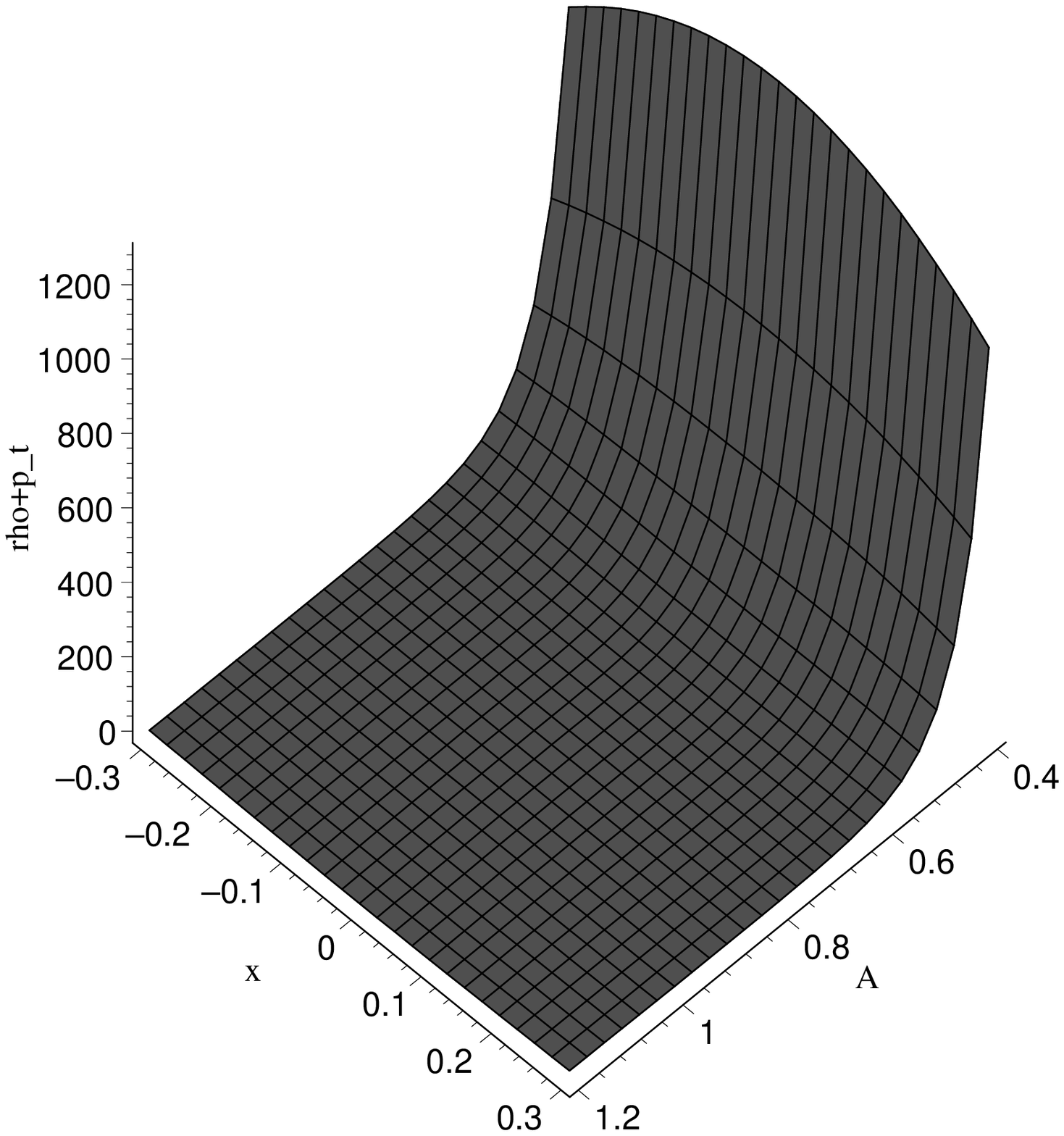}}
\end{tabular}
\end{center}
\vspace{-0.1cm}
\caption{\small{ $R^{3}$ contribution for $x_{0}=1$, $g_{tt}$ \emph{increasing}, and varying throat radius. For negative $\alpha_{3}$ the graphs should be flipped around the plane $z=0$.}}
\label{fig:rcubedAconcup}
\end{framed}
\end{figure}

\paragraph{{\small Non-zero tidal force:}} 
The analysis of the non-zero tidal forces proceeds in a similar order as the $n=2$ case. Due to the interest in keeping the paper of reasonable length, we present all the graphs (figs.~\ref{fig:rcubedalphaconcdown} - \ref{fig:rcubedAconcup}) for this case first, and then briefly comment on the results afterward in table~1. (Please refer to figure captions for details.)

\vspace{-1.4cm}

\vspace{4cm}
\FloatBarrier
We summarize the results for $n=3$ in table~1.

\begin{table}[!ht]
\newcommand\Ta{\rule{0pt}{2.6ex}}
\newcommand\Bb{\rule[-1.2ex]{0pt}{0pt}}
\begin{center}
\caption{\small{Summary of $\alpha_{3}R^{3}$ contributions.}\vspace{-0.2cm}} 
\begin{tabular}{|c|c|c|c|}
\hline
\Ta  \parbox[t]{2.5cm}{Parameter studied}& \parbox[t]{3.8cm}{zero tidal-force}  &  \parbox[t]{3.8cm}{$g_{tt}(x)$ concave-down} & \parbox[t]{3.8cm}{$g_{tt}(x)$ concave-up} \\[0.5cm]
\hline\hline

\parbox[t]{2.5cm}{$x_{0}$} & \parbox[t]{3.8cm}{{\small Possible to make energy conditions positive.}}\Ta & \parbox[t]{3.8cm}{{\small Tendency to make positive contribution to all energy conditions for all range of $x_{0}$ studied near the throat for positive $\alpha_{3}$, and negative contribution for negative $\alpha_{3}$.}} & \parbox[t]{3.8cm}{{\small Tendency to make positive contribution to all energy conditions for positive $\alpha_{3}$, and negative contribution for negative $\alpha_{3}$. Shows insensitivity to form of $\gtt$.}}  \\[0.07cm]
\hline
\parbox[t]{2.5cm}{Throat radius} & \parbox[t]{3.8cm}{{\small Possible to make energy conditions positive.}}\Ta & \parbox[t]{3.8cm}{{\small Energy conditions more positive for small throat radius (with $\alpha_{3}>0$. Reverse reported results for $\alpha_{3}<0$).}} & \parbox[t]{3.8cm}{{\small Energy conditions more positive for small throat radius (with $\alpha_{3}>0$. Reverse reported results for $\alpha_{3}<0$).}}  \\[0.07cm]
\hline
\parbox[t]{2.5cm}{Minimum anisotropy for $R+\alpha_{3}R^{3}$} &\parbox[t]{3.8cm}{{\small Not studied.}}\Ta & \parbox[t]{3.8cm}{{\small $\alpha_{3} < 0$ minimizes anisotropy.}} & \parbox[t]{3.8cm}{{\small $\alpha_{3} < 0$ minimizes anisotropy.}} \\[0.07cm]
\hline
\end{tabular}
\end{center}
\end{table}
\FloatBarrier

\subsubsection{{\normalsize $n=-1$}}

\paragraph{{\small Zero tidal force:}} 
Again we begin with an analytic analysis of the zero tidal-force model. The results are summarized as:
\begin{subequations}\romansubs
{\allowdisplaybreaks\begin{align}
&\tilde{\rho}=\frac{\alpha_{-1}\qzero^{2}}{4\left(2\qzero\qzeropp-1\right)^{4}} \left[1-2\qzero\qzeropp + 4\qzero^{3}\qzeropppp +24 \qzero^{4} (\qzeroppp)^2 +4\qzero^{2}(\qzeropp)^{2}\right. \nonumber \\
&\qquad\quad\quad \left. -24 \qzero^{3}(\qzeropp)^{3} -8\qzero^{4}\qzeropppp\qzeropp +32\qzero^{4}(\qzeropp)^{4}\right] + \mathcal{O}(x)\,, \label{eq:minusonezerorho}\\[0.2cm]
&\rhot+\pr=\frac{\alpha_{-1}\qzero^{3}}{2\left(2\qzero\qzeropp-1\right)^{4}} \left[3\qzeropp+2\qzero^{2}\qzeropppp +12 \qzero^{3}(\qzeroppp)^{2} -8\qzero(\qzeropp)^{2} \right. \nonumber \\
&\qquad\quad\quad \left. -4\qzero^{2}(\qzeropp)^{3} -4\qzero^{3}\qzeropp\qzeropppp + 16 \qzero^{3}(\qzeropp)^{4}  \right] + \mathcal{O}(x)\,,  \label{eq:minusonezeroecond1}\\[0.2cm]
&\rhot+\pt= \frac{\alpha_{-1}\qzero^{2}}{4\left(2\qzero\qzeropp-1\right)^{2}}\left[\qzero\qzeropp-1\right] + \mathcal{O}(x)\,.\label{eq:minusonezeroecond2}
\end{align}}
\end{subequations}
Here it can be seen that with inverse powers of $R$ a peculiar singularity occurs. The singularity occurs when $2QQ^{\prime\prime}=1$ and corresponds to the vanishing of the Ricci scalar at these points. This is not unexpected and it is not a curvature singularity, but does herald a problem with the equations of motion when $R=0$. One common way to attempt remedy this situation in studies of inverse-$R$ gravity is to postulate that the (gravitational) vacuum state of the theory is not Minkowski space-time, but is instead deSitter or anti-deSitter space-time \cite{ref:carroll}. However, when one is not far away from sources (as in the study here) there can be curves or surfaces in the space-time on which $R=0$ and these must be excluded. Hence the parameter space studied here will not include parameters near these pathologies, which we excise. 

Here the vanishing of $\qzeropp$ does not yield an energy condition respecting throat, unlike in the positive $n$ cases. (Either $\rhot+\pt <0$ for $\alpha_{-1}>0$ or else $\rhot+\pr$ and $\rhot<0$ for $\alpha_{-1}<0$.) Note that in principle it may be possible to respect energy conditions near the throat. One can say that if $\qzero\qzeropp >1$ and furthermore if $\qzeropppp$ is large then it may be possible to respect energy conditions, but the situation is not obvious. Hence here the numerical results are needed, which we present for the more general non-zero tidal force scenarios.

\paragraph{{\small Non-zero tidal force:}} 
If one is considering the region far from the throat, then $g_{tt}(x)$ should, as mentioned above, asymptote to deSitter or anti-deSitter space-time. Our choices for $g_{tt}(x)$ can accommodate this, but we are only interested in the near-throat region, so strictly speaking it is not a requirement. 

\begin{figure}[!ht]
\begin{framed}
\begin{center}
\vspace{-0,5cm}
\begin{tabular}{cc}
\subfloat[$g_{tt}(x)$]{\includegraphics[width=45mm,height=45mm,clip]{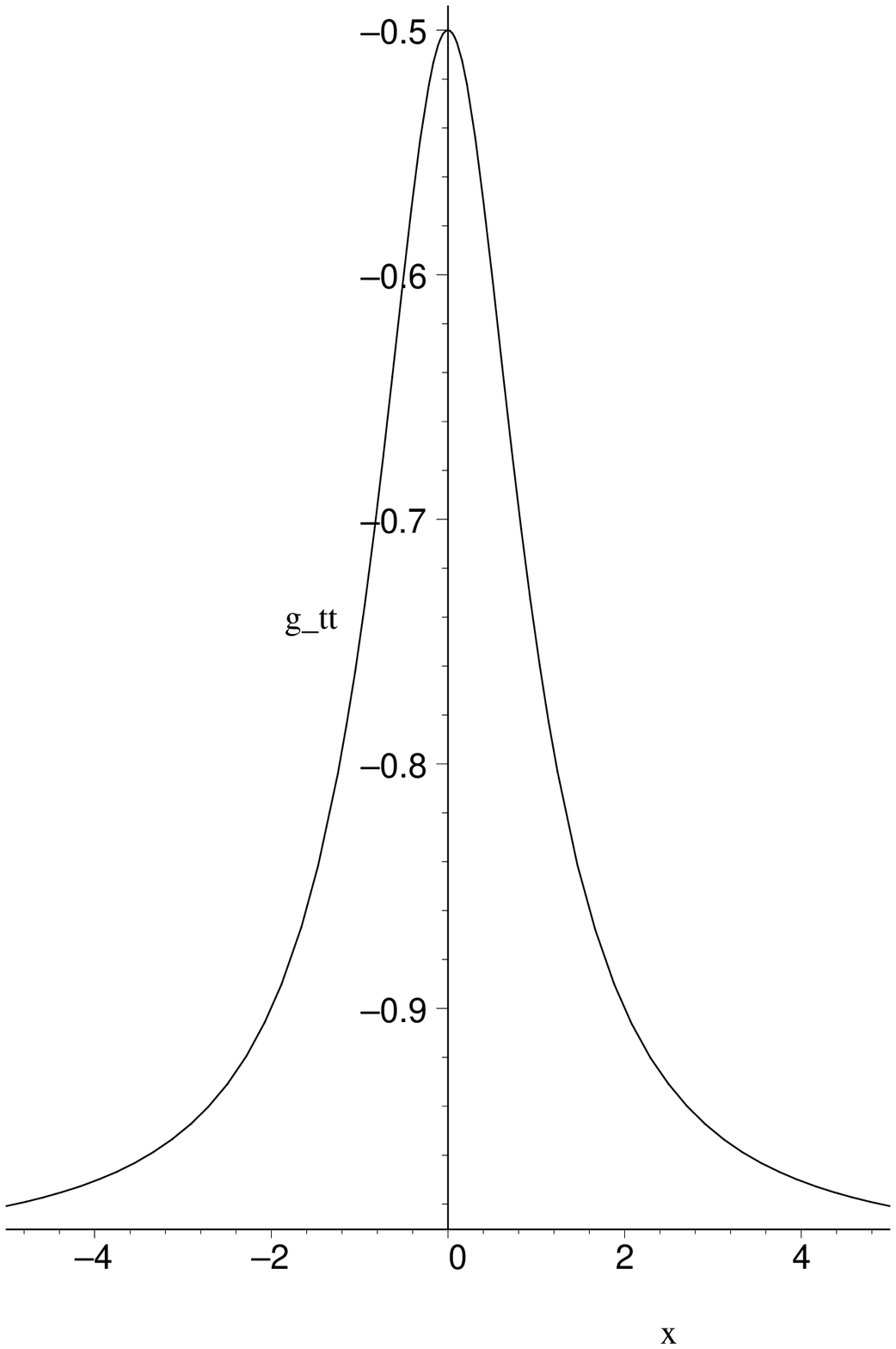}}&\hspace{0.5cm}
\subfloat[$\tilde{\rho}(x)$]{\includegraphics[width=45mm,height=45mm,clip]{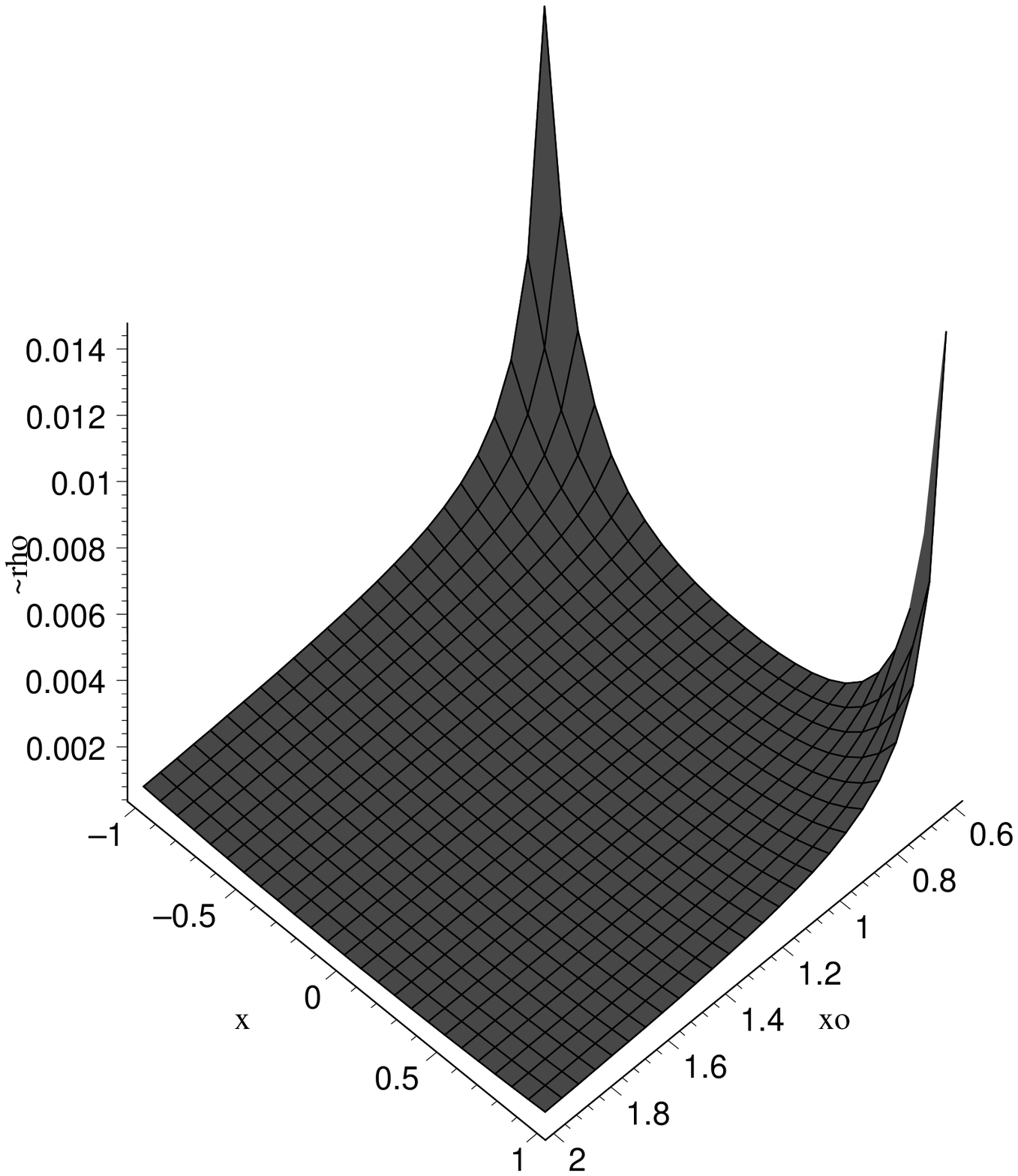}} \\
\subfloat[$\tilde{\rho}+\tilde{p}_{r}$]{\includegraphics[width=45mm,height=45mm,clip]{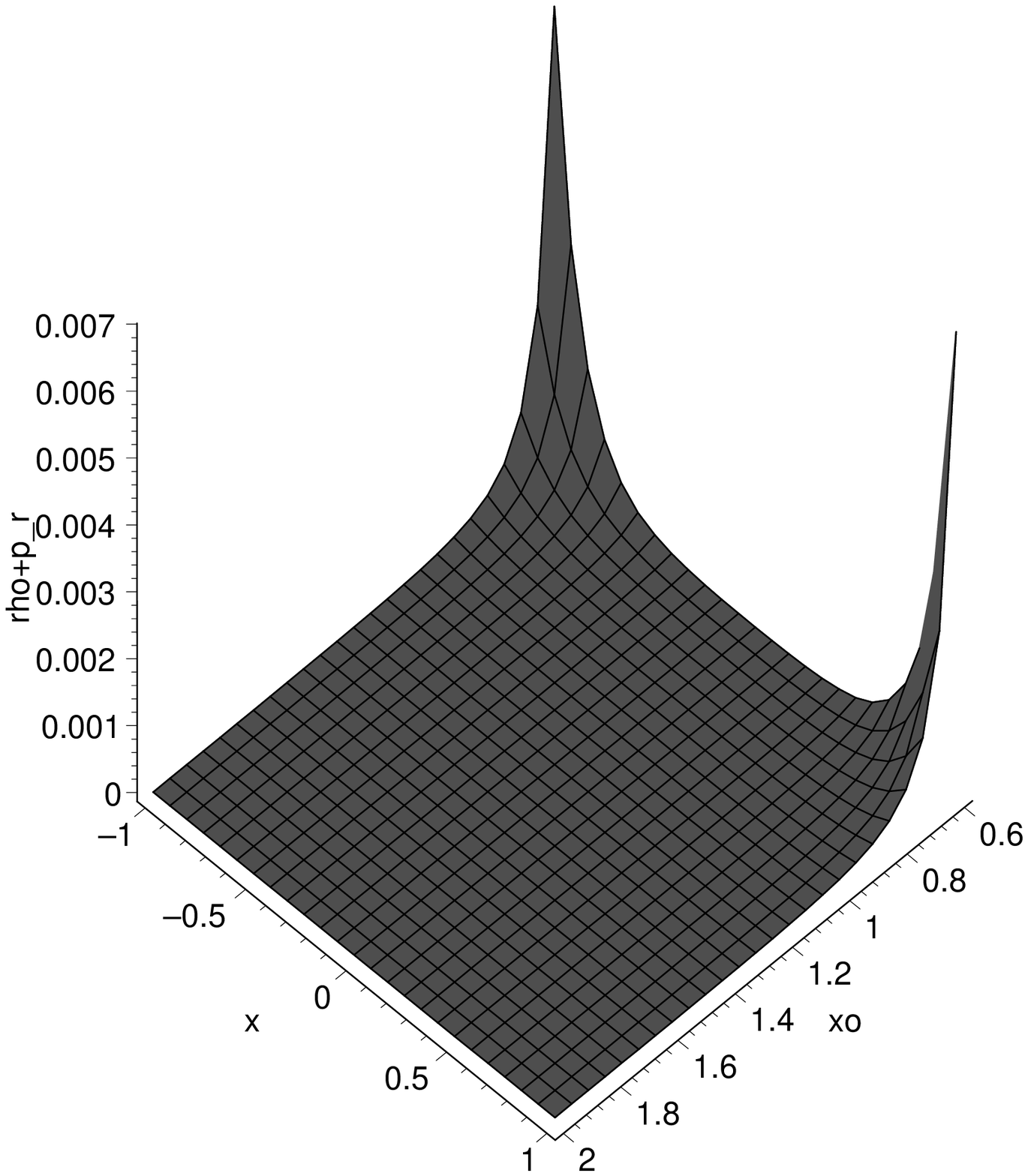}}&\hspace{0.5cm}
\subfloat[$\tilde{\rho}+\tilde{p}_{t}$]{\includegraphics[width=45mm,height=45mm,clip]{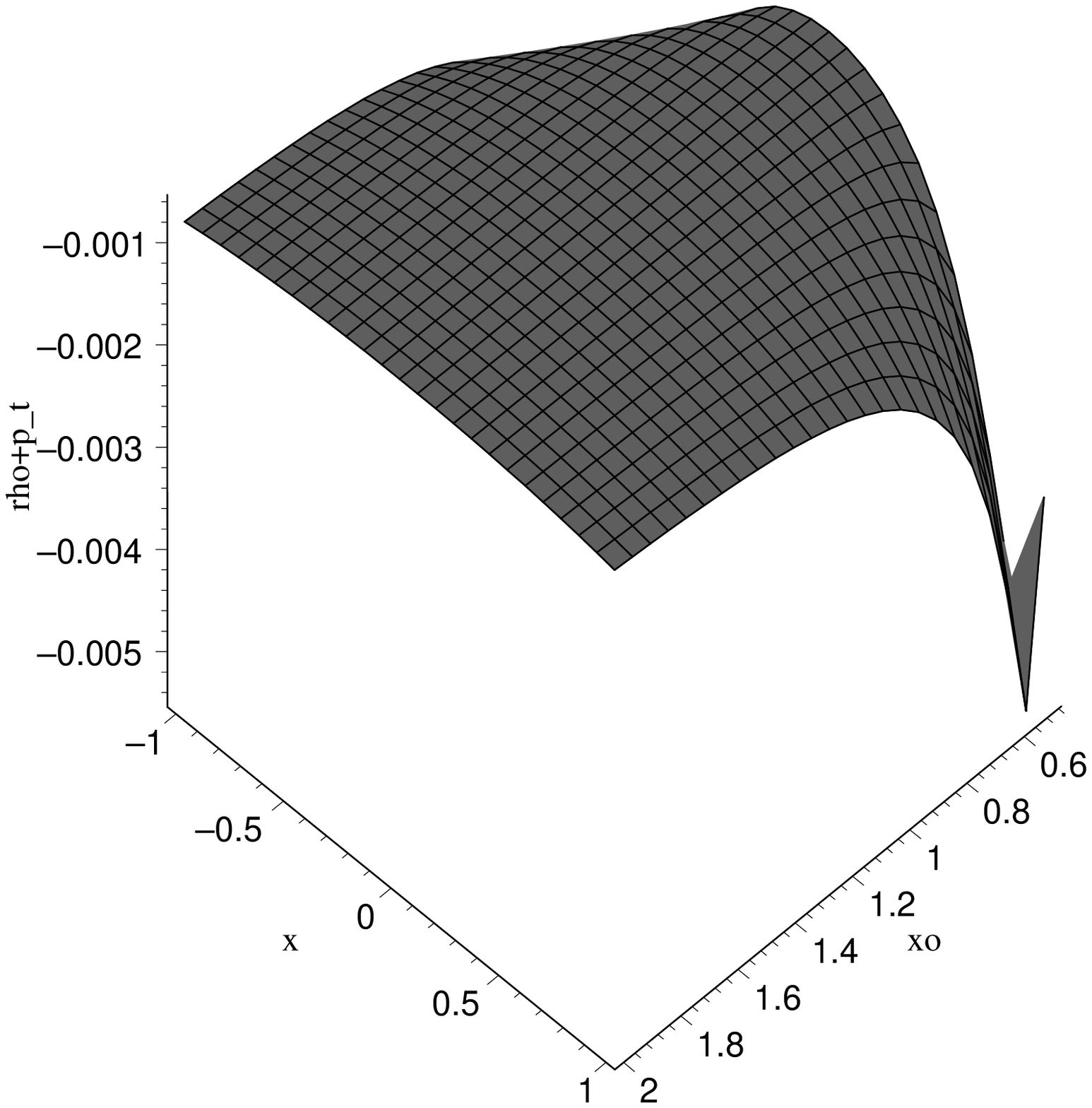}}
\end{tabular}
\end{center}
\vspace{-0.1cm}
\caption{\small{$R^{-1}$ contribution, $g_{tt}$ \emph{decreasing}, throat radius=0.05, and varying $x_{0}$.}}
\label{fig:rminusonealphaconcdown}
\end{framed}
\end{figure}

\begin{figure}[!ht]
\begin{framed}
\begin{center}
\vspace{0.0cm}
%\fbox{
\includegraphics[width=60mm,height=45mm, clip]{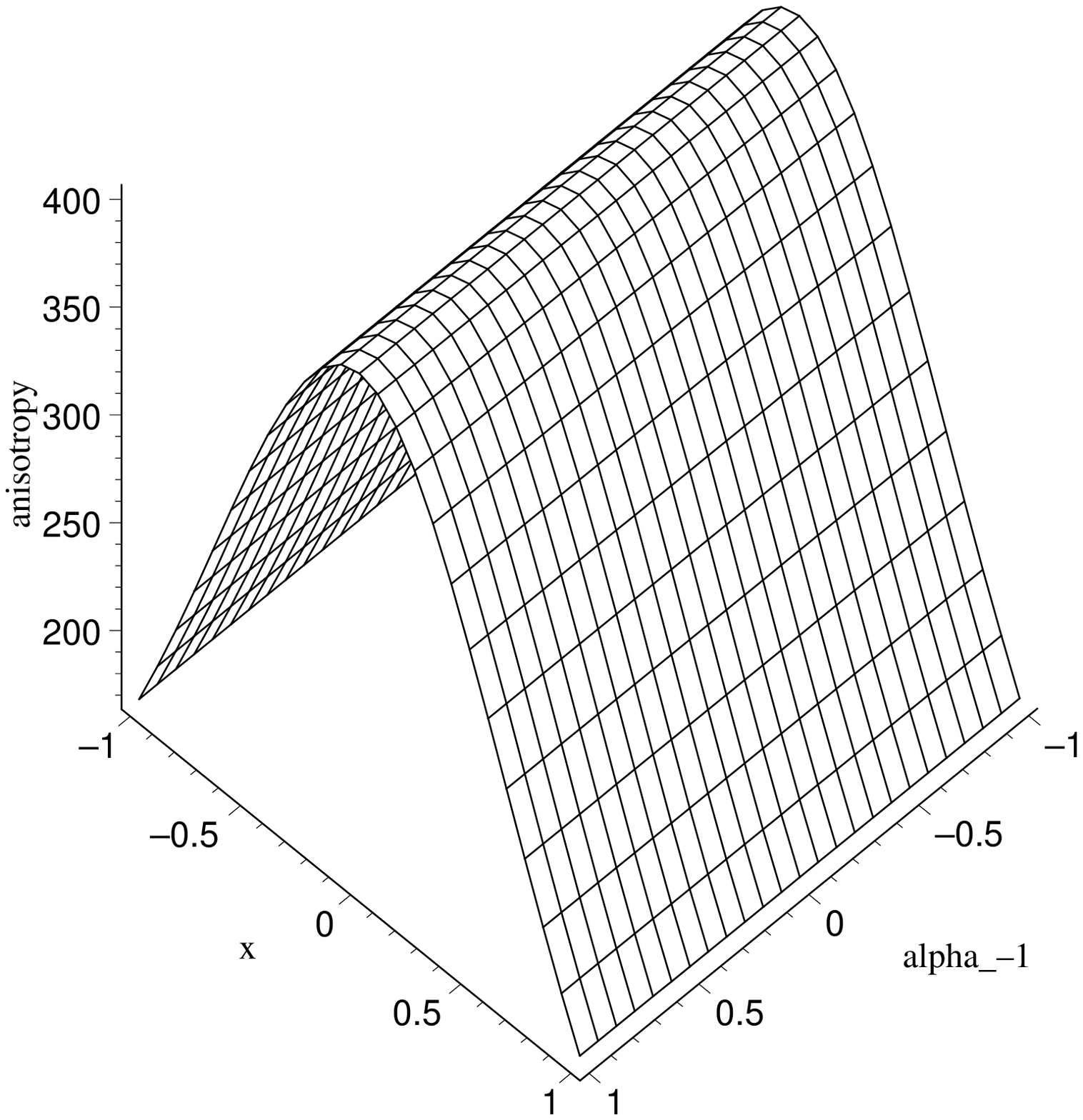}
%}
\end{center}\vspace{-0.1cm}
\caption{\small{Anisotropy of throat region: $R+\alpha_{-1}R^{-1}$ contribution, $g_{tt}$ \emph{decreasing}, throat radius=0.05, $x_{0}=1$, and varying $\alpha_{-1}$.}}
\label{fig:rminusonealphaconcdownaniso}
\end{framed}
\end{figure}

\begin{figure}[!ht]
\begin{framed}
\begin{center}
\vspace{-0.5cm}
\begin{tabular}{ccc}
%%\subfloat[$g_{tt}(x)$]{\includegraphics[width=45mm,height=45mm,clip]{squared_gtt_c_down.eps}}&
\subfloat[$\tilde{\rho}(x)$]{\includegraphics[width=42mm,height=42mm,clip]{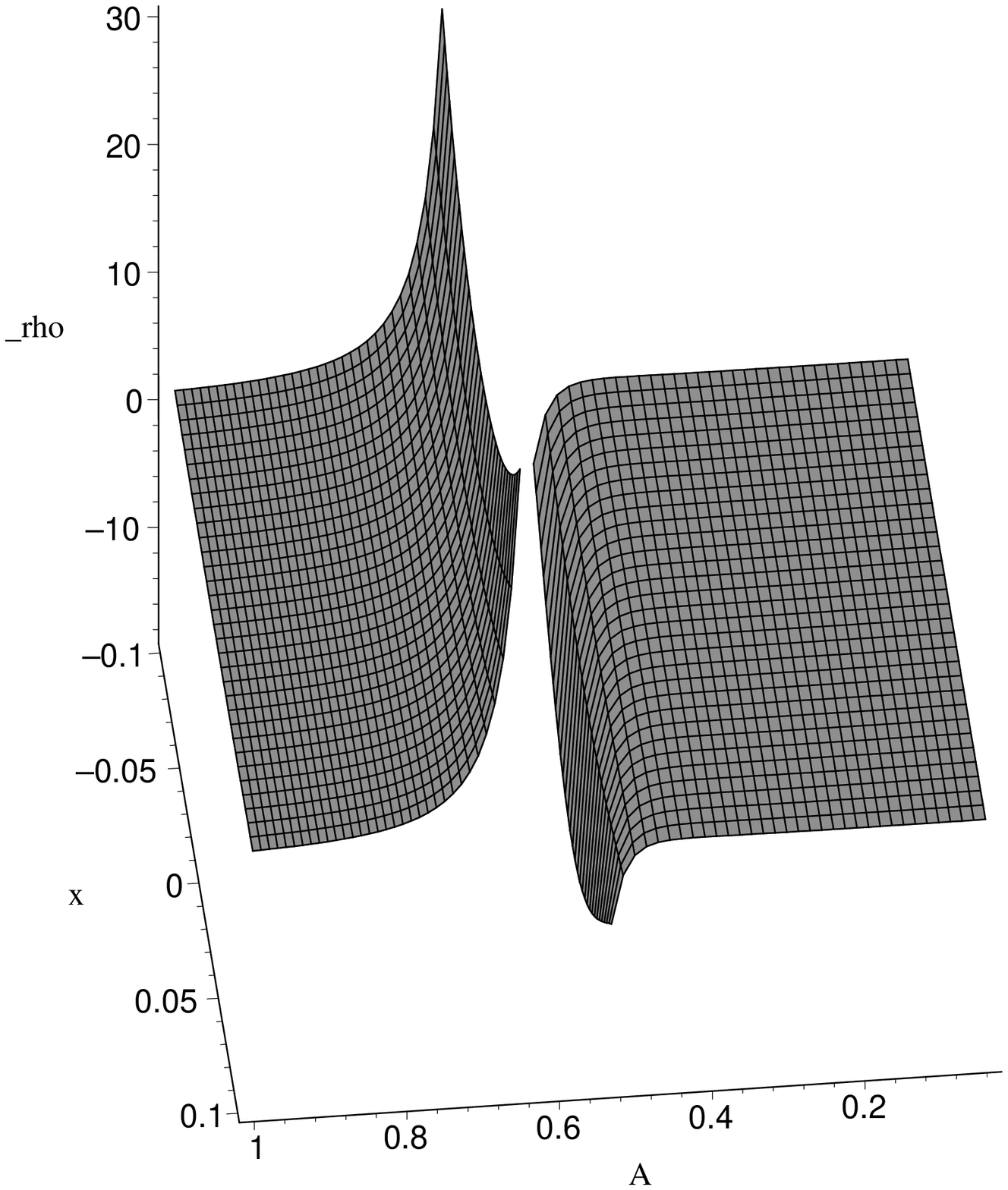}} &
\subfloat[$\tilde{\rho}+\tilde{p}_{r}$]{\includegraphics[width=42mm,height=42mm,clip]{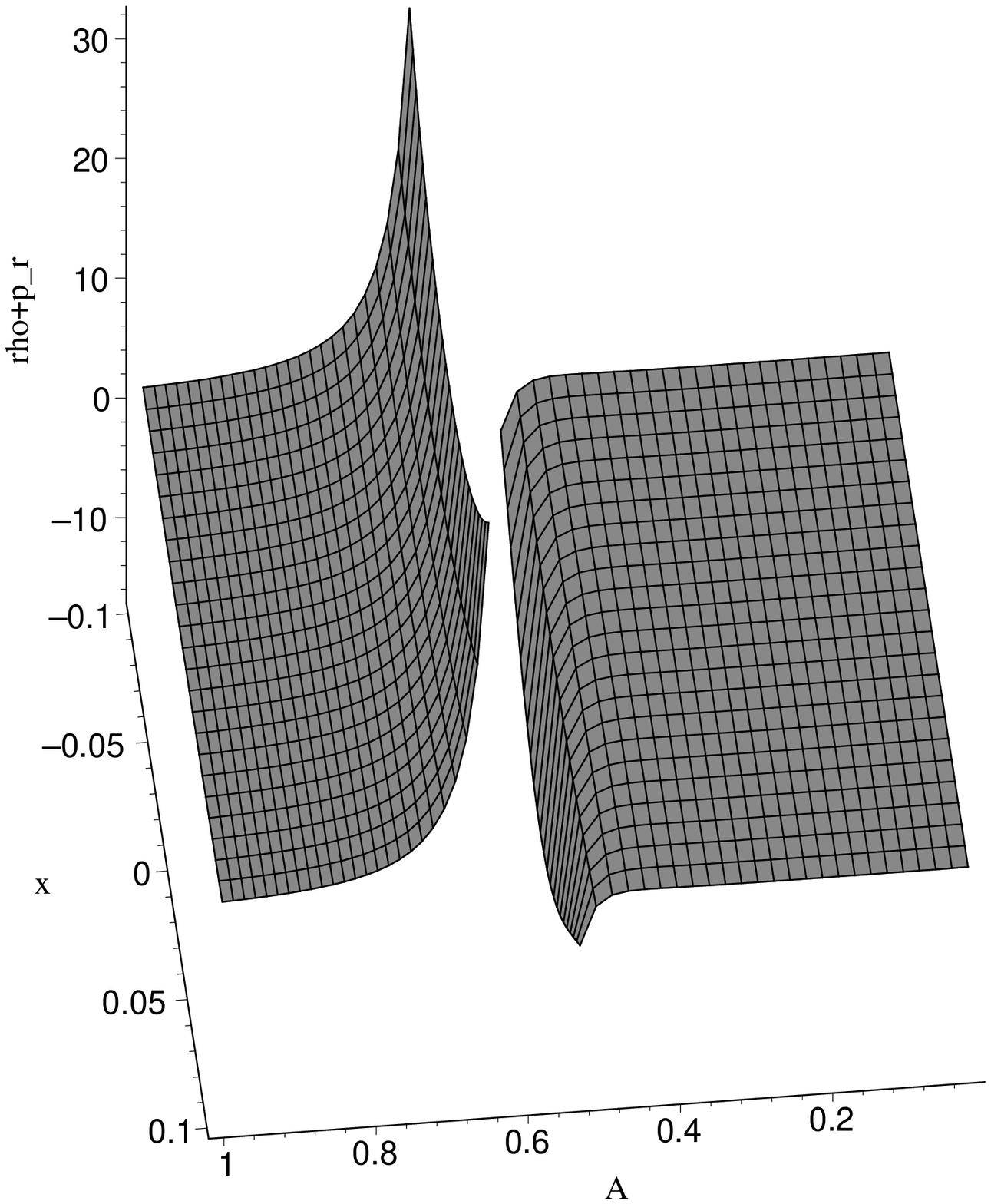}}&
\subfloat[$\tilde{\rho}+\tilde{p}_{t}$]{\includegraphics[width=42mm,height=42mm,clip]{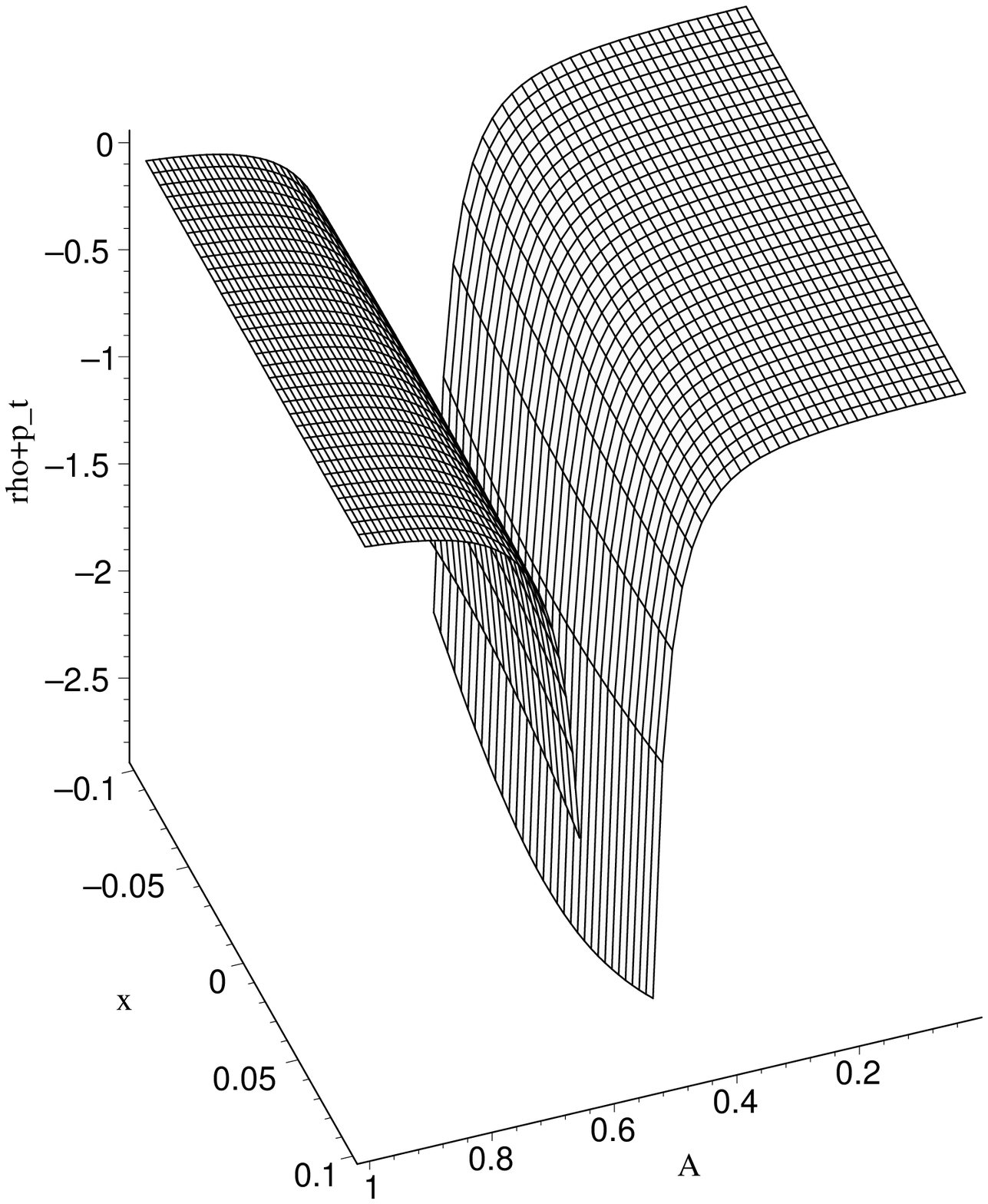}}
\end{tabular}
\end{center}
\vspace{-0.1cm}
\caption{\small{$R^{-1}$ contribution for $x_{0}=1$, $g_{tt}$ \emph{decreasing}, and varying throat radius. For negative $\alpha_{-1}$ the graphs should be flipped around the plane $z=0$. The region around the singularity has been omitted.}}
\label{fig:rminusoneAconcdown}
\end{framed}
\end{figure}

\begin{figure}[!ht]
\begin{framed}
\begin{center}
\vspace{-0.5cm}
\begin{tabular}{cc}
\subfloat[$g_{tt}(x)$]{\includegraphics[width=45mm,height=45mm,clip]{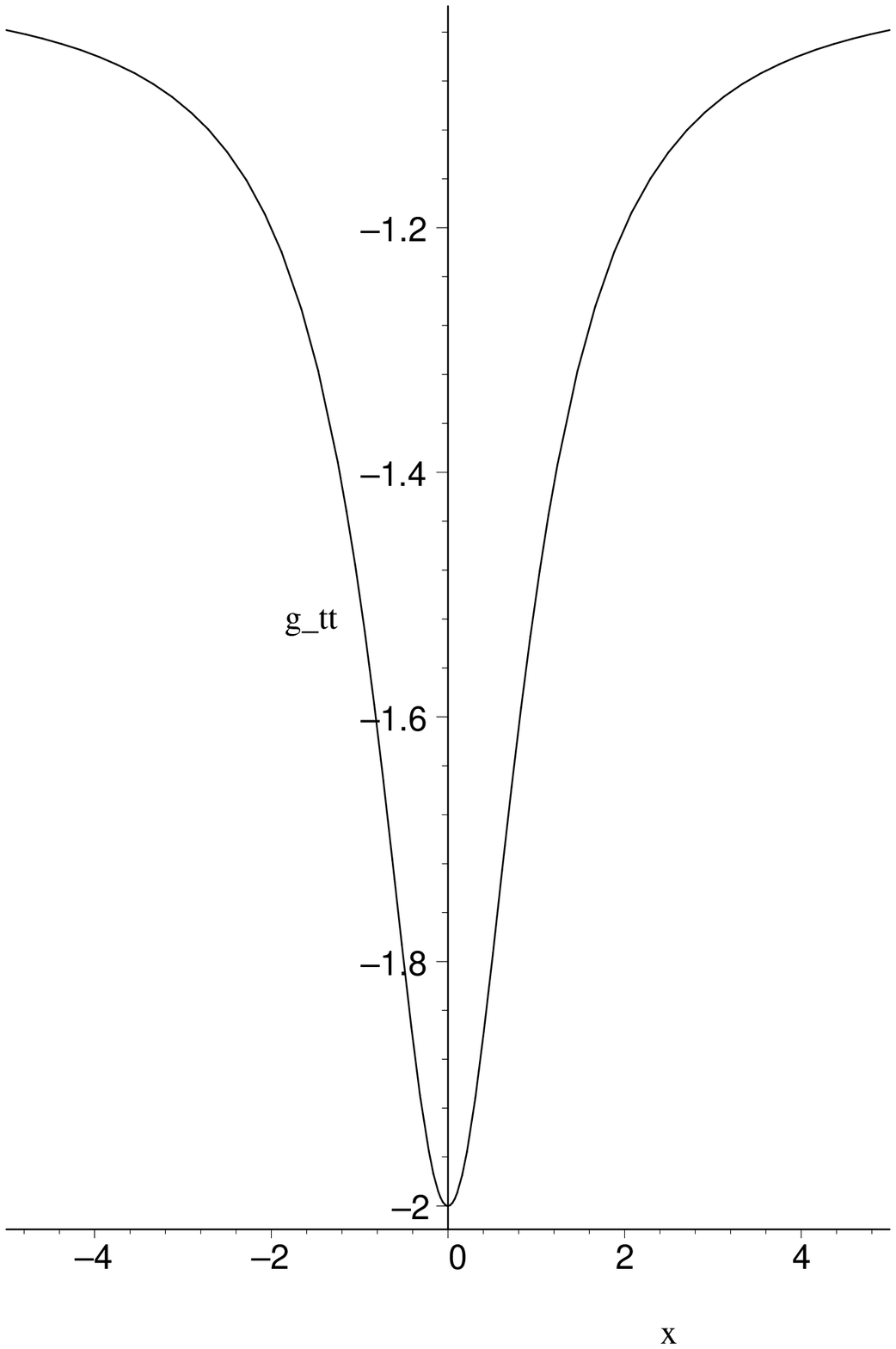}}&\hspace{0.5cm}
\subfloat[$\tilde{\rho}(x)$]{\includegraphics[width=45mm,height=45mm,clip]{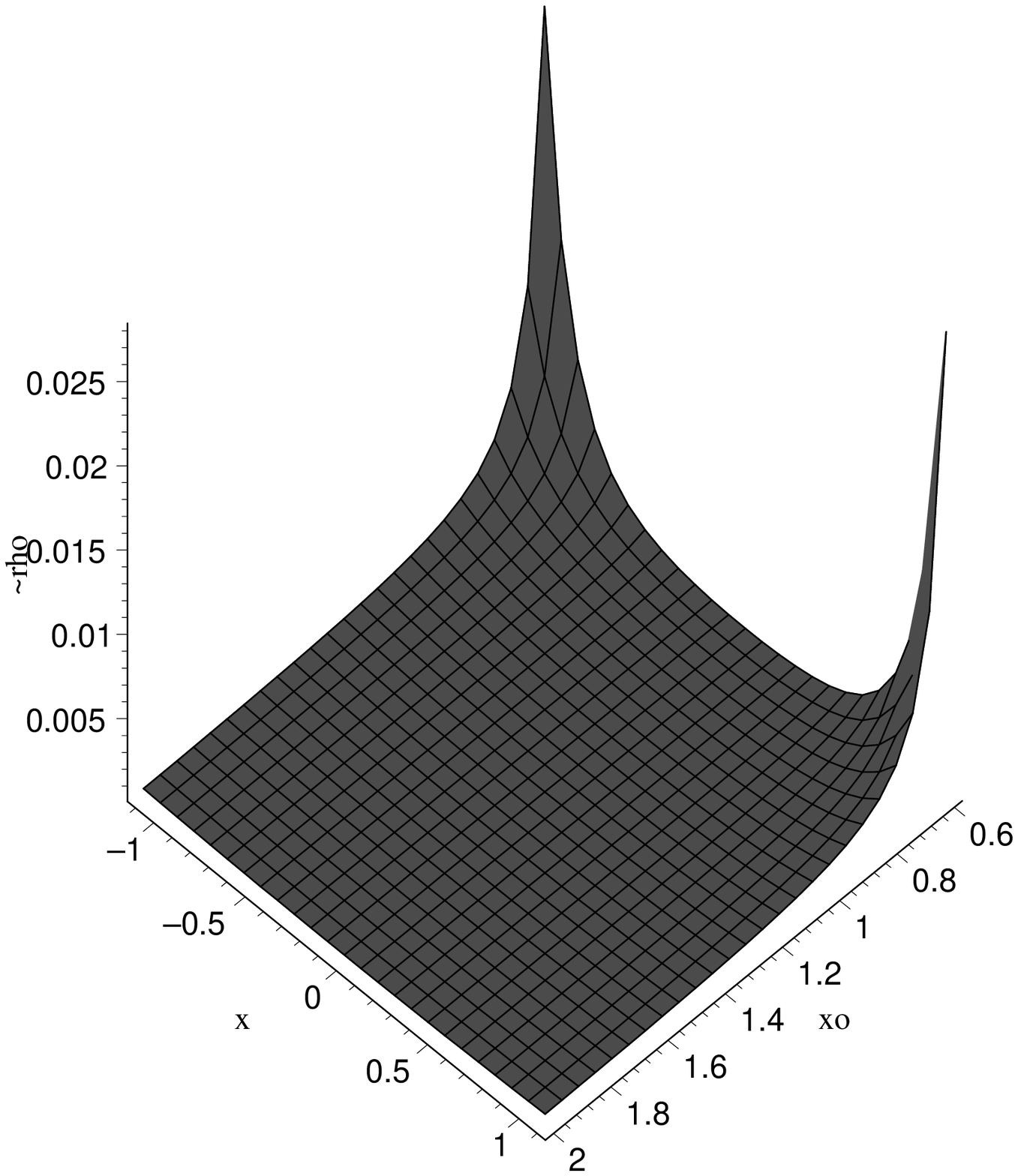}} \\
\subfloat[$\tilde{\rho}+\tilde{p}_{r}$]{\includegraphics[width=45mm,height=45mm,clip]{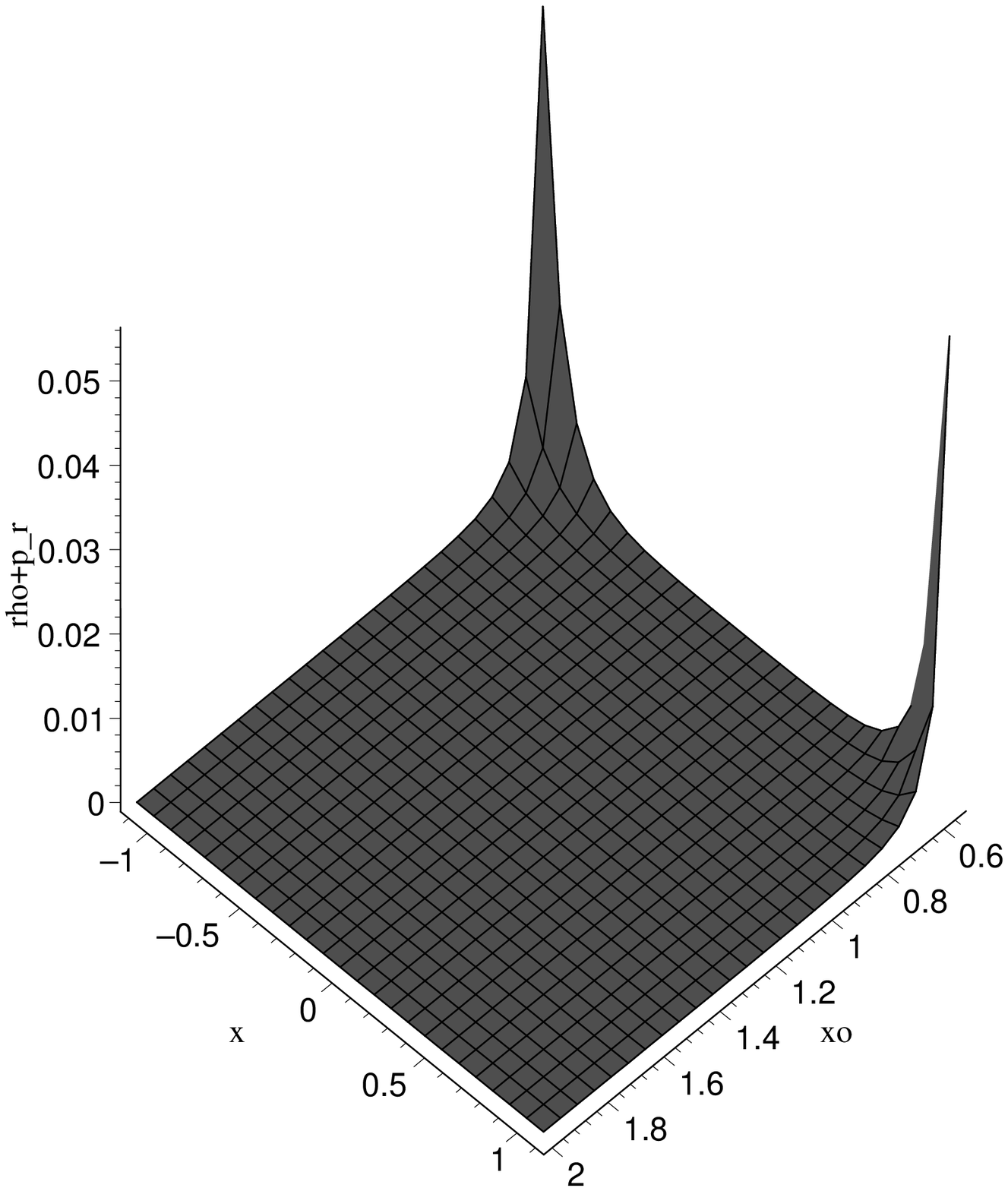}}&\hspace{0.5cm}
\subfloat[$\tilde{\rho}+\tilde{p}_{t}$]{\includegraphics[width=45mm,height=45mm,clip]{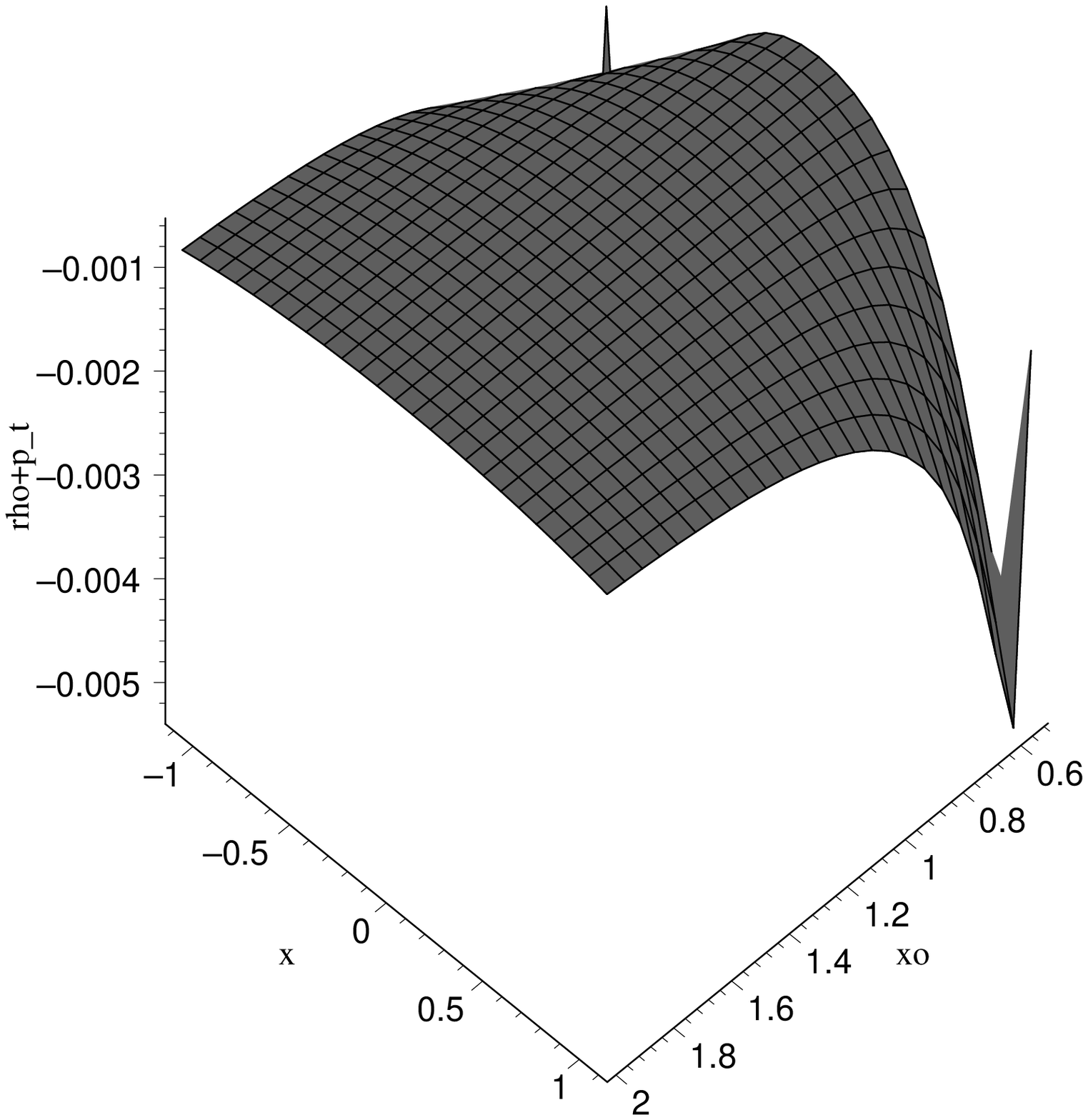}}
\end{tabular}
\end{center}
\vspace{-0.1cm}
\caption{\small{$R^{-1}$ contribution, $g_{tt}$ \emph{increasing}, throat radius=0.05, and varying $x_{0}$.}}
\label{fig:rminusonealphaconcup}
\end{framed}
\end{figure}

\begin{figure}[!ht]
\begin{framed}
\begin{center}
\vspace{0.0cm}
%\fbox{
\includegraphics[width=60mm,height=45mm, clip]{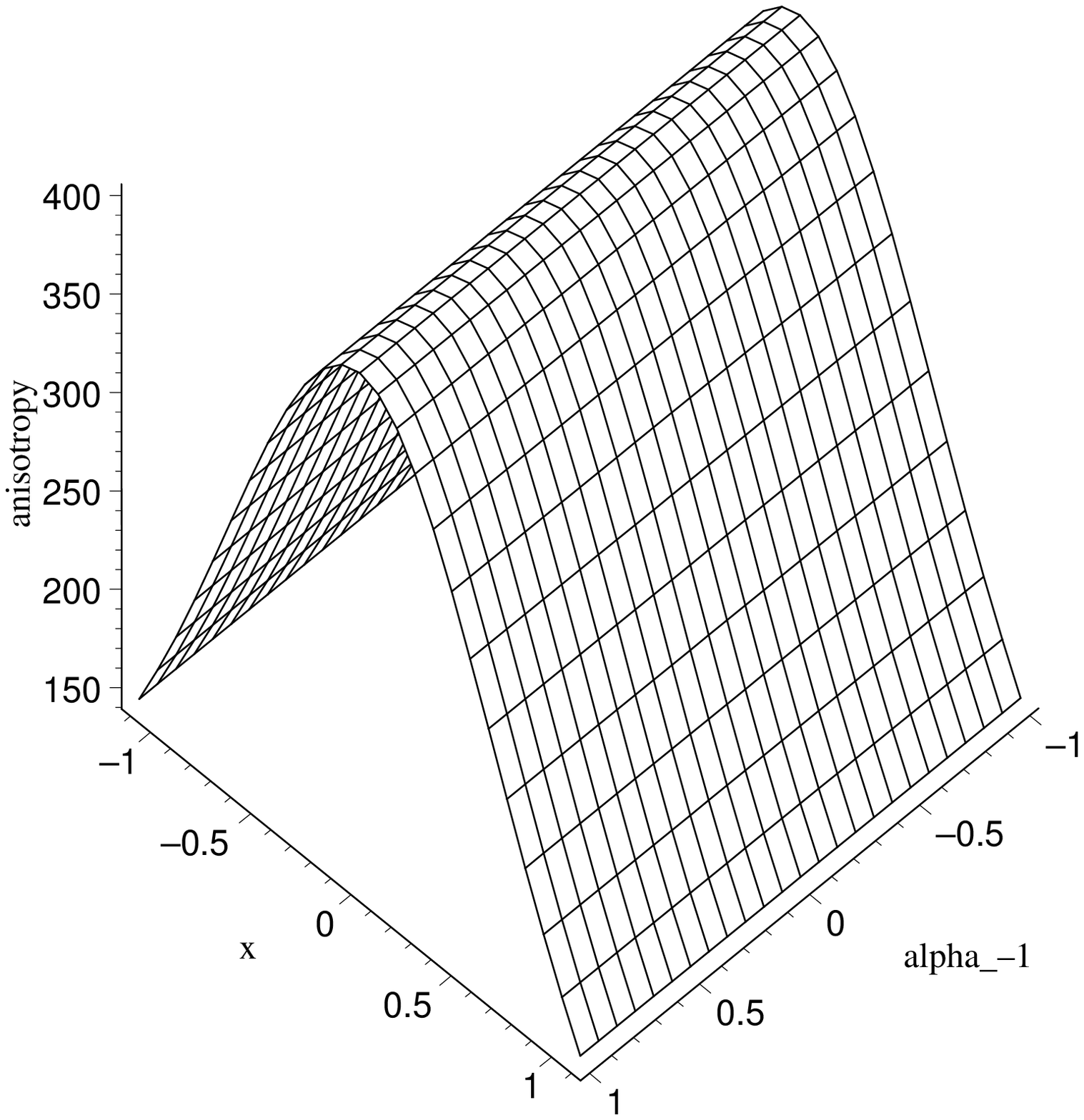}
%}
\end{center}\vspace{-0.1cm}
\caption{\small{Anisotropy of throat region: $R+\alpha_{-1}R^{-1}$ contribution, $g_{tt}$ \emph{increasing}, throat radius=0.05, $x_{0}=1$, and varying $\alpha_{-1}$.}}
\label{fig:rminusonealphaconcupaniso}
\end{framed}
\end{figure}

\begin{figure}[H]
\begin{framed}
\begin{center}
\vspace{-0.5cm}
\begin{tabular}{ccc}
%%\subfloat[$g_{tt}(x)$]{\includegraphics[width=45mm,height=45mm,clip]{squared_gtt_c_down.eps}}&
\subfloat[$\tilde{\rho}(x)$]{\includegraphics[width=42mm,height=42mm,clip]{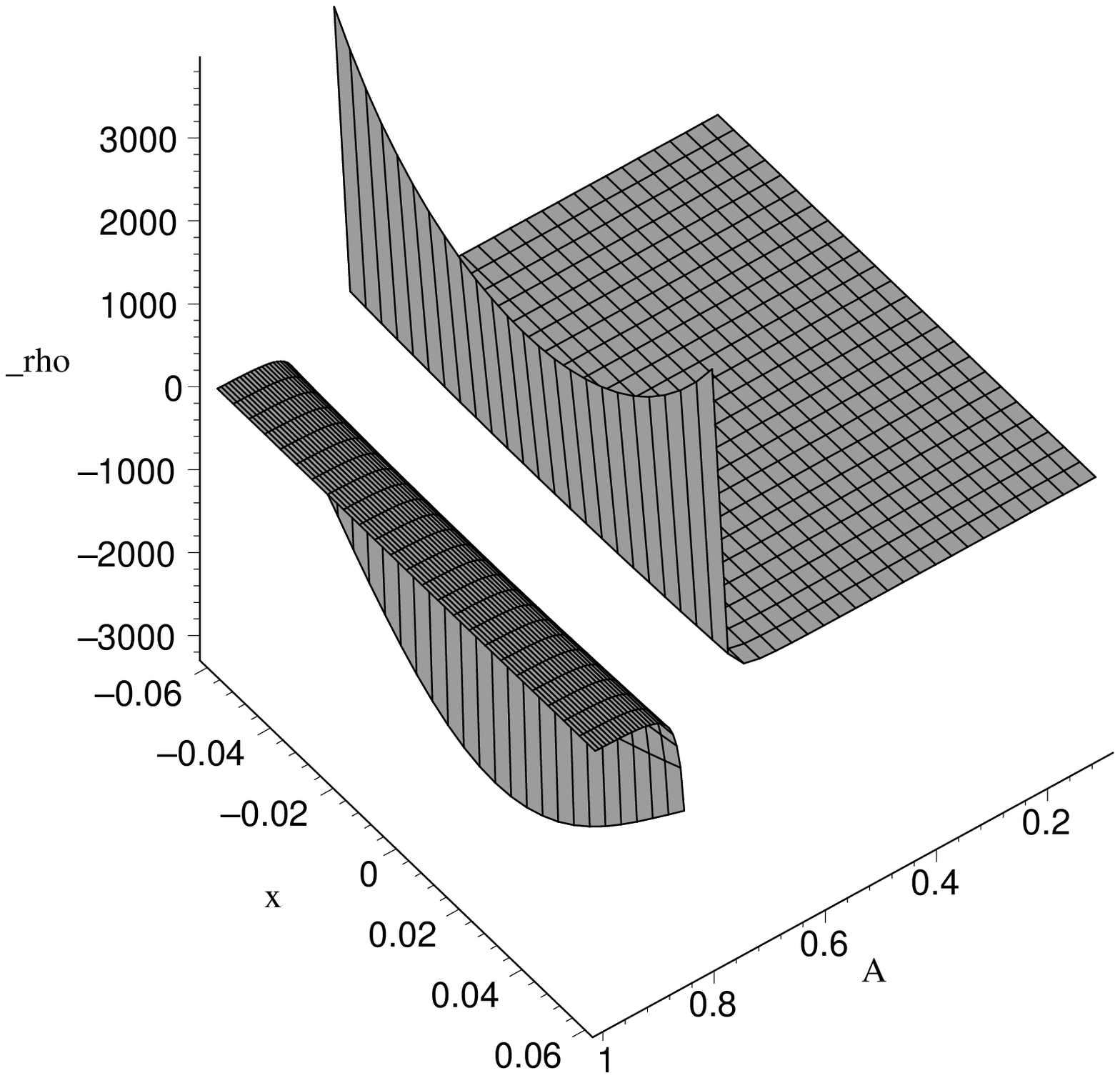}} &
\subfloat[$\tilde{\rho}+\tilde{p}_{r}$]{\includegraphics[width=42mm,height=42mm,clip]{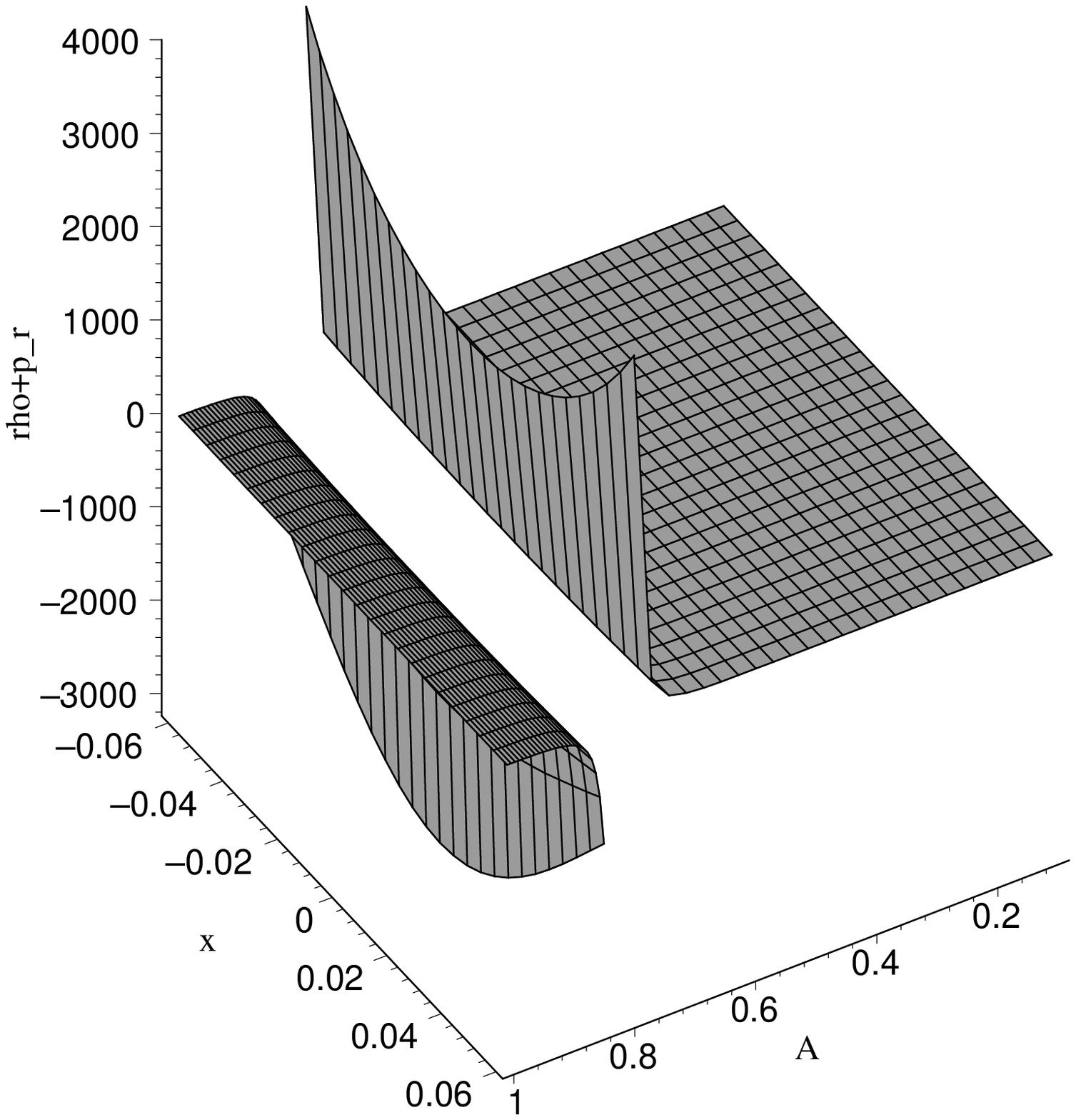}}&
\subfloat[$\tilde{\rho}+\tilde{p}_{t}$]{\includegraphics[width=42mm,height=42mm,clip]{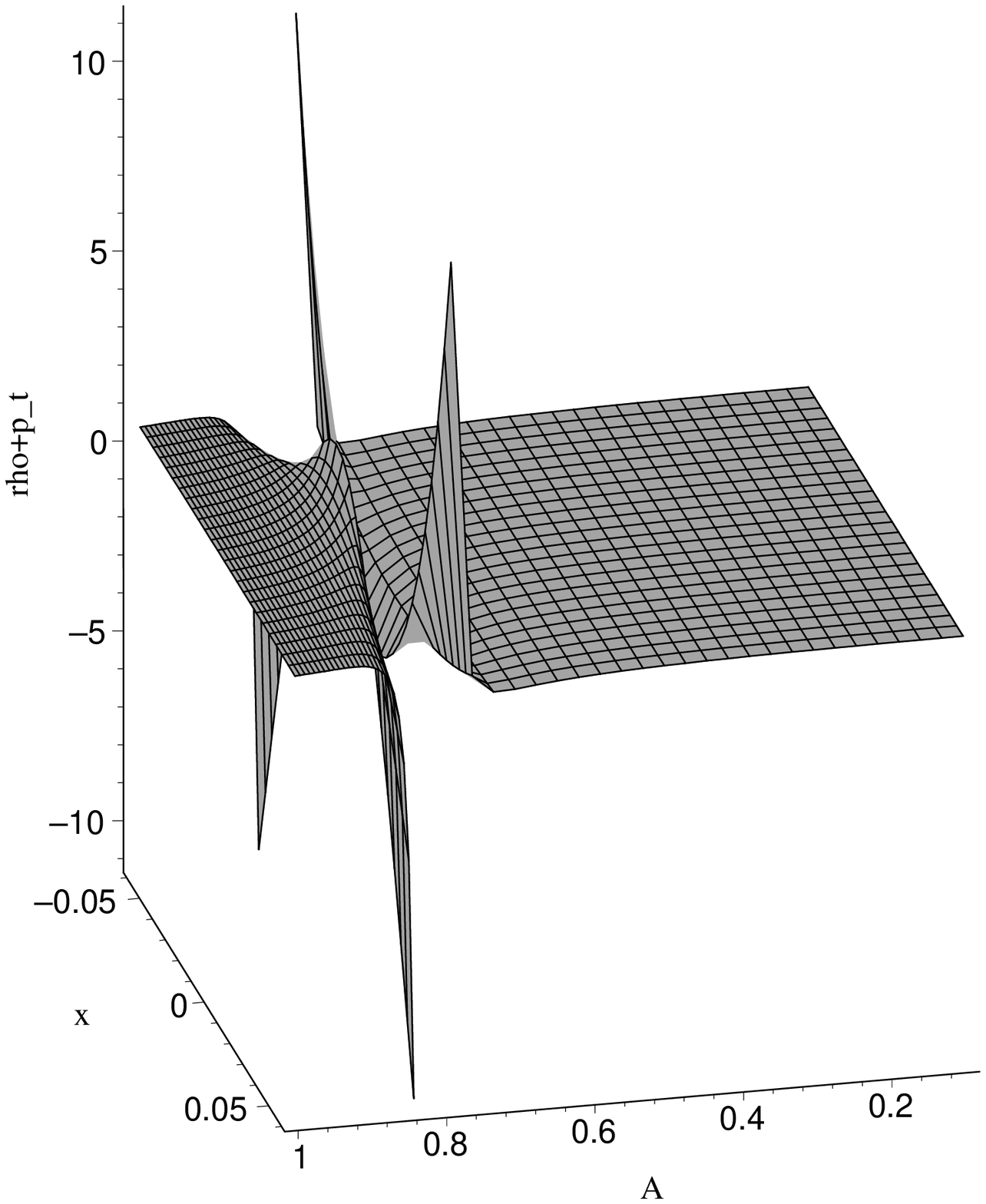}}
\end{tabular}
\end{center}
\vspace{-0.1cm}
\caption{\small{$R^{-1}$ contribution for $x_{0}=1$, $g_{tt}$ \emph{increasing}, and varying throat radius. For negative $\alpha_{-1}$ the graphs should be flipped around the plane $z=0$. The region around the singularity has been omitted.}}
\label{fig:rminusoneAconcup}
\end{framed}
\end{figure}

Again, due to the interest in keeping the length reasonable, we present all the graphs for this case first (figs.~\ref{fig:rminusonealphaconcdown} - \ref{fig:rminusoneAconcup}), and then briefly comment on the results afterward in a summary table (table~2).
\vspace{-0.2cm}

\clearpage
%%\begin{figure}[!ht]
\FloatBarrier
We summarize the results for $n=-1$ in table 2.

\begin{table}[!ht]
\newcommand\Ta{\rule{0pt}{2.6ex}}
\newcommand\Bb{\rule[-1.2ex]{0pt}{0pt}}
\begin{center}
\caption{\small{Summary of $\alpha_{-1}R^{-1}$ contributions.}\vspace{-0.2cm}} 
\begin{tabular}{|c|c|c|c|}
\hline
\Ta  \parbox[t]{2.5cm}{Parameter studied}& \parbox[t]{3.8cm}{zero tidal-force}  &  \parbox[t]{3.8cm}{$g_{tt}(x)$ concave-down} & \parbox[t]{3.8cm}{$g_{tt}(x)$ concave-up} \\[0.5cm]
\hline\hline

\parbox[t]{2.5cm}{$x_{0}$} & \parbox[t]{3.8cm}{{\small Respecting energy conditions seems difficult at best.}}\Ta & \parbox[t]{3.8cm}{{\small Little sensitivity very near the throat on variations of $x_{0}$.}} & \parbox[t]{3.8cm}{{\small Little sensitivity very near the throat on variations of $x_{0}$.}}  \\[0.07cm]
\hline
\parbox[t]{2.5cm}{Throat radius} & \parbox[t]{3.8cm}{{\small Respecting energy conditions seems difficult at best.}}\Ta & \parbox[t]{3.8cm}{{\small Possible to respect energy conditions near singular region for negative $\alpha_{-1}$ for a small section of parameter space. (Only on one side of the singular region.)}} & \parbox[t]{3.8cm}{{\small Seems possible to make small positive contribution to all energy conditions (see general $n$ section) for very small region of parameter space.}}  \\[0.07cm]
\hline
\parbox[t]{2.5cm}{Minimum anisotropy for $R+\alpha_{-1}R^{-1}$} &\parbox[t]{3.8cm}{{\small Not studied.}}\Ta & \parbox[t]{3.8cm}{{\small Magnitude of anisotropy larger than corresponding pure Einstein case regardless of $\alpha_{-1}$.}} & \parbox[t]{3.8cm}{{\small Magnitude of anisotropy larger than corresponding pure Einstein case regardless of $\alpha_{-1}$.}}  \\[0.07cm]
\hline
\end{tabular}
\end{center}
\end{table}
\FloatBarrier

\subsubsection{{\normalsize $n=-2$}}
\begin{figure}[!ht]
\begin{framed}
\begin{center}
\vspace{-0.5cm}
\begin{tabular}{cc}
\subfloat[$g_{tt}(x)$]{\includegraphics[width=45mm,height=45mm,clip]{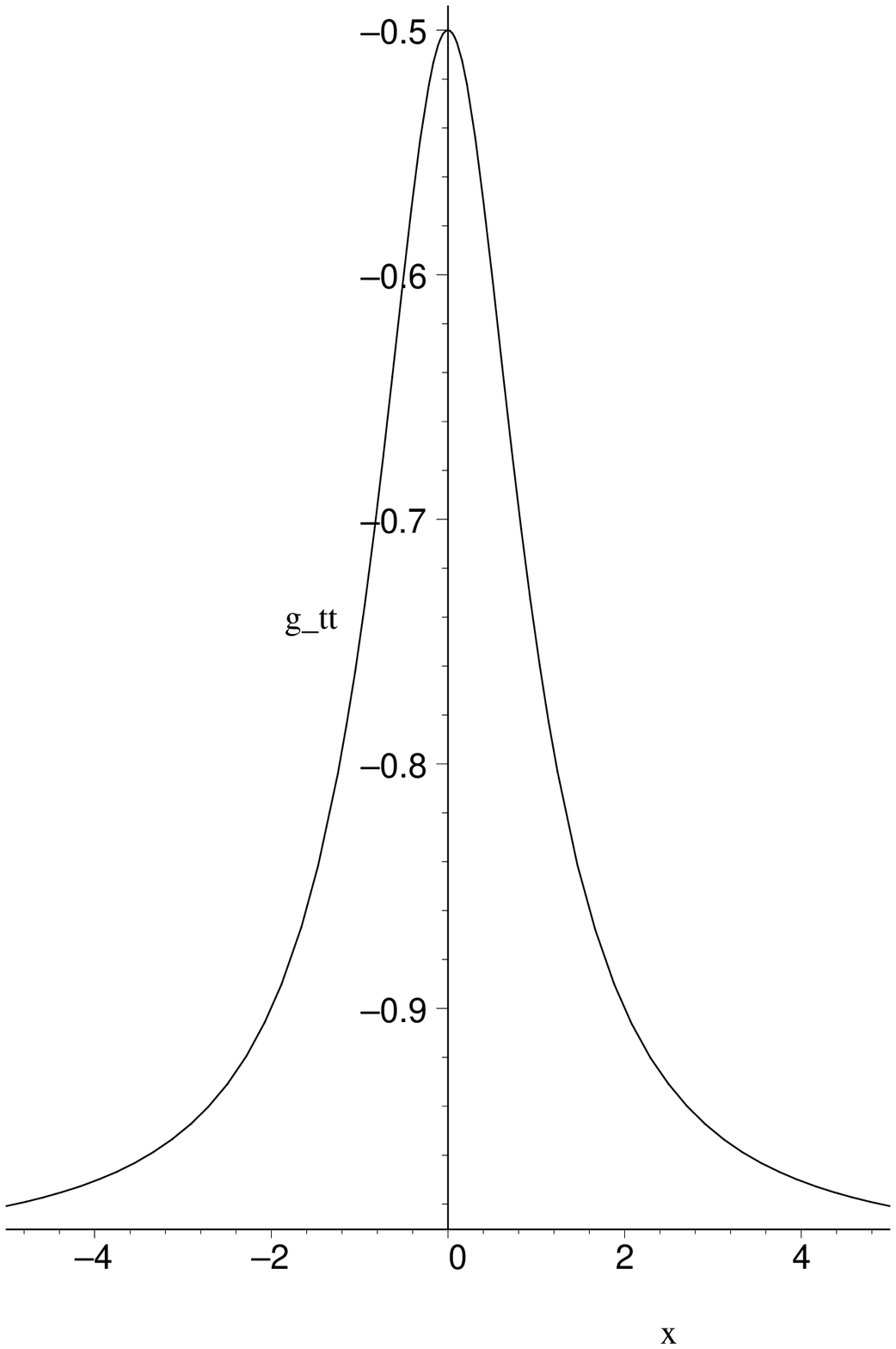}}&\hspace{0.5cm}
\subfloat[$\tilde{\rho}(x)$]{\includegraphics[width=45mm,height=45mm,clip]{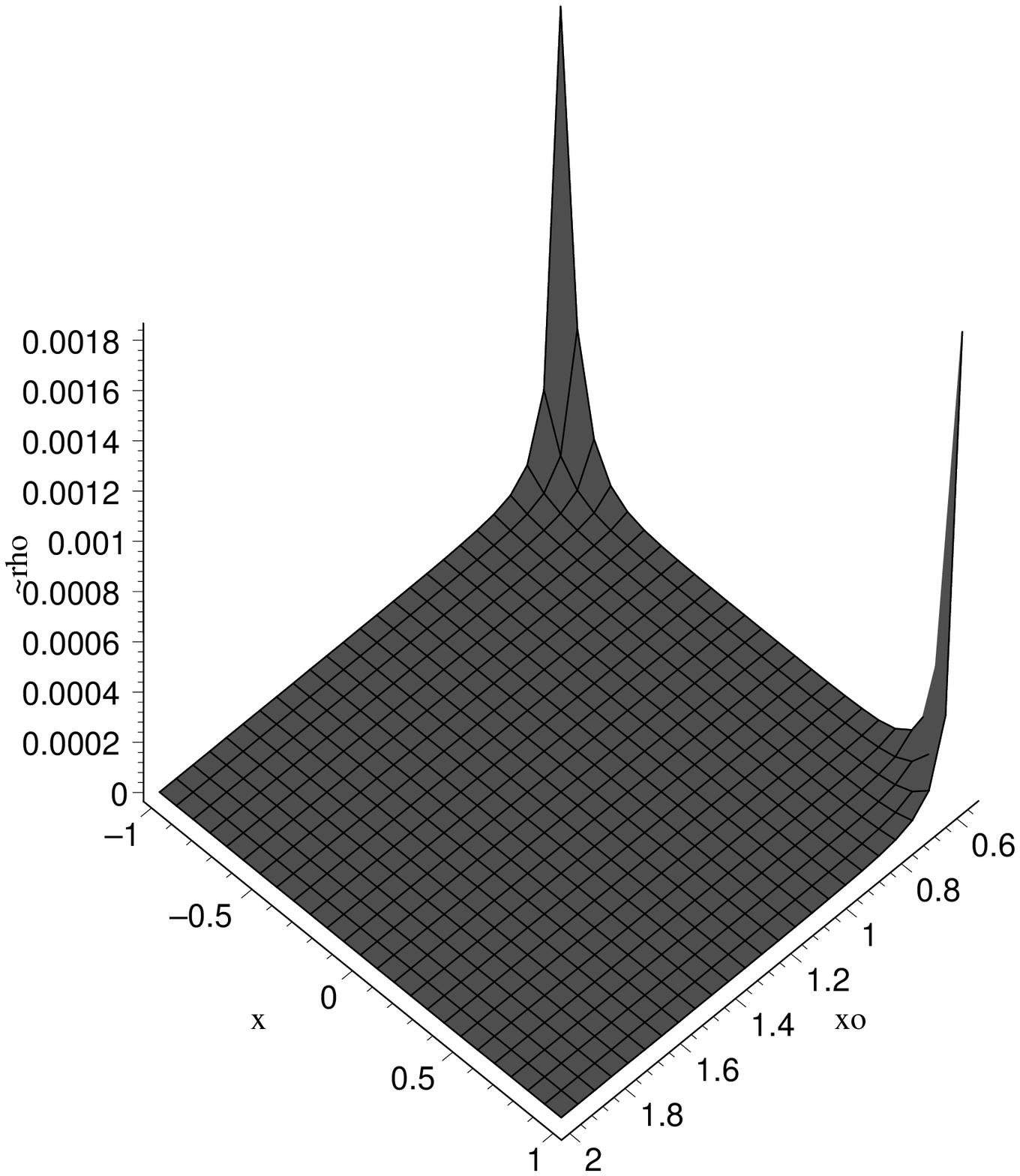}} \\
\subfloat[$\tilde{\rho}+\tilde{p}_{r}$]{\includegraphics[width=45mm,height=45mm,clip]{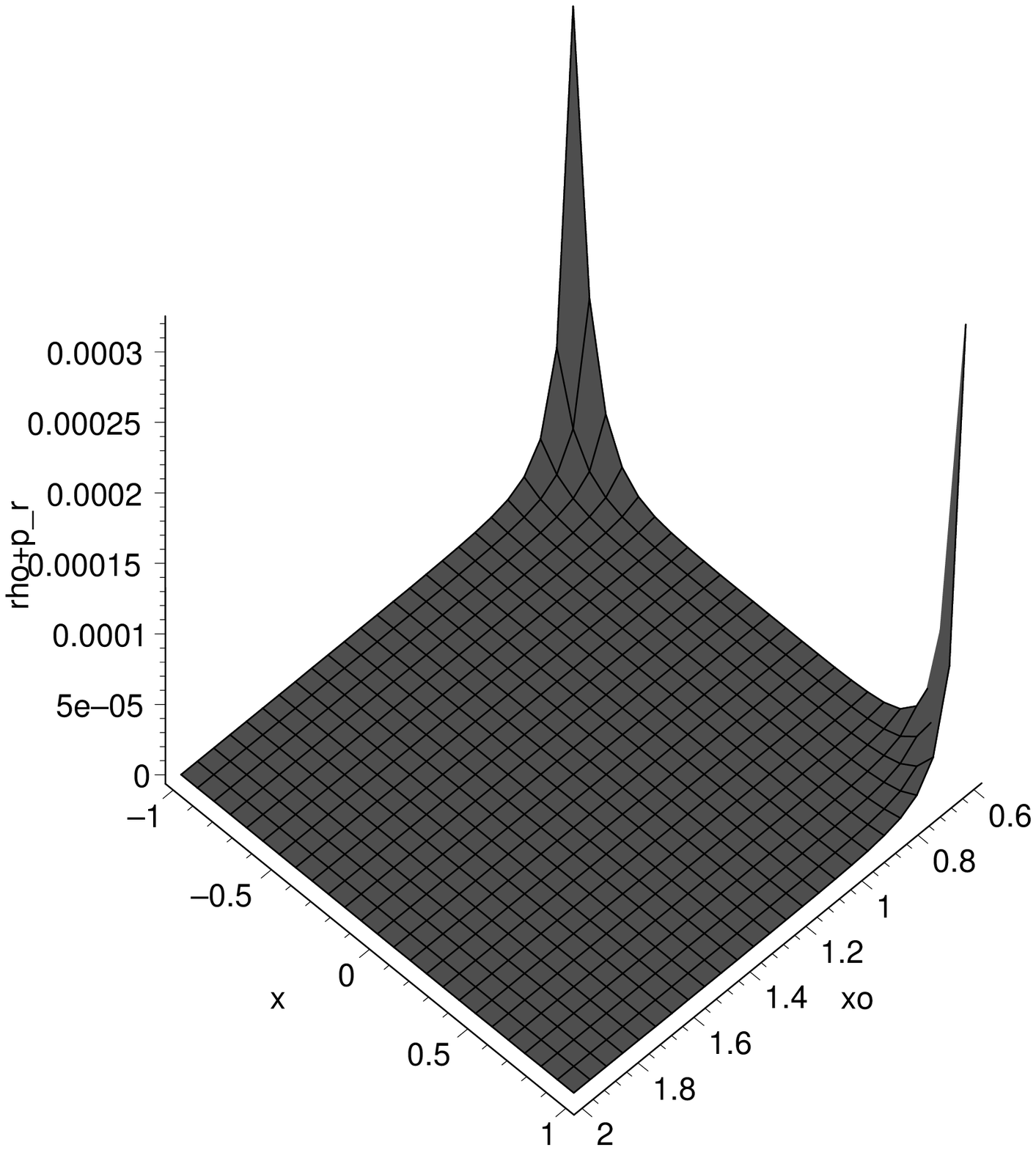}}&\hspace{0.5cm}
\subfloat[$\tilde{\rho}+\tilde{p}_{t}$]{\includegraphics[width=45mm,height=45mm,clip]{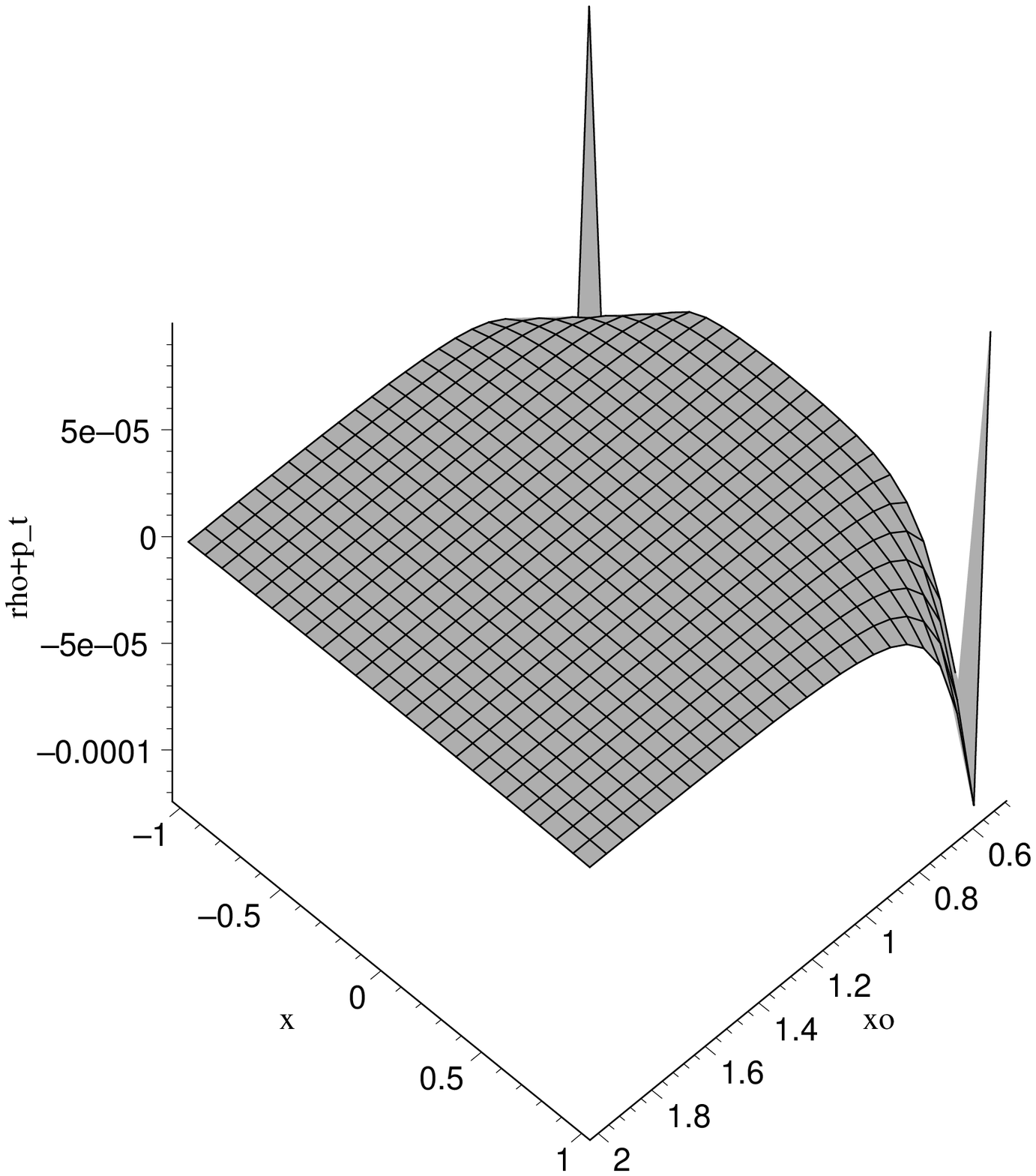}}
\end{tabular}
\end{center}
\vspace{-0.1cm}
\caption{\small{$R^{-2}$ contribution, $g_{tt}$ \emph{decreasing}, throat radius=0.05, and varying $x_{0}$. For negative $\alpha_{-2}$ the graphs should be flipped around the plane $z=0$.}}
\label{fig:rminustwoalphaconcdown}
\end{framed}
\end{figure}

\begin{figure}[!ht]
\begin{framed}
\begin{center}
\vspace{-0.22cm}
%\fbox{
\includegraphics[width=60mm,height=45mm, clip]{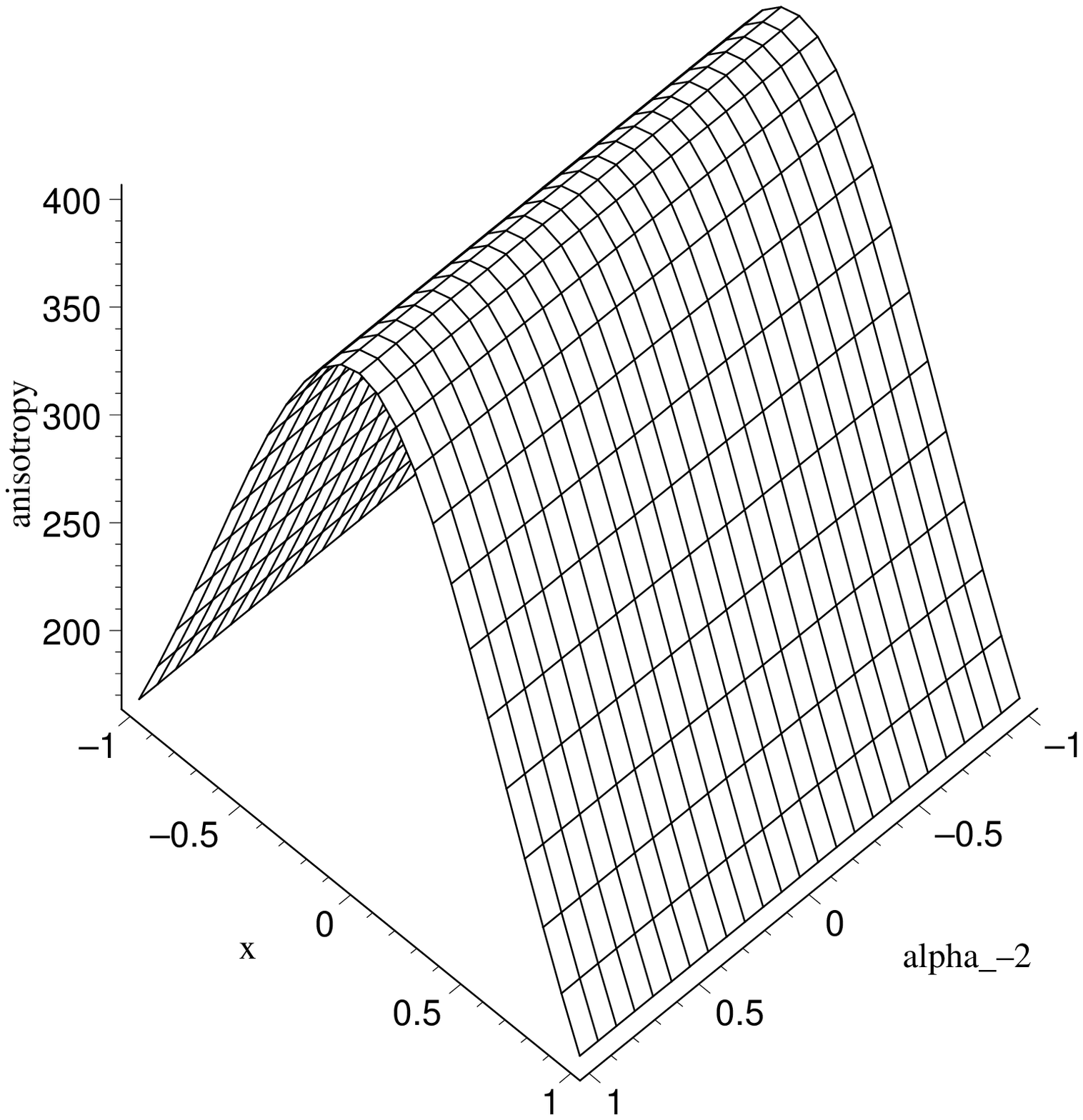}
%}
\end{center}
\vspace{-0.1cm}
\caption{\small{Anisotropy of throat region: $R+\alpha_{-2}R^{-2}$ contribution, $g_{tt}$ \emph{decreasing}, throat radius=0.05, $x_{0}=1$, and varying $\alpha_{-2}$.}}
\label{fig:rminustwoalphaconcdownaniso}\vspace{-0.1cm}
\end{framed}
\end{figure}

\begin{figure}[!ht]
\begin{framed}
\begin{center}
\vspace{-0.5cm}
\begin{tabular}{ccc}
%%\subfloat[$g_{tt}(x)$]{\includegraphics[width=45mm,height=45mm,clip]{squared_gtt_c_down.eps}}&
\subfloat[$\tilde{\rho}(x)$]{\includegraphics[width=42mm,height=42mm,clip]{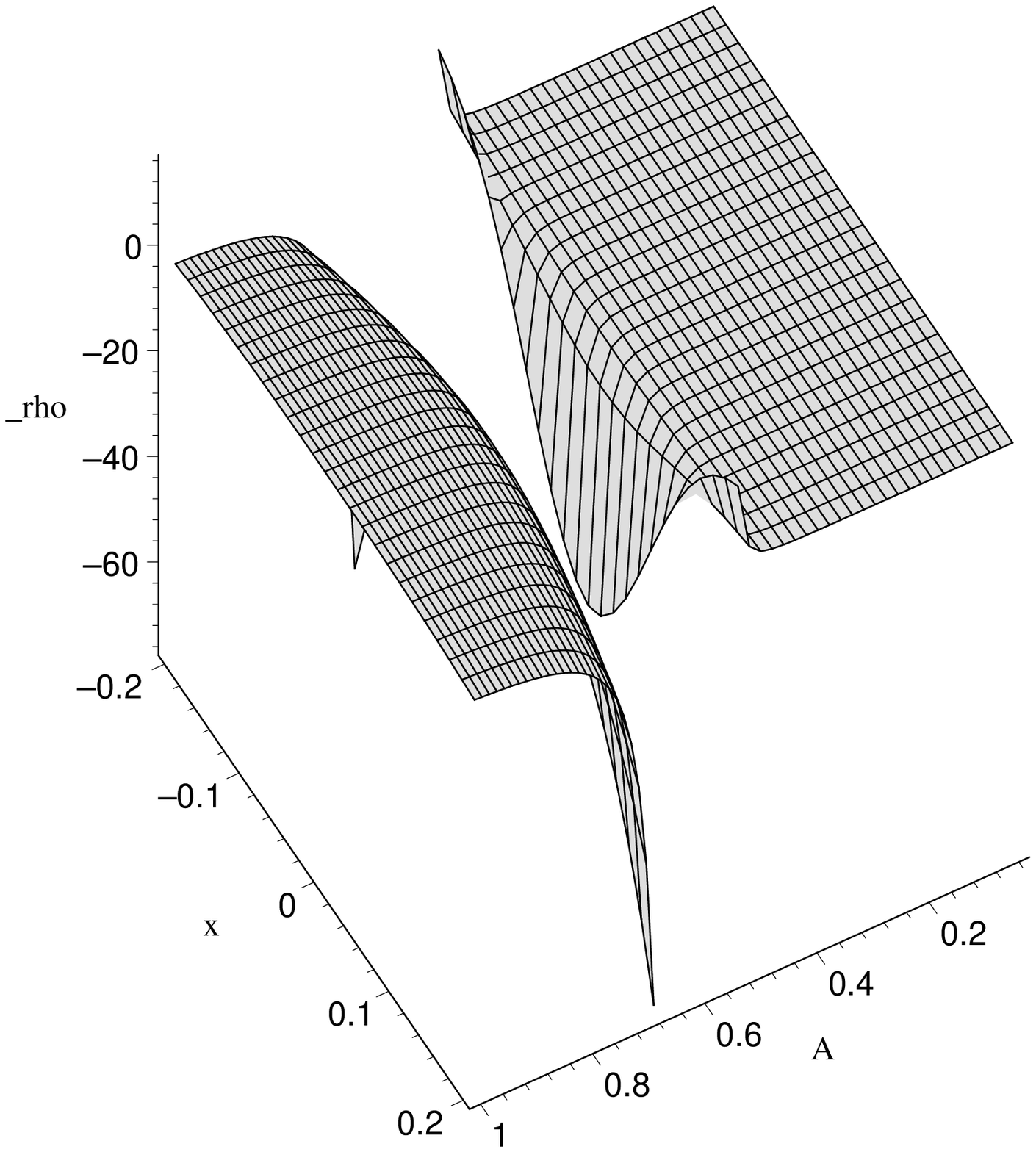}} &
\subfloat[$\tilde{\rho}+\tilde{p}_{r}$]{\includegraphics[width=42mm,height=42mm,clip]{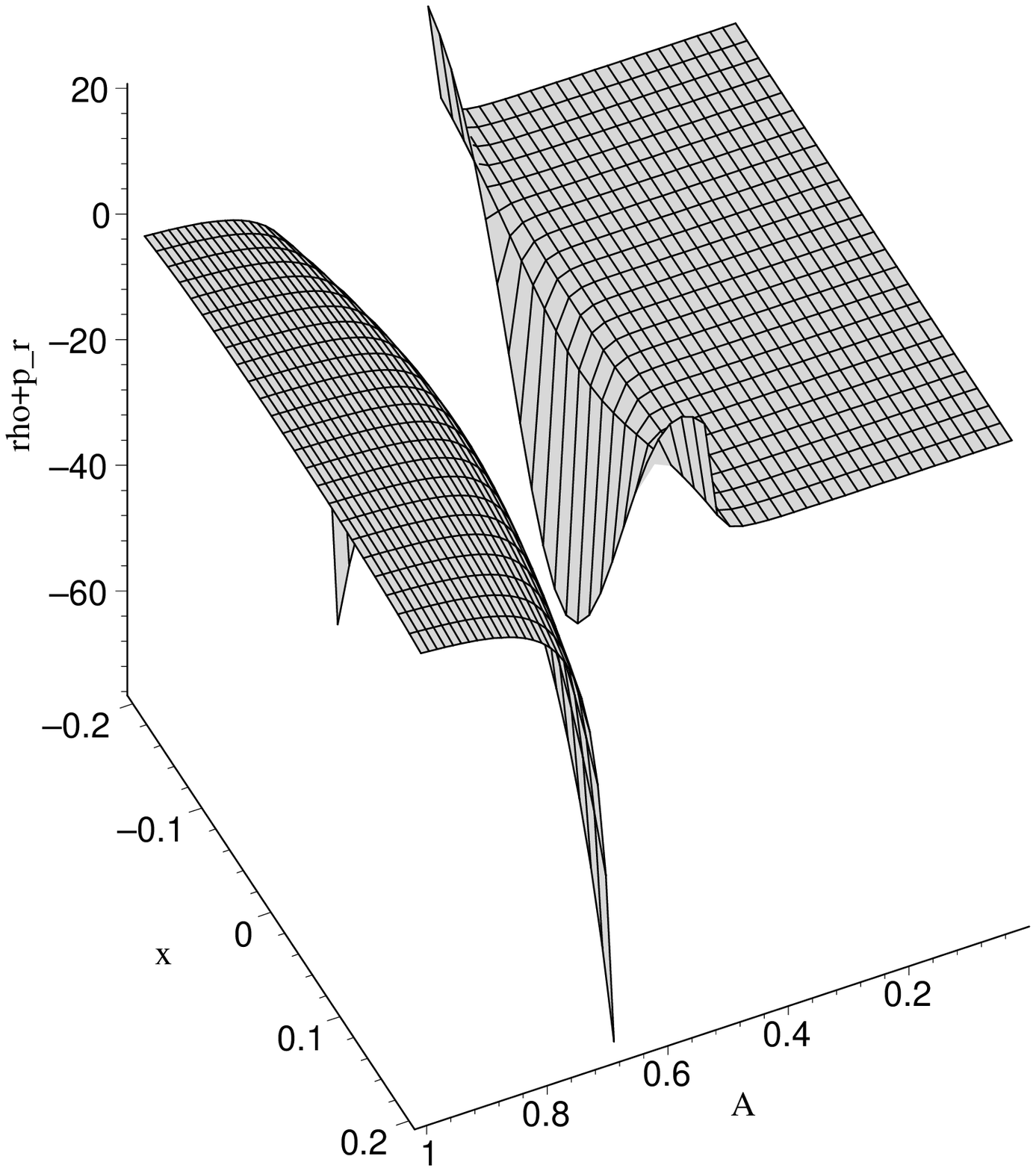}}&
\subfloat[$\tilde{\rho}+\tilde{p}_{t}$]{\includegraphics[width=42mm,height=42mm,clip]{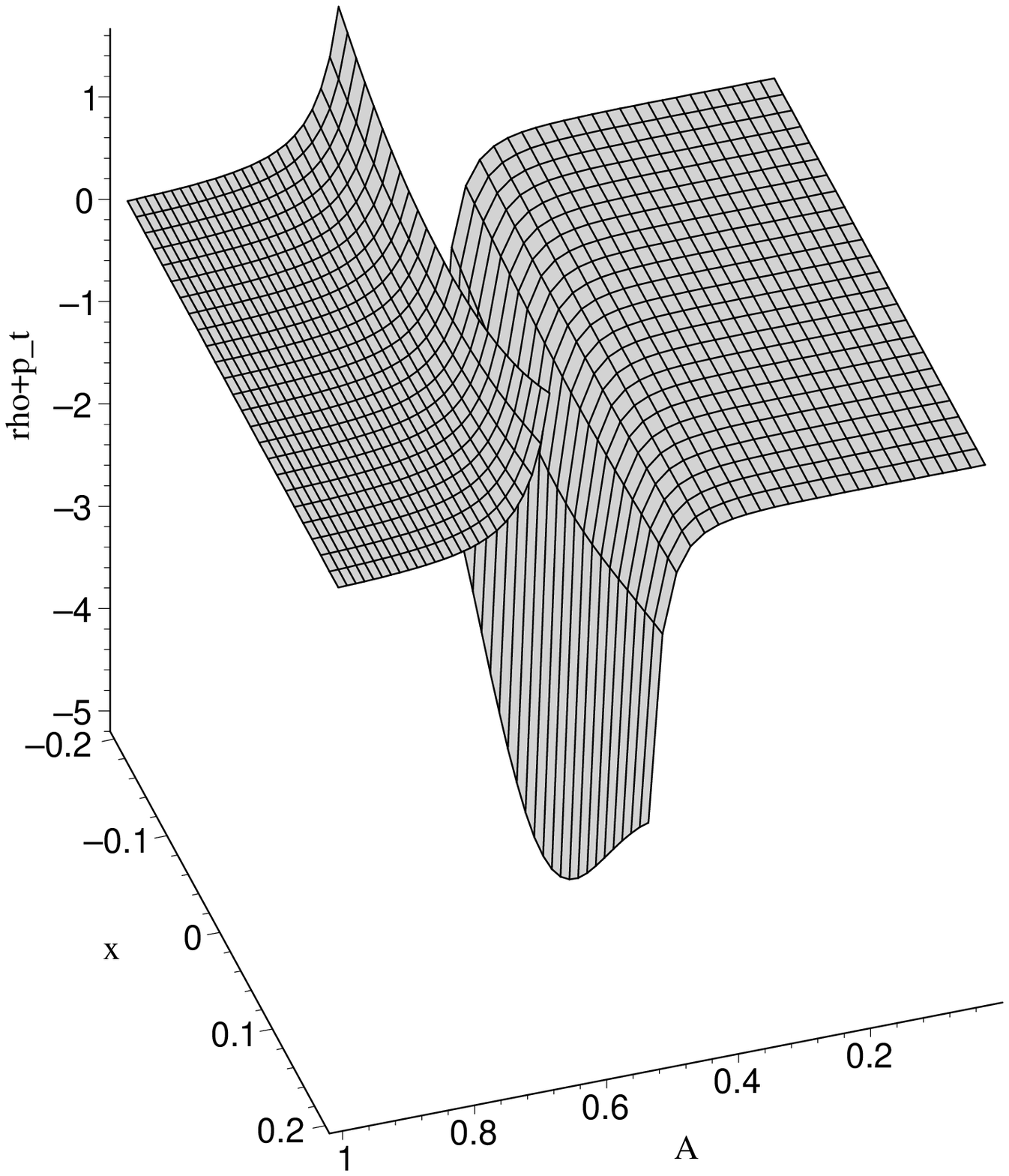}}
\end{tabular}
\end{center}
\vspace{-0.1cm}
\caption{\small{$R^{-2}$ contribution for $x_{0}=1$, $g_{tt}$ \emph{decreasing}, and varying throat radius. For negative $\alpha_{-2}$ the graphs should be flipped around the plane $z=0$. The region around the singularity has been omitted.}}
\label{fig:rminustwoAconcdown}
\end{framed}
\end{figure}

\begin{figure}[!ht]
\begin{framed}
\begin{center}
\vspace{-0.5cm}
\begin{tabular}{cc}
\subfloat[$g_{tt}(x)$]{\includegraphics[width=45mm,height=45mm,clip]{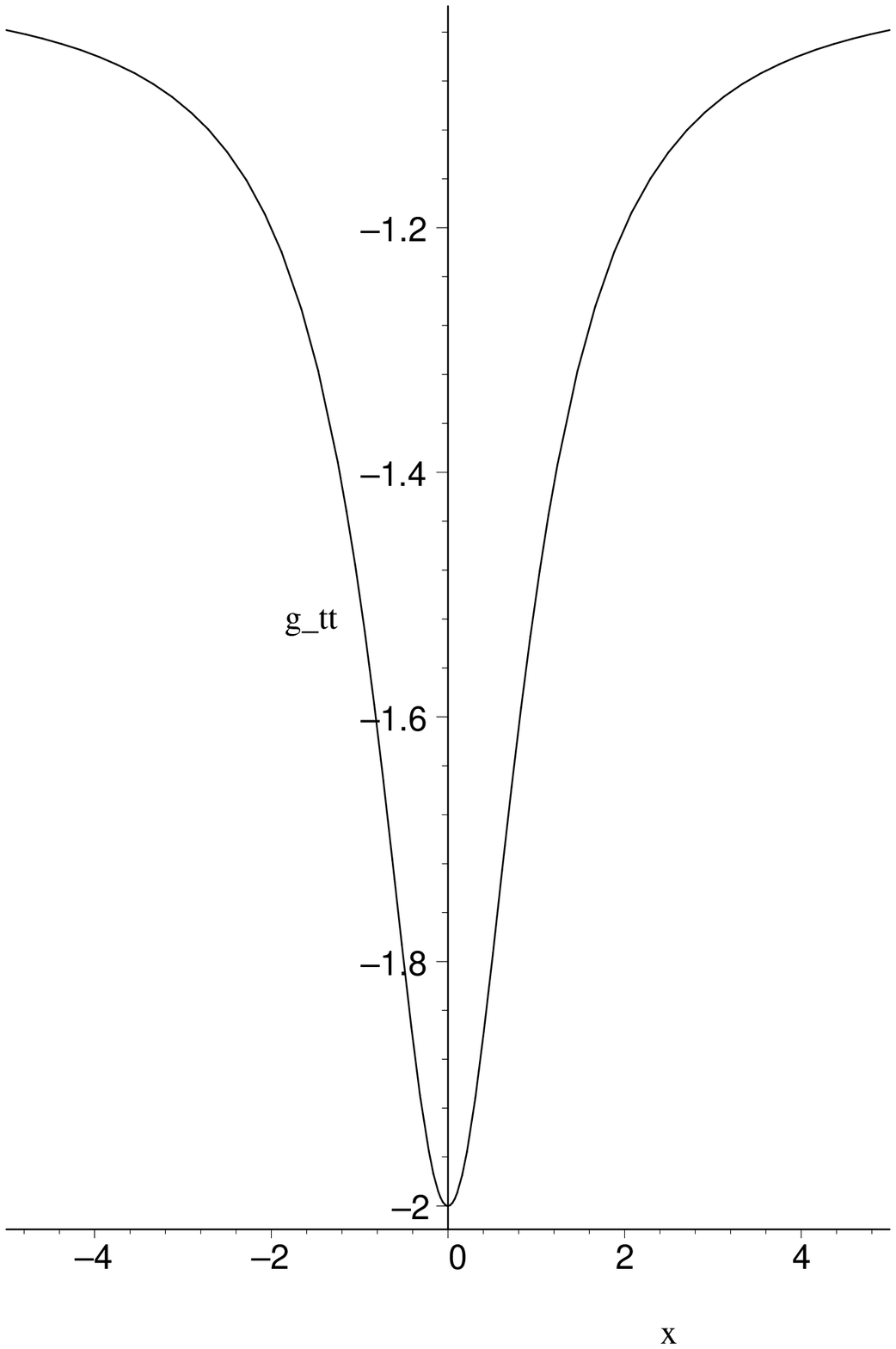}}&\hspace{0.5cm}
\subfloat[$\tilde{\rho}(x)$]{\includegraphics[width=45mm,height=45mm,clip]{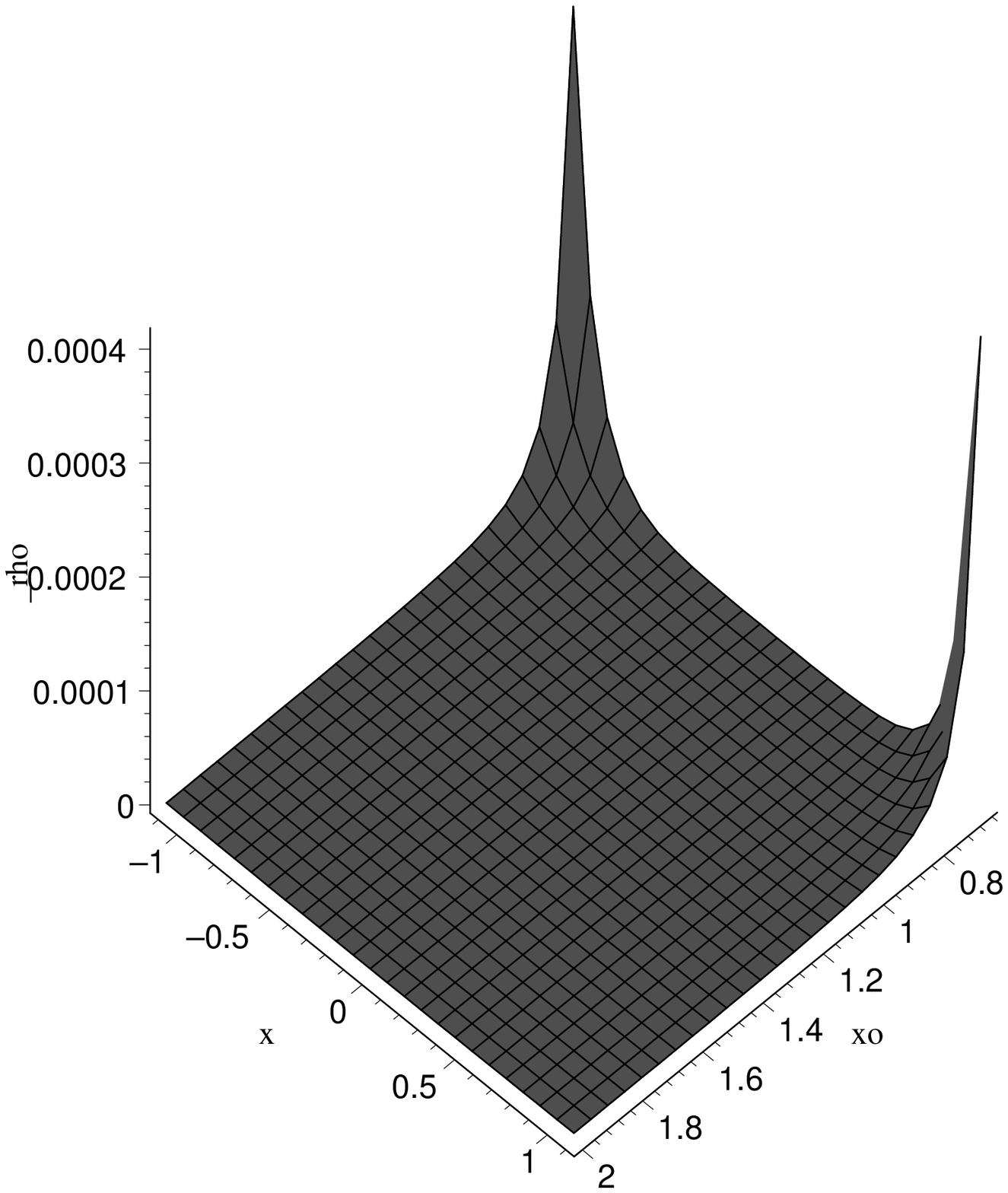}} \\
\subfloat[$\tilde{\rho}+\tilde{p}_{r}$]{\includegraphics[width=45mm,height=45mm,clip]{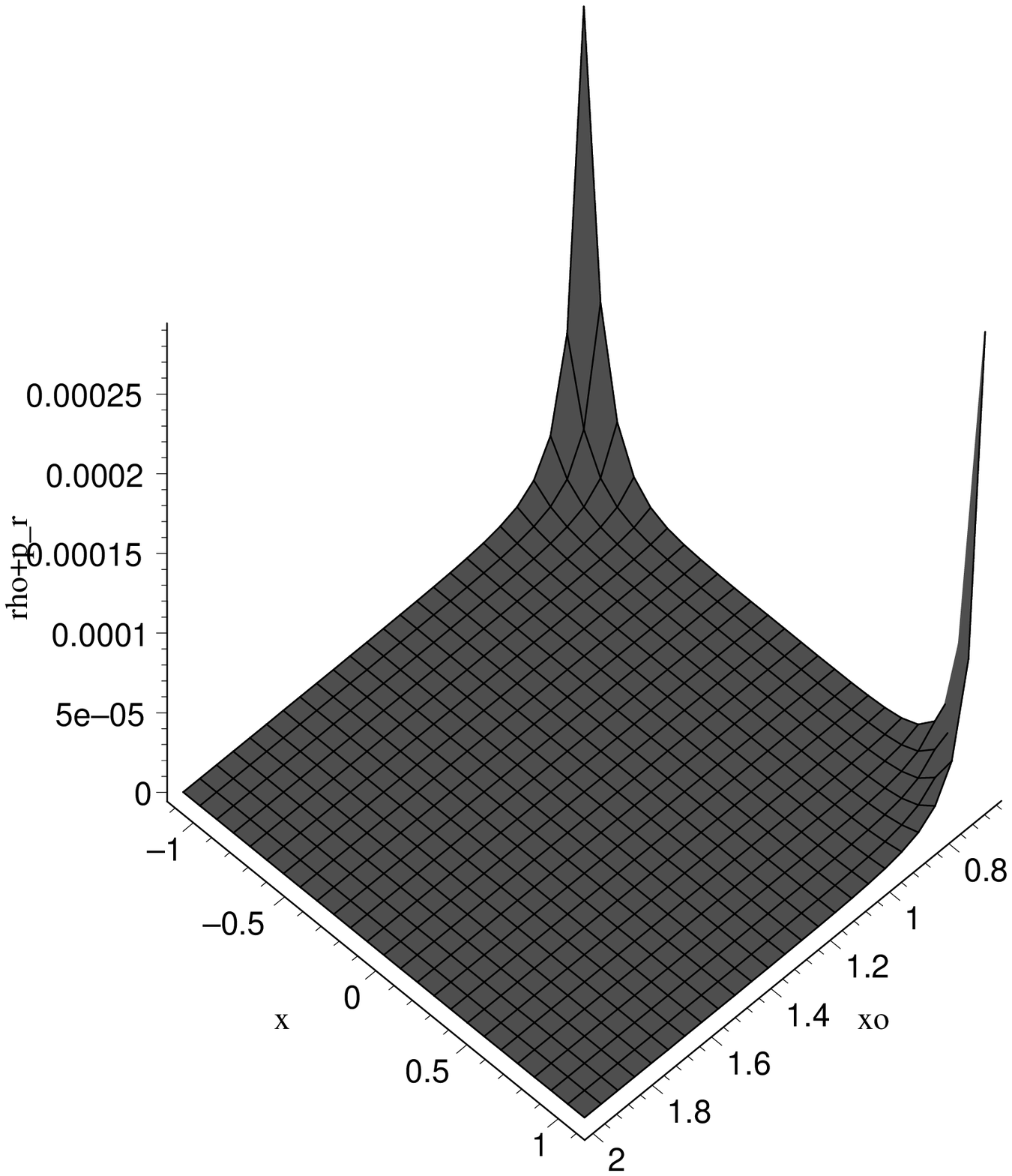}}&\hspace{0.5cm}
\subfloat[$\tilde{\rho}+\tilde{p}_{t}$]{\includegraphics[width=45mm,height=45mm,clip]{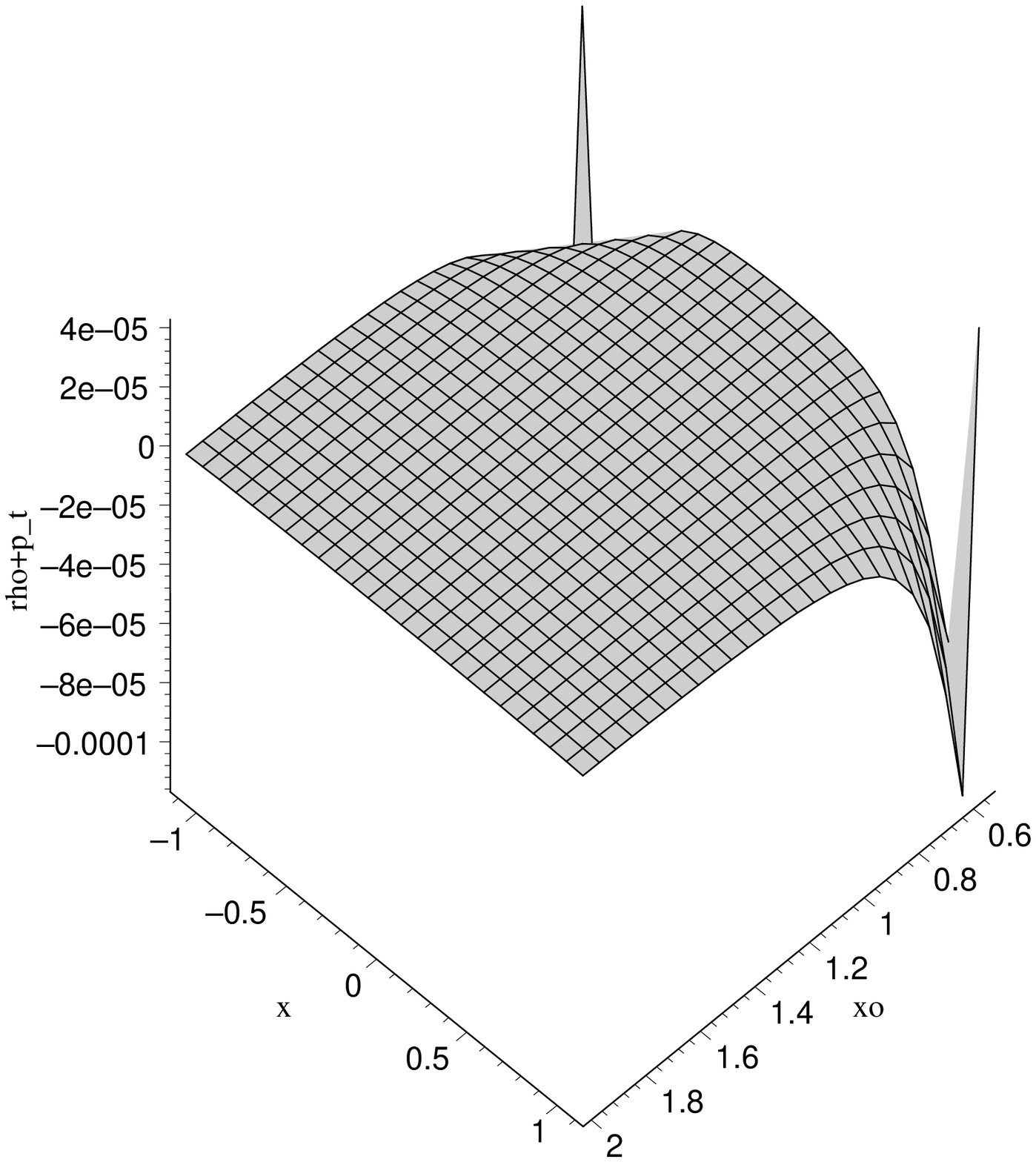}}
\end{tabular}
\end{center}
\vspace{-0.1cm}
\caption{\small{$R^{-2}$ contribution, $g_{tt}$ \emph{increasing}, throat radius=0.05, and varying $x_{0}$. For negative $\alpha_{-2}$ the graphs should be flipped around the plane $z=0$.}}
\label{fig:rminustwoalphaconcup}
\end{framed}
\end{figure}

\begin{figure}[!ht]
\begin{framed}
\begin{center}
\vspace{0.0cm}
%\fbox{
\includegraphics[width=60mm,height=45mm, clip]{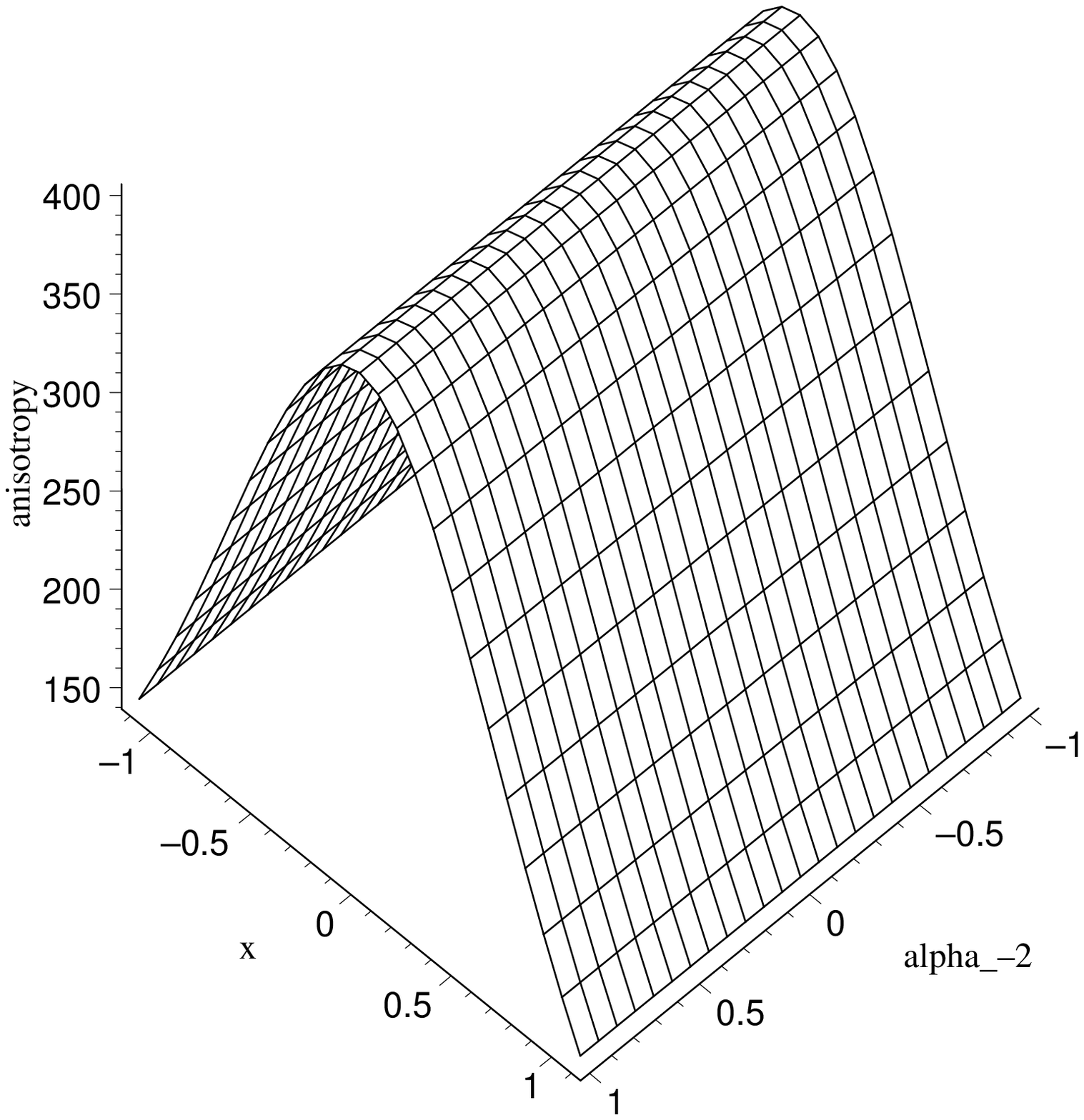}
%}
\end{center}\vspace{-0.1cm}
\caption{\small{Anisotropy of throat region: $R+\alpha_{-2}R^{-2}$ contribution, $g_{tt}$ \emph{increasing}, throat radius=0.05, $x_{0}=1$, and varying $\alpha_{-2}$.}}
\label{fig:rminustwoalphaconcupaniso}
\end{framed}
\end{figure}

\begin{figure}[H]
\begin{framed}
\begin{center}
\vspace{-0.5cm}
\begin{tabular}{ccc}
%%\subfloat[$g_{tt}(x)$]{\includegraphics[width=45mm,height=45mm,clip]{squared_gtt_c_down.eps}}&
\subfloat[$\tilde{\rho}(x)$]{\includegraphics[width=42mm,height=42mm,clip]{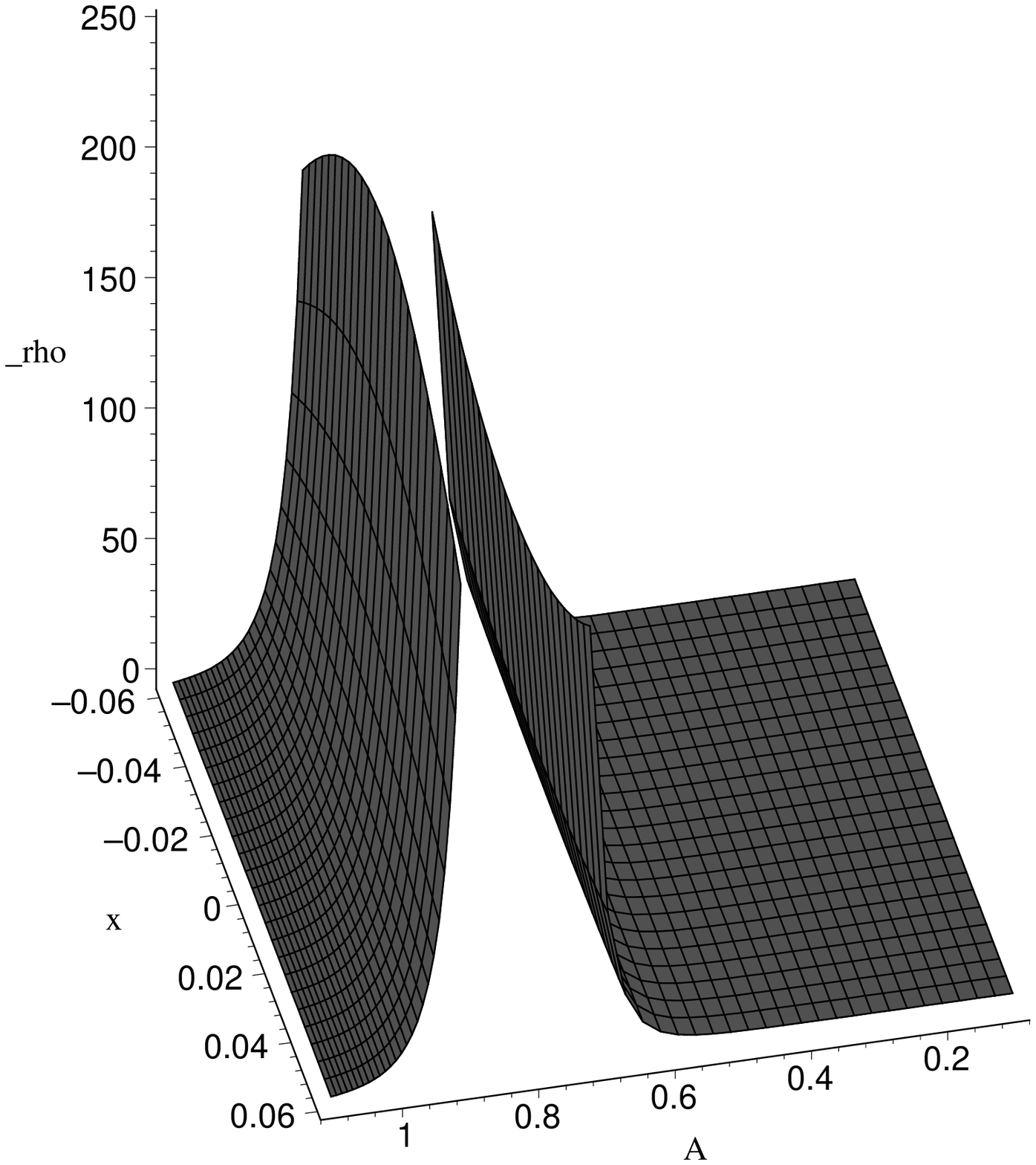}} &
\subfloat[$\tilde{\rho}+\tilde{p}_{r}$]{\includegraphics[width=42mm,height=42mm,clip]{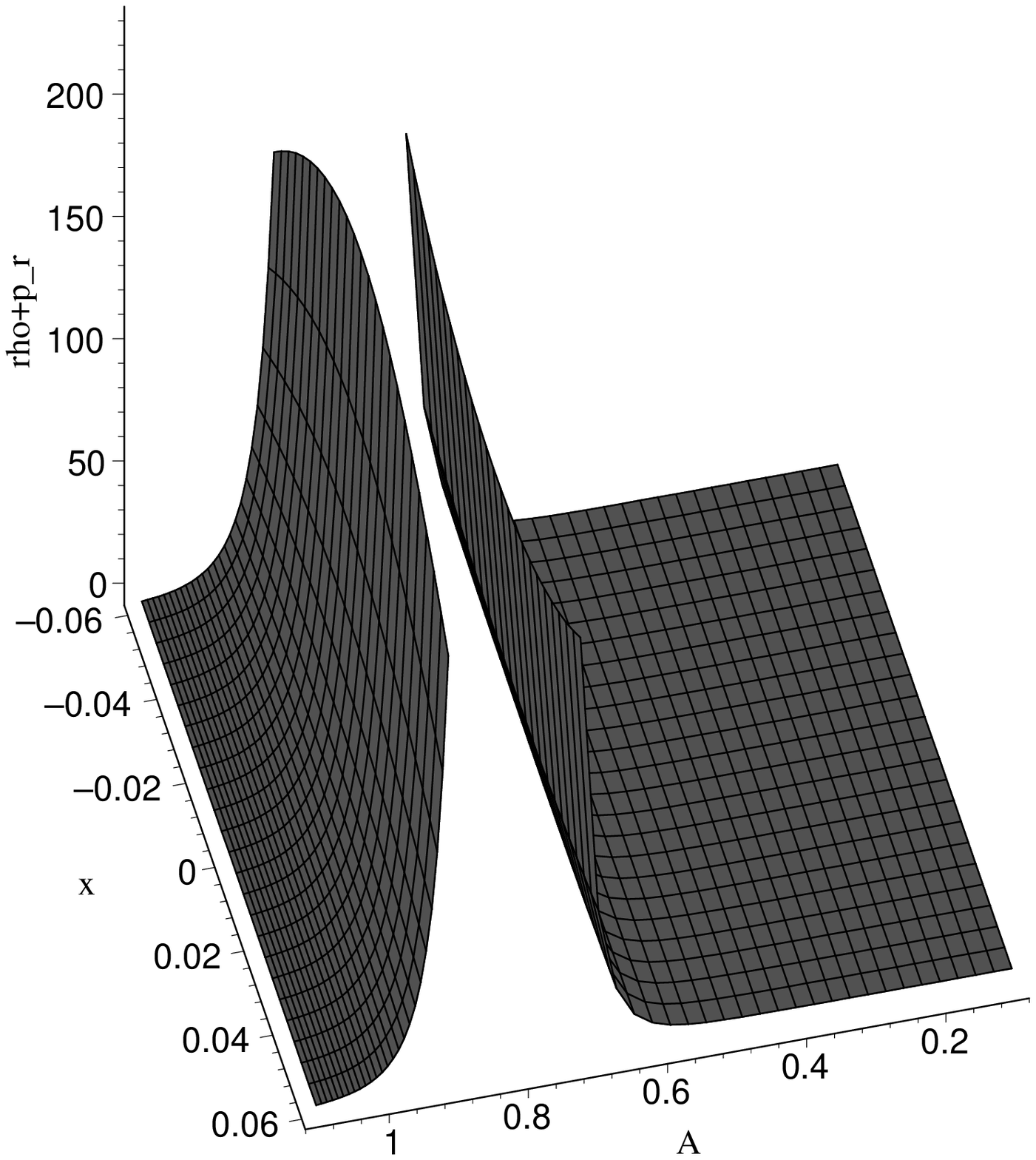}}&
\subfloat[$\tilde{\rho}+\tilde{p}_{t}$]{\includegraphics[width=42mm,height=42mm,clip]{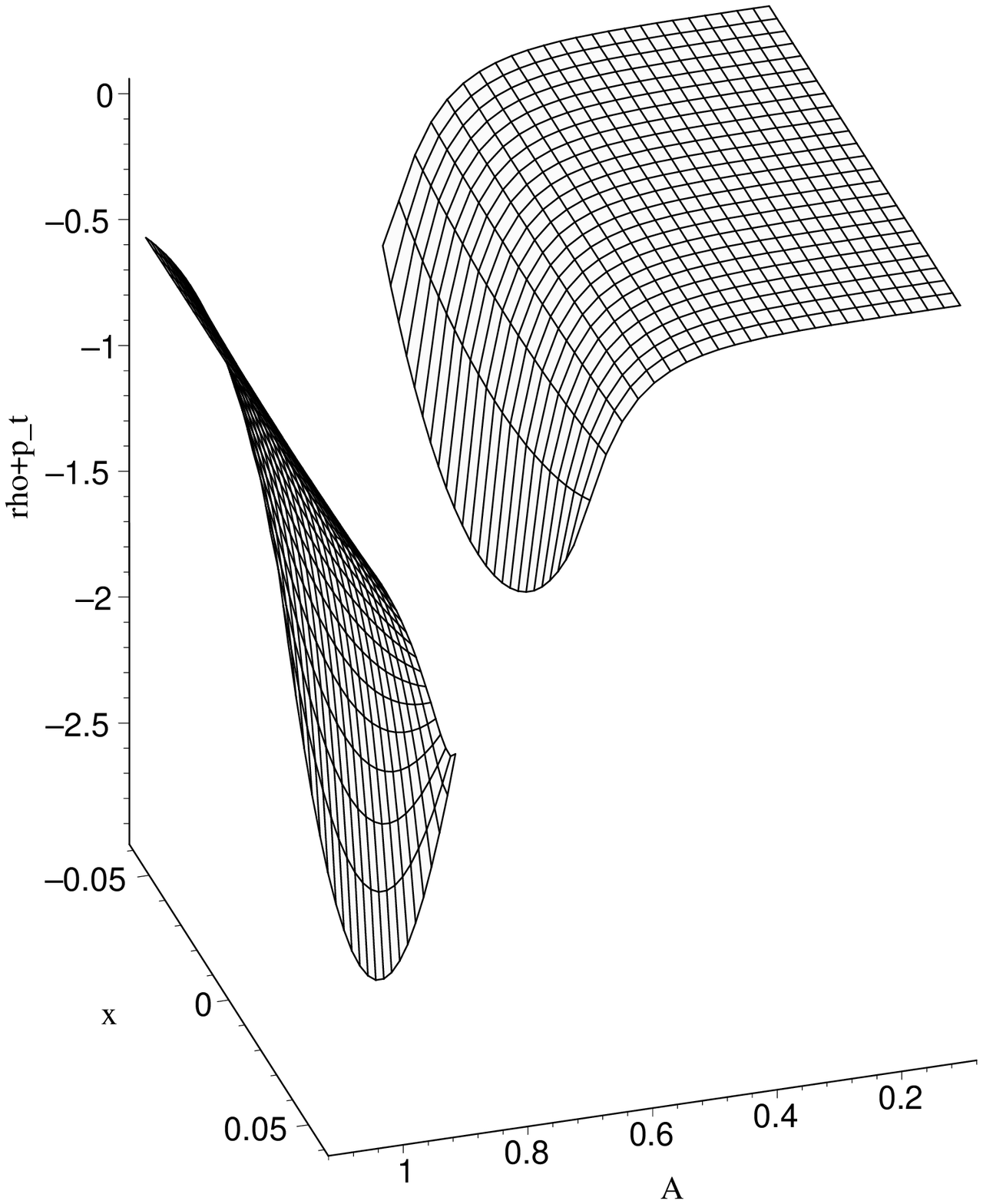}}
\end{tabular}
\end{center}
\vspace{-0.1cm}
\caption{\small{$R^{-2}$ contribution for $x_{0}=1$, $g_{tt}$ \emph{increasing}, varying throat radius. For negative $\alpha_{-2}$ the graphs should be flipped around the plane $z=0$. The region around the singularity has been omitted.}}
\label{fig:rminustwoAconcup}
\end{framed}
\end{figure}

Finally we present here the contribution from $\alpha_{-2}R^{-2}$. Much of this scenario mimics the $\alpha_{-1}R^{-1}$ contribution and hence we also summarize the results in a table (table~3) after all the graphs (figs.~\ref{fig:rminustwoalphaconcdown} - \ref{fig:rminustwoAconcup}).

\paragraph{{\small Zero tidal force:}} At the throat, the relevant quantities possess the following values:
\begin{subequations}\romansubs
{\allowdisplaybreaks\begin{align}
&\tilde{\rho}= \frac{\alpha_{-2}\qzero^{4}}{8\left(2\qzero\qzeropp-1\right)^{5}}\left[56\qzero^{3}(\qzeropp)^{3} -6\qzero\qzeropp -12\qzero^{3}\qzeropppp -96\qzero^{4}(\qzeroppp)^{2} -96\qzero^{4}(\qzeropp)^{4} \right. \nonumber \\
&\qquad\quad\quad \left. +12\qzero^{2}(\qzeropp)^{2} +24\qzero^{4}\qzeropp\qzeropppp-1\right] +\mathcal{O}(x)\,, \label{eq:minustwozerorho}\\[0.2cm]
&\rhot+\pr=\frac{\alpha_{-2}\qzero^{5}}{2\left(2\qzero\qzeropp-1\right)^{5}}\left[8\qzero^{2}(\qzeropp)^{3} - 4\qzeropp -3\qzero^{2}\qzeropppp -24\qzero^{3}(\qzeroppp)^{2} \right. \nonumber \\
&\qquad\quad\quad \left. -24\qzero^{3}(\qzeropp)^{4}+10\qzero(\qzeropp)^{2} +6\qzero^{3}\qzeropp\qzeropppp\right] + \mathcal{O}(x)\,,  \label{eq:minustwozeroecond1}\\[0.2cm]
&\rhot+\pt= \frac{\alpha_{-2}\qzero^{4}}{4\left(2\qzero\qzeropp-1\right)^{3}}\left[1-\qzero\qzeropp\right] + \mathcal{O}(x)\,.\label{eq:minustwozeroecond2}
\end{align}}
\end{subequations}
Again it can be noted that the equations of motion become singular where the Ricci scalar vanishes. As expected, the degree of the singularity has increased in comparison to the $n=-1$ case. The situation regarding energy conditions here, like for $n=-1$, does not seem promising in the zero tidal force regime.
 
\paragraph{{\small Non-zero tidal force:}} The following figs.~\ref{fig:rminustwoalphaconcdown} - \ref{fig:rminustwoAconcup} display the results of the non-zero tidal force scenarios.
\FloatBarrier

\clearpage
%%\begin{figure}[!ht]
\FloatBarrier
%%\clearpage
We summarize the results for $n=-2$ in table 3.

\begin{table}[!ht]
\newcommand\Ta{\rule{0pt}{2.6ex}}
\newcommand\Bb{\rule[-1.2ex]{0pt}{0pt}}
\begin{center}
\caption{\small{Summary of $\alpha_{-2}R^{-2}$ contributions.}\vspace{-0.2cm}} 
\begin{tabular}{|c|c|c|c|}
\hline
\Ta  \parbox[t]{2.5cm}{Parameter studied}& \parbox[t]{3.8cm}{zero tidal-force}  &  \parbox[t]{3.8cm}{$g_{tt}(x)$ concave-down} & \parbox[t]{3.8cm}{$g_{tt}(x)$ concave-up} \\[0.5cm]
\hline\hline

\parbox[t]{2.5cm}{$x_{0}$} & \parbox[t]{3.8cm}{{\small Respecting energy conditions seems difficult at best.}}\Ta & \parbox[t]{3.8cm}{{\small Energy condition respecting model possible near throat. For most of positive parameter space, small values are found, which may be dwarfed by negative value of Einstein term).}} & \parbox[t]{3.8cm}{{\small Energy condition respecting model not found near the throat.}}  \\[0.07cm]
\hline
\parbox[t]{2.5cm}{Throat radius} & \parbox[t]{3.8cm}{{\small Respecting energy conditions seems difficult at best.}}\Ta & \parbox[t]{3.8cm}{{\small Seems possible to make positive contribution to energy conditions for very small parameter space near the singular region for $\alpha_{-2}<0$. (Only on one side of the singular region.)}} & \parbox[t]{3.8cm}{{\small Does not seem possible to respect energy conditions.}}  \\[0.07cm]
\hline
\parbox[t]{2.5cm}{Minimum anisotropy for $R+\alpha_{-2}R^{-2}$} &\parbox[t]{3.8cm}{{\small Not studied}} \Ta & \parbox[t]{3.8cm}{{\small Magnitude of anisotropy larger than corresponding pure Einstein case regardless of $\alpha_{-2}$.}} & \parbox[t]{3.8cm}{{\small Magnitude of anisotropy larger than corresponding pure Einstein case regardless of $\alpha_{-2}$.}}  \\[0.07cm]
\hline
\end{tabular}
\end{center}
\end{table}
\FloatBarrier
\subsubsection{{\normalsize General $n$}}
Finally, for general $n$ we first quote the throat values in the zero tidal-force case. The quantities are as follows:
\begin{subequations}\romansubs
{\allowdisplaybreaks\begin{align}
\rhot=& 2^{n-1}\frac{\alpha_{n}}{\qzero^{2n}}\left(1-2\qzero\qzeropp\right)^{n-3} \left[12n^{2}\qzero^{4} (\qzeroppp)^{2} +16n^{2}\qzero^{4}(\qzeropp)^{4} +2n^{2}\qzero^{3}\qzeropppp +2n^{2}\qzero\qzeropp \right. \nonumber \\[0.1cm]
&-4n^{2}\qzero^{2}(\qzeropp)^2 -8\qzero^{3}(\qzeropp)^{3} -6\qzero\qzeropp +1 -8n^{2}\qzero^{3}(\qzeropp)^{3} +4n\qzero^{4}\qzeropp\qzeropppp\nonumber \\[0.1cm]
&-4n^{3}\qzero^{4}(\qzeroppp)^{2} + 8n\qzero^{3}(\qzeropp)^{3} -2n\qzero\qzeropp -16n\qzero^{4} (\qzeropp)^{4} +12\qzero^{2}(\qzeropp)^{2}  \nonumber \\[0.1cm]
&+4n\qzero^{2}(\qzeropp)^{2}-8n\qzero^{4}(\qzeroppp)^{2} -2n\qzero^{3}\qzeropppp -4n^{2}\qzero^{4}\qzeropp\qzeropppp  \left.\right] +\mathcal{O}(x)\,,\label{eq:gennrho}\\[0.2cm]
\rhot+\pr=& 2^{n} \frac{n\alpha_{n}}{\qzero^{2n-1}} \left(1-2\qzero\qzeropp\right)^{n-3} \left[2\qzero^{3}\qzeropppp\qzeropp +6n\qzero^{3}(\qzeroppp)^{2} +8n\qzero^{3}(\qzeropp)^{4}\right. \nonumber \\[0.1cm]
& +n\qzero^{2} \qzeropppp -2\qzeropp +6\qzero(\qzeropp)^{2} - 2n^{2}\qzero^{3}(\qzeroppp)^{2} -4n\qzero^{2}(\qzeropp)^{3}+n\qzeropp -8\qzero^{3}(\qzeropp)^{4} \nonumber \\[0.1cm]
&-2n\qzero(\qzeropp)^{2} -4\qzero^{3}(\qzeroppp)^{2} -\qzero^{2}\qzeropppp -2n\qzero^{3}\qzeropp\qzeropppp\left.\right] +\mathcal{O}(x)\,,\label{eq:gennecond1}\\[0.2cm]
\rhot+\pt=&2^{n-1}\frac{n\alpha_{n}}{\qzero^{2n}} \left[1-\qzero\qzeropp\right]\left[1-2\qzero\qzeropp\right]^{n-1}+\mathcal{O}(x) \,.\label{eq:gennecond2}
\end{align}}
\end{subequations}
These conditions represent the contribution to the energy conditions for a particular value of (arbitrary) $n$. The general energy conditions would then constitute the sum of these conditions summed over the $n$ values which contribute to the gravitational action.

\begin{figure}[H]
\begin{framed}
\begin{center}
\vspace{-0.2cm}
\begin{tabular}{ccc}
\subfloat[$\tilde{\rho}(x)$]{\includegraphics[width=42mm,height=42mm,clip]{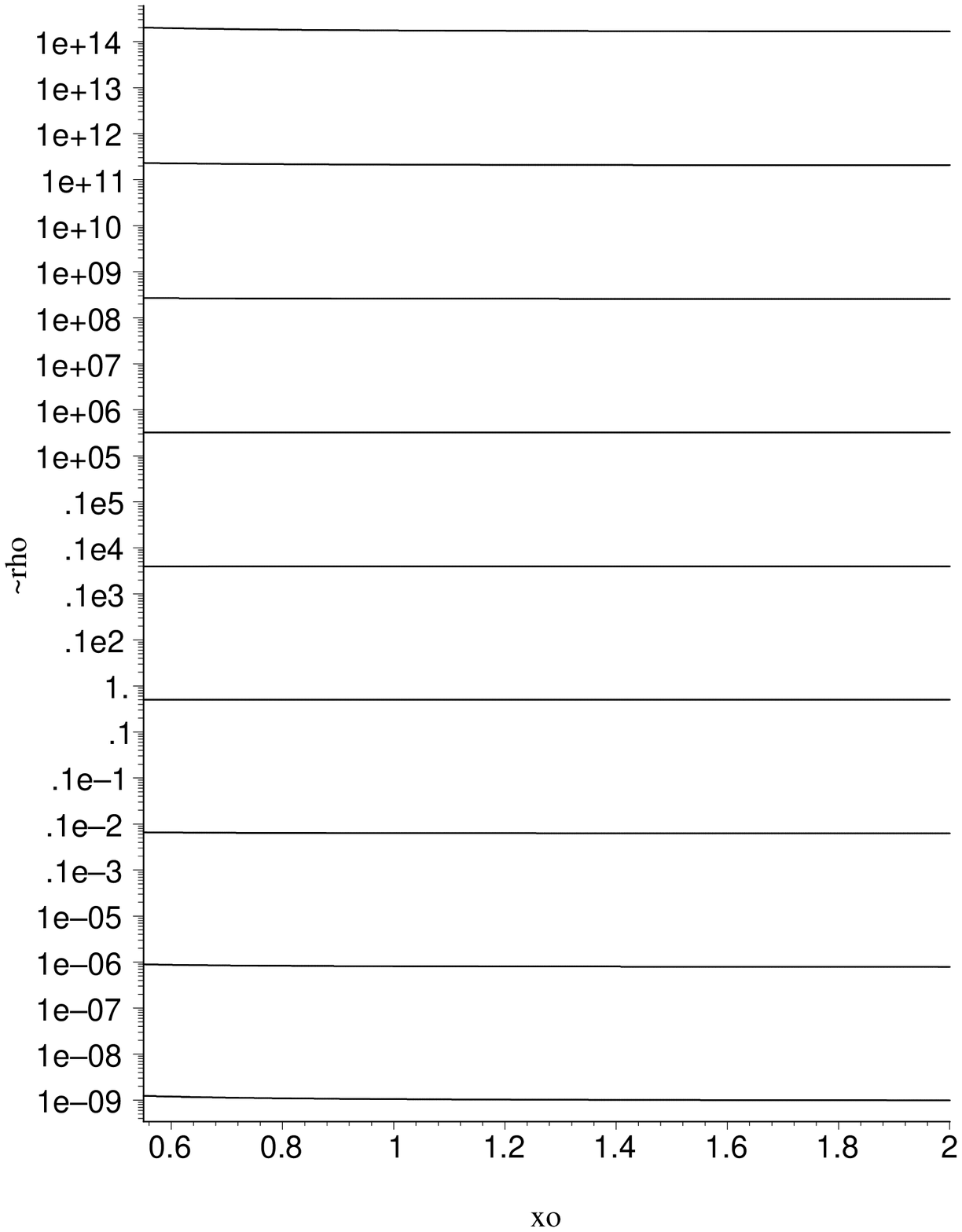}} &
\subfloat[$\tilde{\rho}+\tilde{p}_{r}$]{\includegraphics[width=42mm,height=42mm,clip]{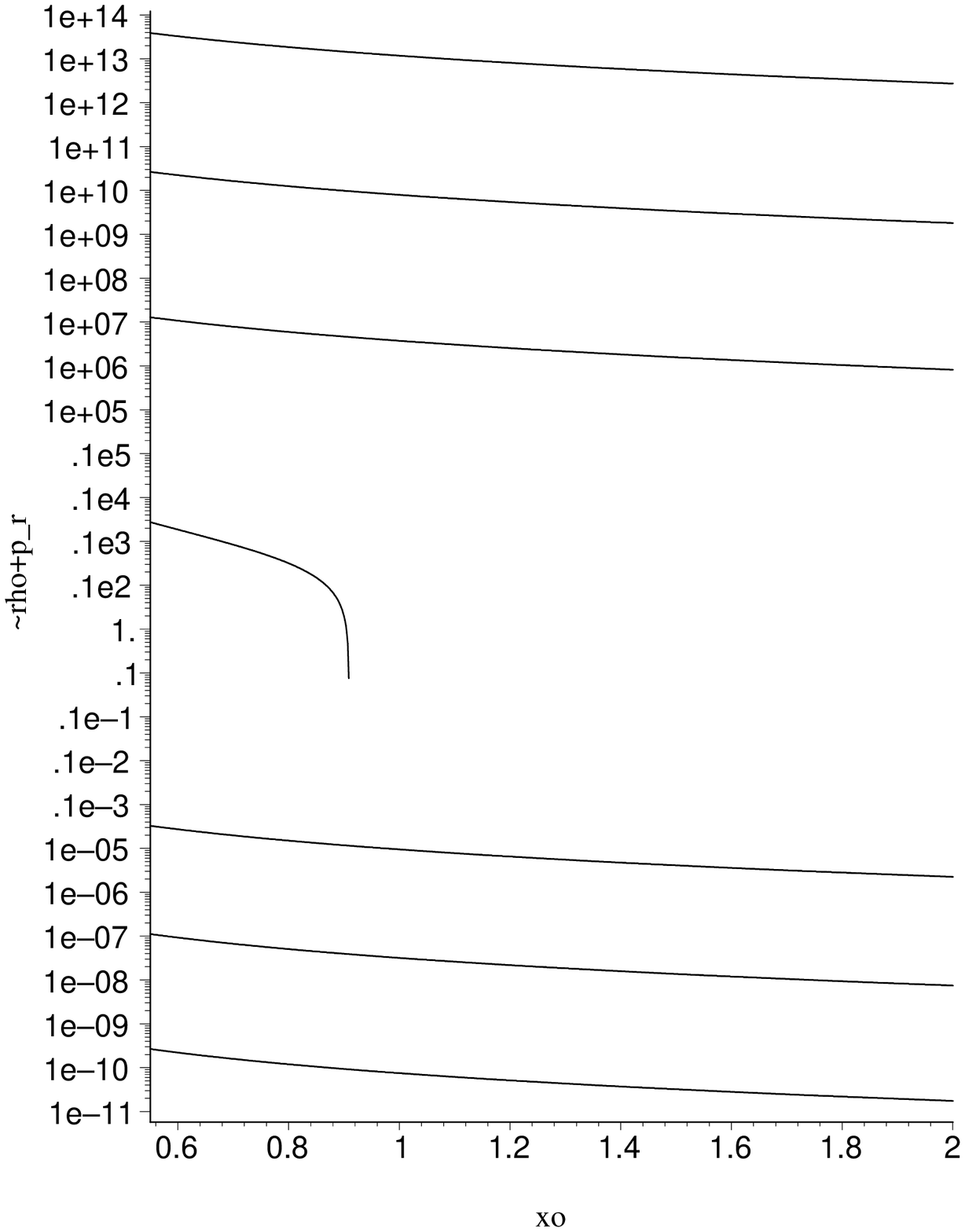}}&
\subfloat[$\tilde{\rho}+\tilde{p}_{t}$]{\includegraphics[width=42mm,height=42mm,clip]{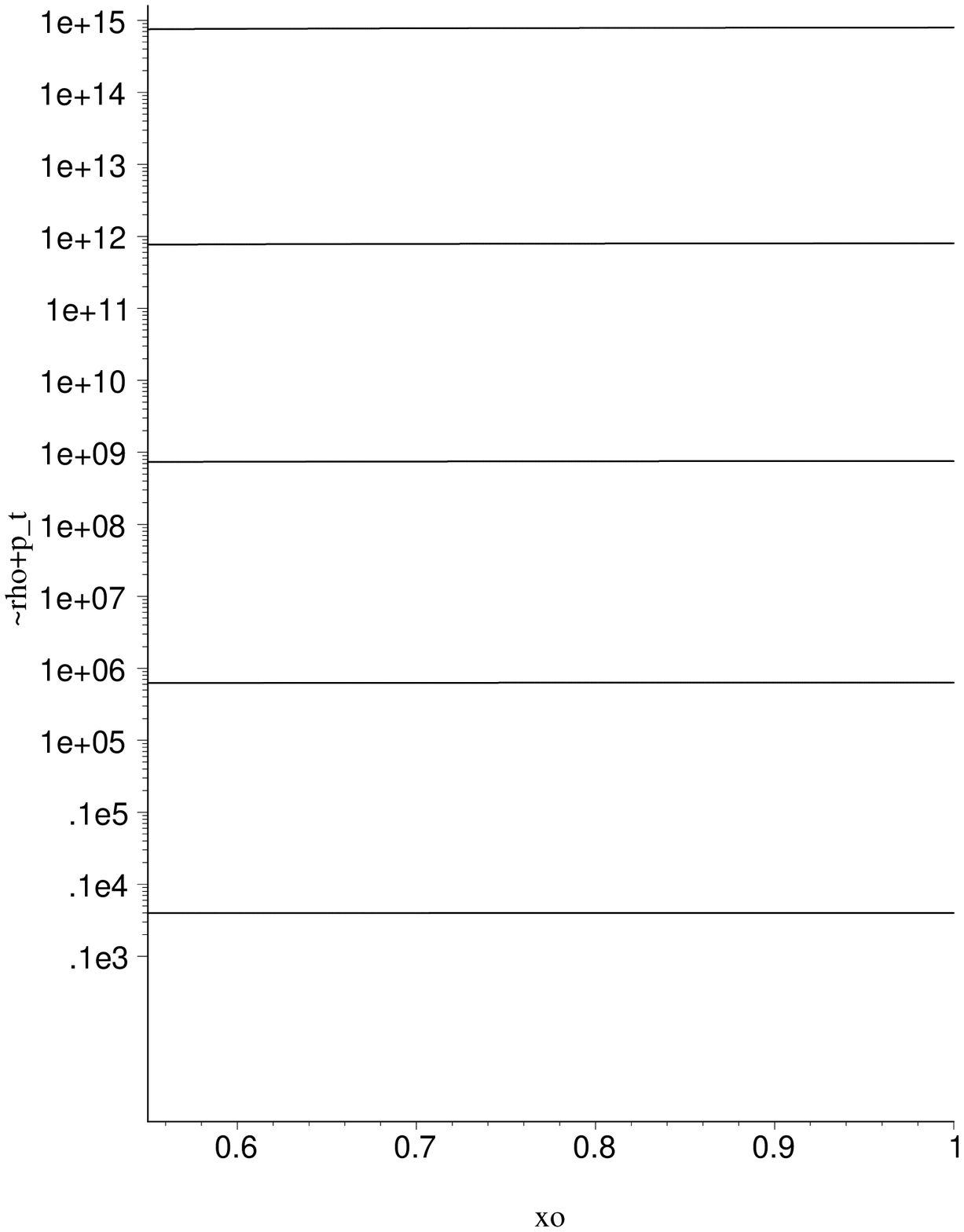}}
\end{tabular}
\end{center}
\vspace{-0.1cm}
\caption{\small{$R^{n}$ contribution at $x=0$ for $A=0.05$ and $\alpha_{n}=1$, $g_{tt}$ \emph{decreasing}, and varying $x_{0}$. In the far-left graph the plots correspond to $n=-3\rightarrow 5$ (bottom to top). The middle graph corresponds $n=-3,\,-2,\,-1,\,2,\,3,\,4,\,5$ (bottom to top). The right graph corresponds to parameters $n=1,\,2,\,3,\,4,\,5$ (bottom to top). Values of $n$ not displayed are negative (note though that these would be positive if the sign of $\alpha_{n}$ was negative).}}
\label{fig:rn_gtt_dec_xo}
\end{framed}
\end{figure}

\begin{figure}[!ht]
\begin{framed}
\begin{center}
\vspace{-0.5cm}
\begin{tabular}{ccc}
%%\subfloat[$g_{tt}(x)$]{\includegraphics[width=45mm,height=45mm,clip]{squared_gtt_c_down.eps}}&
\subfloat[$\tilde{\rho}(x)$]{\includegraphics[width=42mm,height=42mm,clip]{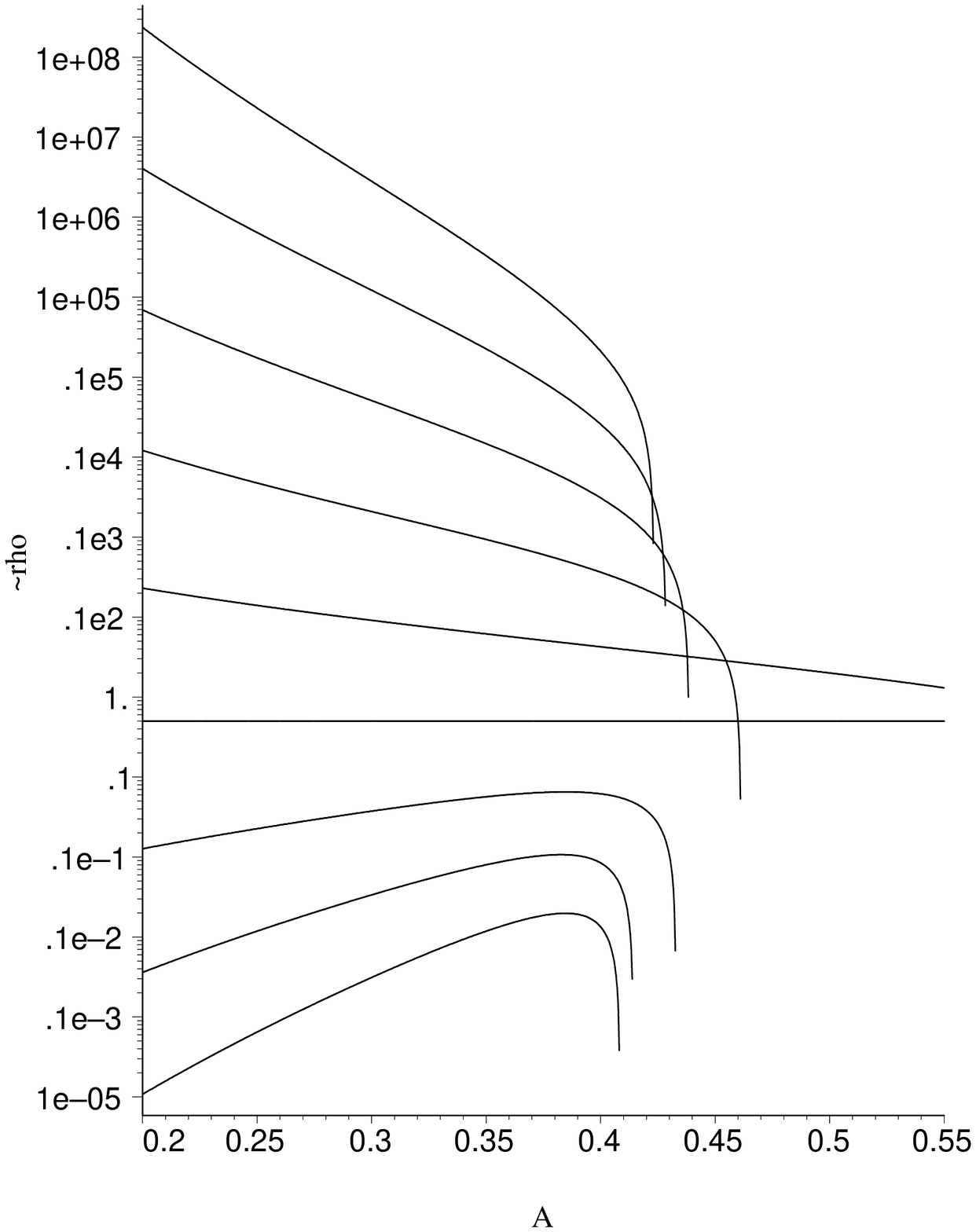}} &
\subfloat[$\tilde{\rho}+\tilde{p}_{r}$]{\includegraphics[width=42mm,height=42mm,clip]{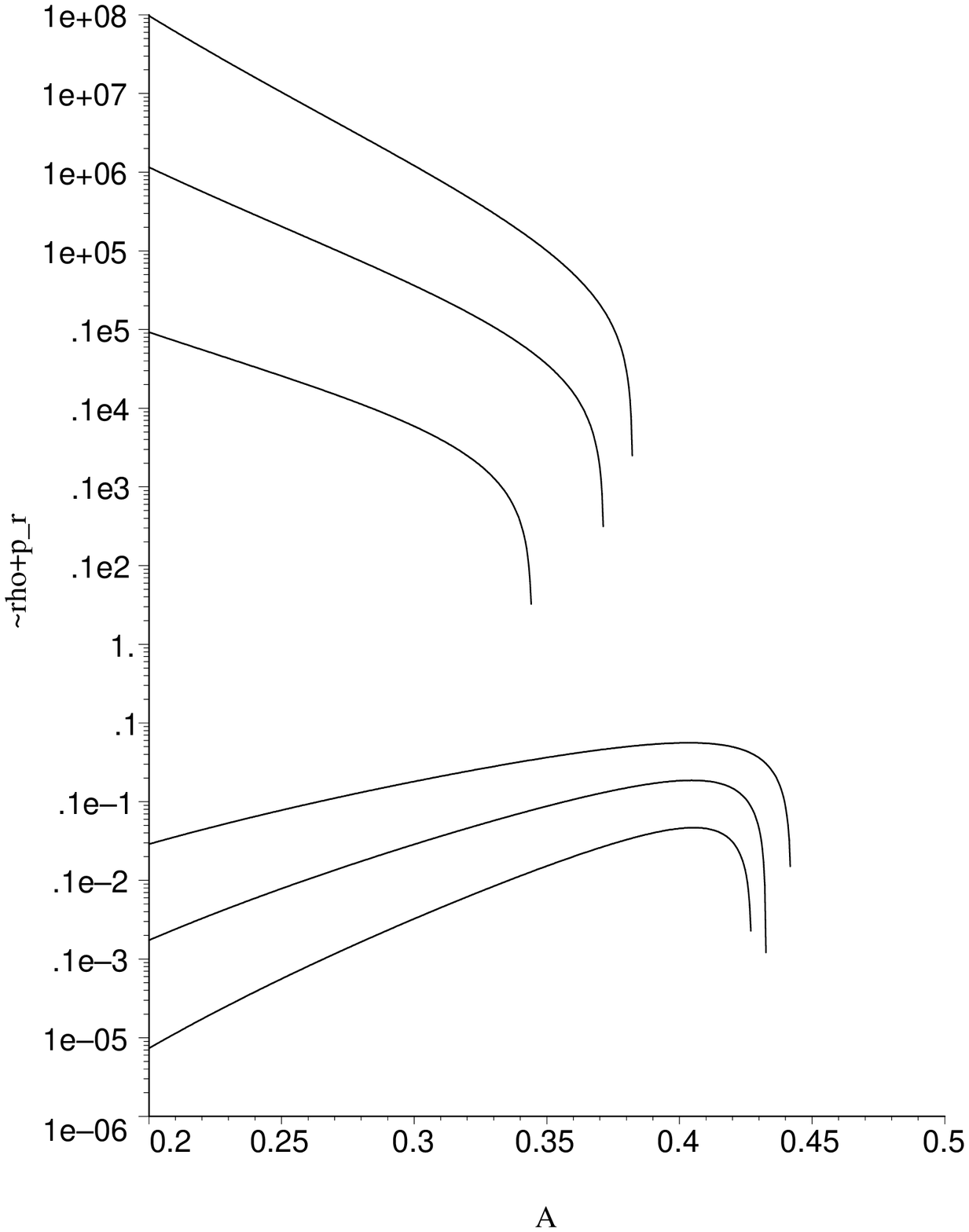}}&
\subfloat[$\tilde{\rho}+\tilde{p}_{t}$]{\includegraphics[width=42mm,height=42mm,clip]{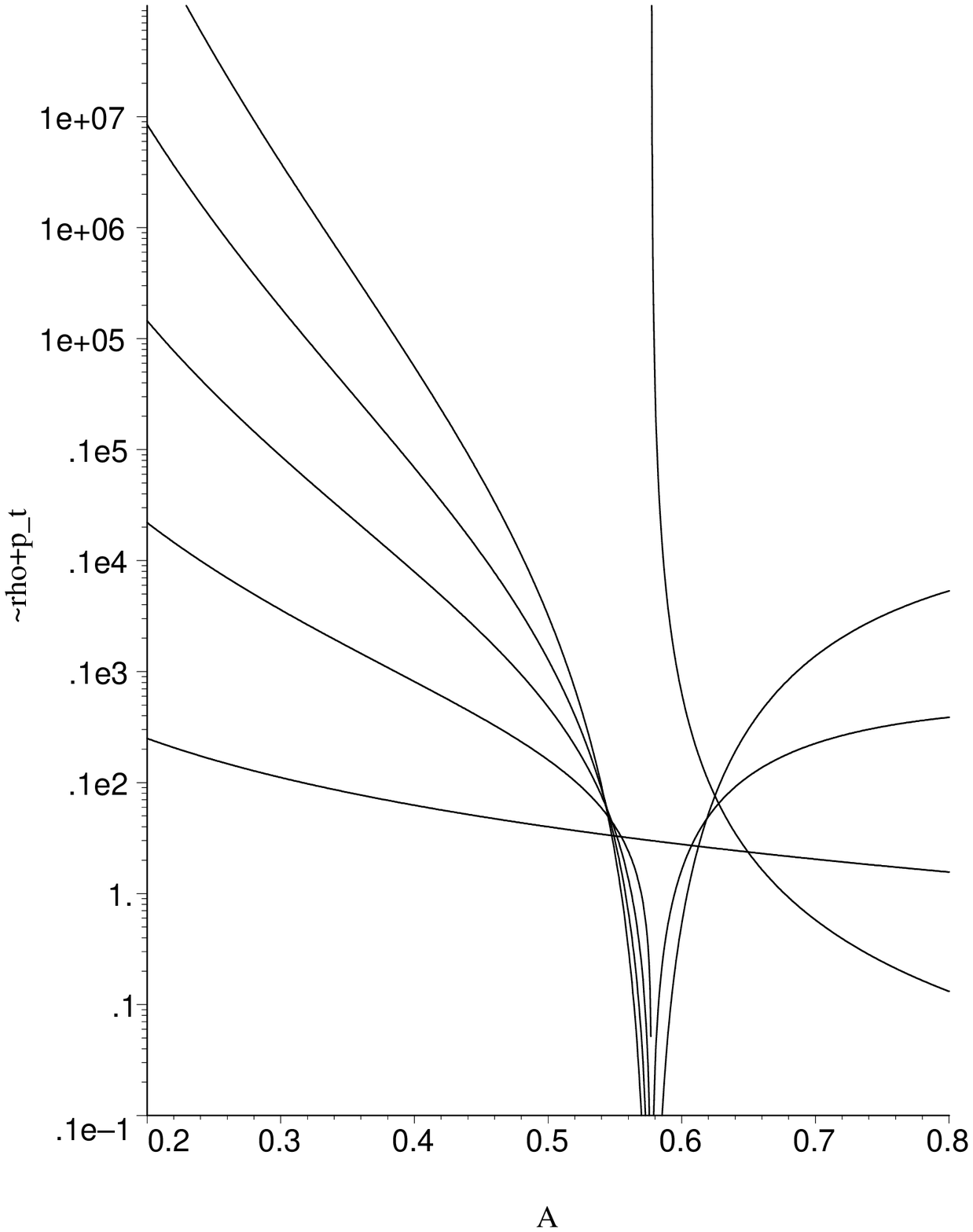}}
\end{tabular}
\end{center}
\vspace{-0.1cm}
\caption{\small{$R^{n}$ contribution at $x=0$ for $x_{0}=1$ and $\alpha_{n}=1$, $g_{tt}$ \emph{decreasing}, and varying throat radius. In the far-left graph, the graphs correspond to $n=-3\rightarrow +5$ from bottom to top. In the middle graph, the plots correspond to $n=-3,\,-2,\,-1,\,3,\,4,\,5$ from bottom to top (the others being either negative or too close to zero to show on the scale). In the far-right plot, the curves approaching the vertical-axis correspond to $n=1\rightarrow 5$ from bottom to top. The curve that does not approach the vertical-axis corresponds to $n=-2$.}}
\label{fig:rn_gtt_dec_A}
\end{framed}
\end{figure}

\begin{figure}[!ht]
\begin{framed}
\begin{center}
\vspace{-0.0cm}
\begin{tabular}{ccc}
\subfloat[$\tilde{\rho}(x)$]{\includegraphics[width=42mm,height=42mm,clip]{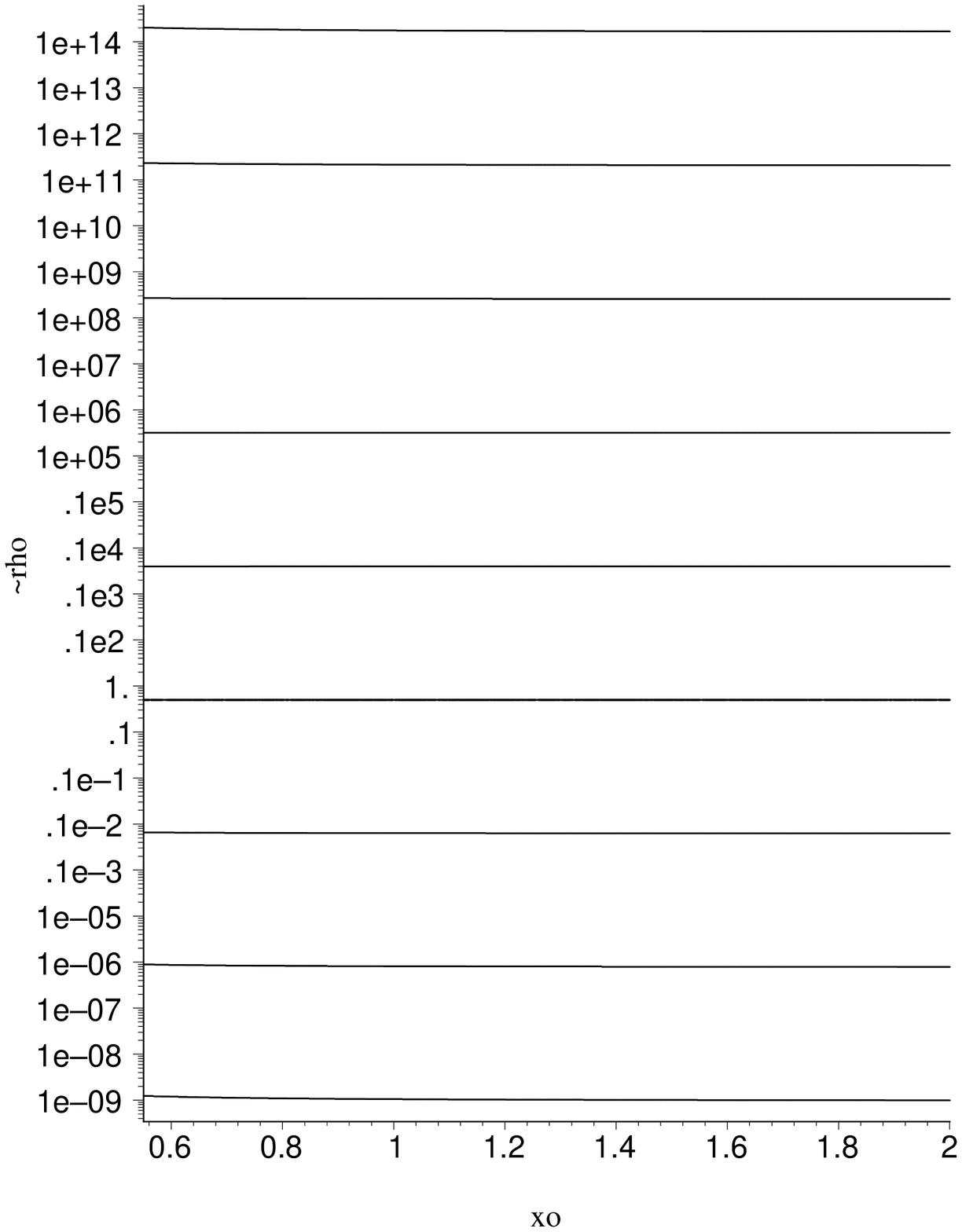}} &
\subfloat[$\tilde{\rho}+\tilde{p}_{r}$]{\includegraphics[width=42mm,height=42mm,clip]{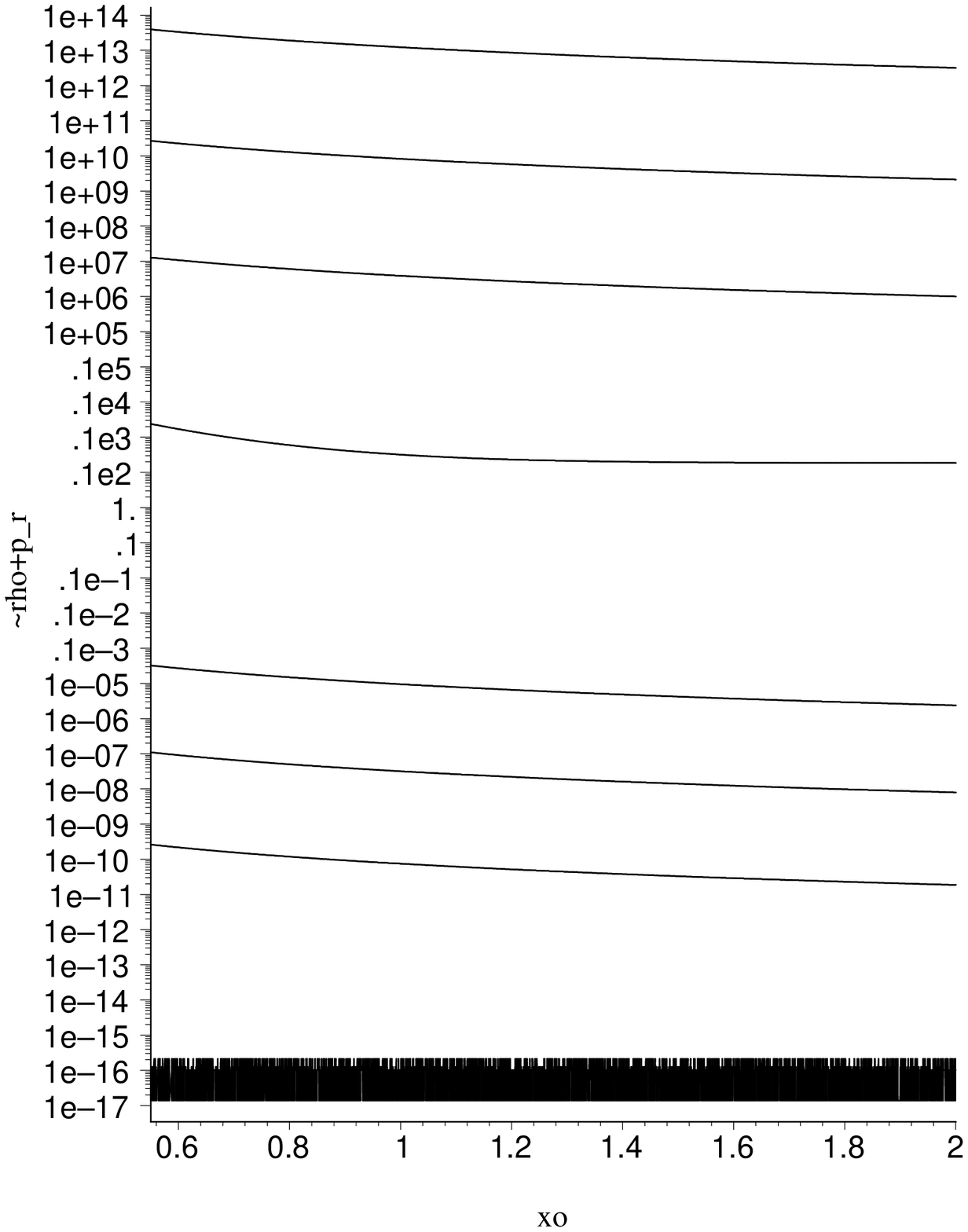}}&
\subfloat[$\tilde{\rho}+\tilde{p}_{t}$]{\includegraphics[width=42mm,height=42mm,clip]{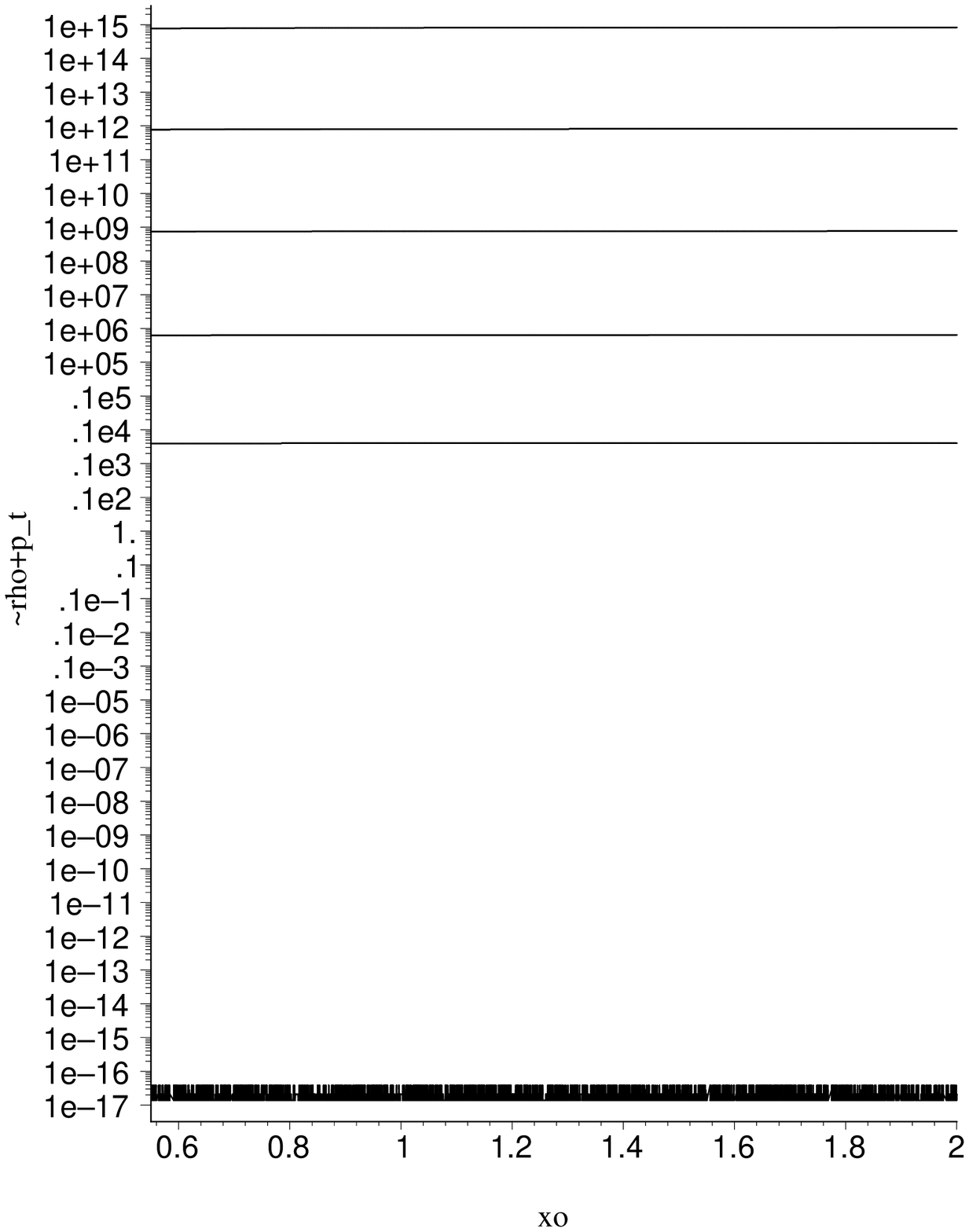}}
\end{tabular}
\end{center}
\vspace{-0.1cm}
\caption{\small{$R^{n}$ contribution at $x=0$ for $A=0.05$ and $\alpha_{n}=1$, $g_{tt}$ \emph{increasing}, and varying $x_{0}$. In the far-left graph the plots correspond to $n=-3\rightarrow 5$ (bottom to top). The middle graph corresponds $n=0,\,-3,\,-2,\,-1,\,2,\,3,\,4,\,5$ from bottom to top. (Note that $n=0$ should give a contribution of zero, as it corresponds to the cosmological constant only scenario. The line at $10^{-16}\approx 0$ is a numerical artefact.) The right graph corresponds to parameters $n=0,\,1,\,2,\,3,\,4,\,5$ (bottom to top, again with the $n=0$ plot being a numerical artefact.) Values of $n$ not displayed are negative (note though that these would be positive if the sign of $\alpha_{n}$ was negative).}}
\label{fig:rn_gtt_inc_xo}
\end{framed}
\end{figure}

\begin{figure}[H]
\begin{framed}
\begin{center}
\vspace{-0.5cm}
\begin{tabular}{ccc}
%%\subfloat[$g_{tt}(x)$]{\includegraphics[width=45mm,height=45mm,clip]{squared_gtt_c_down.eps}}&
\subfloat[$\tilde{\rho}(x)$]{\includegraphics[width=42mm,height=42mm,clip]{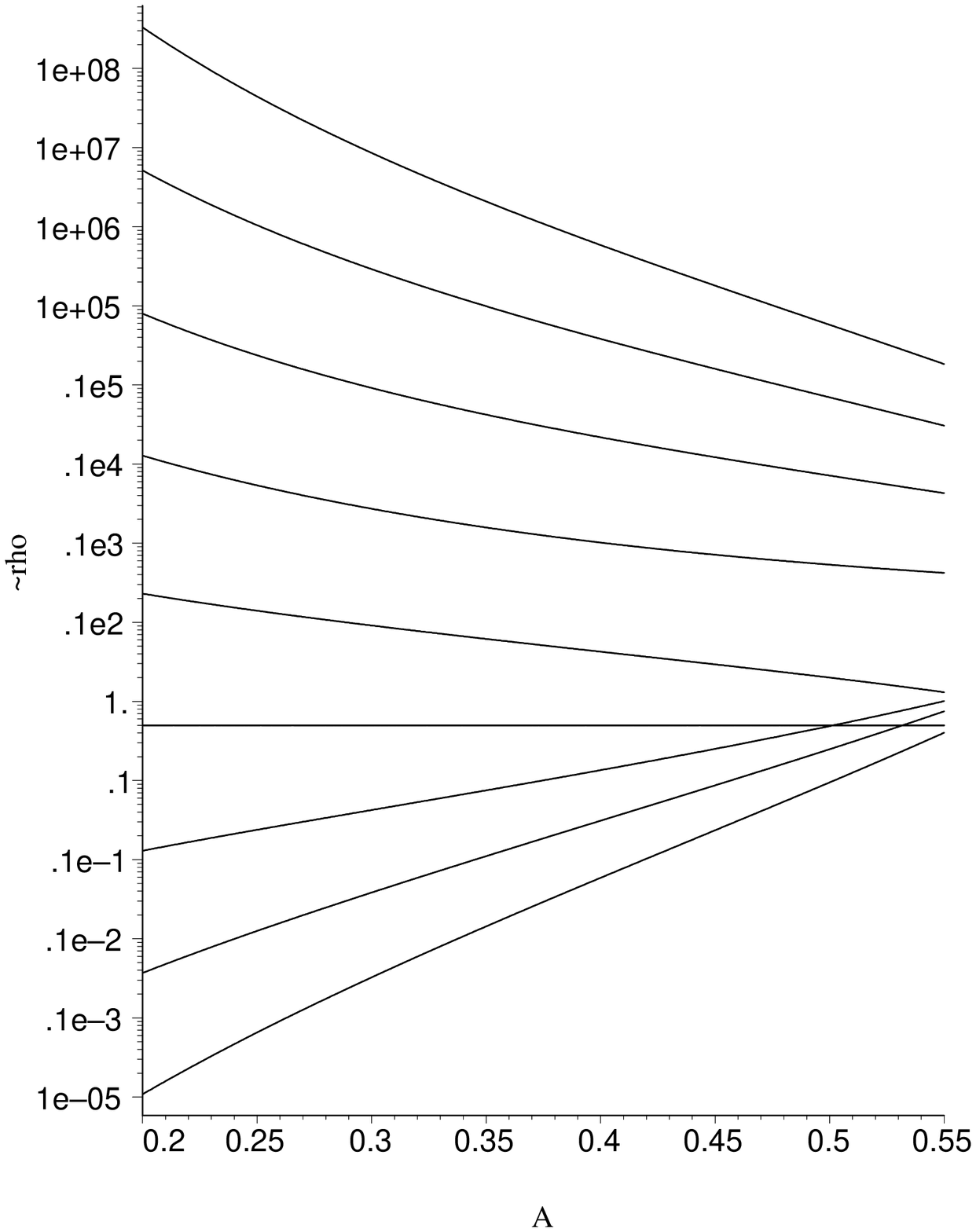}} &
\subfloat[$\tilde{\rho}+\tilde{p}_{r}$]{\includegraphics[width=42mm,height=42mm,clip]{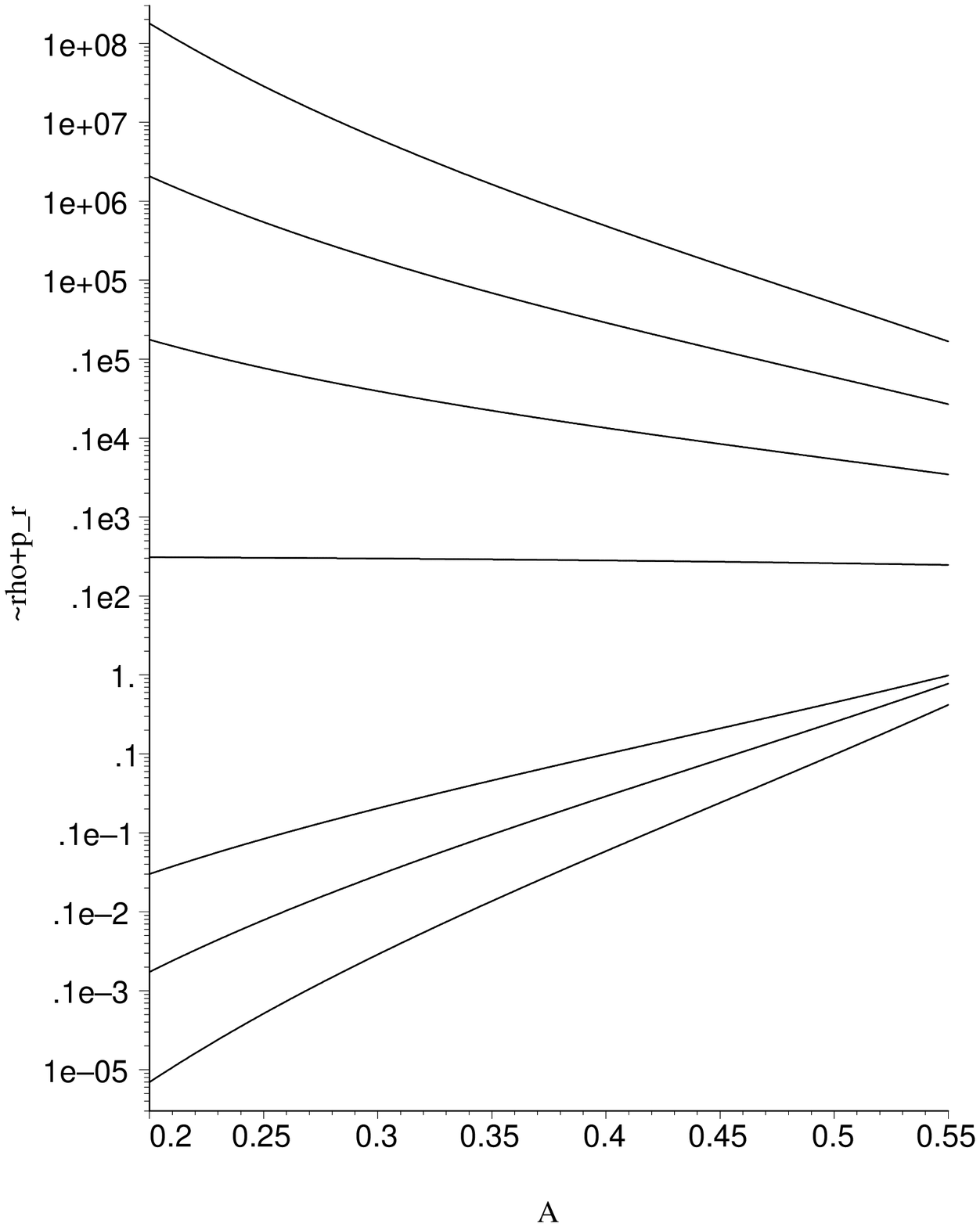}}&
\subfloat[$\tilde{\rho}+\tilde{p}_{t}$]{\includegraphics[width=42mm,height=42mm,clip]{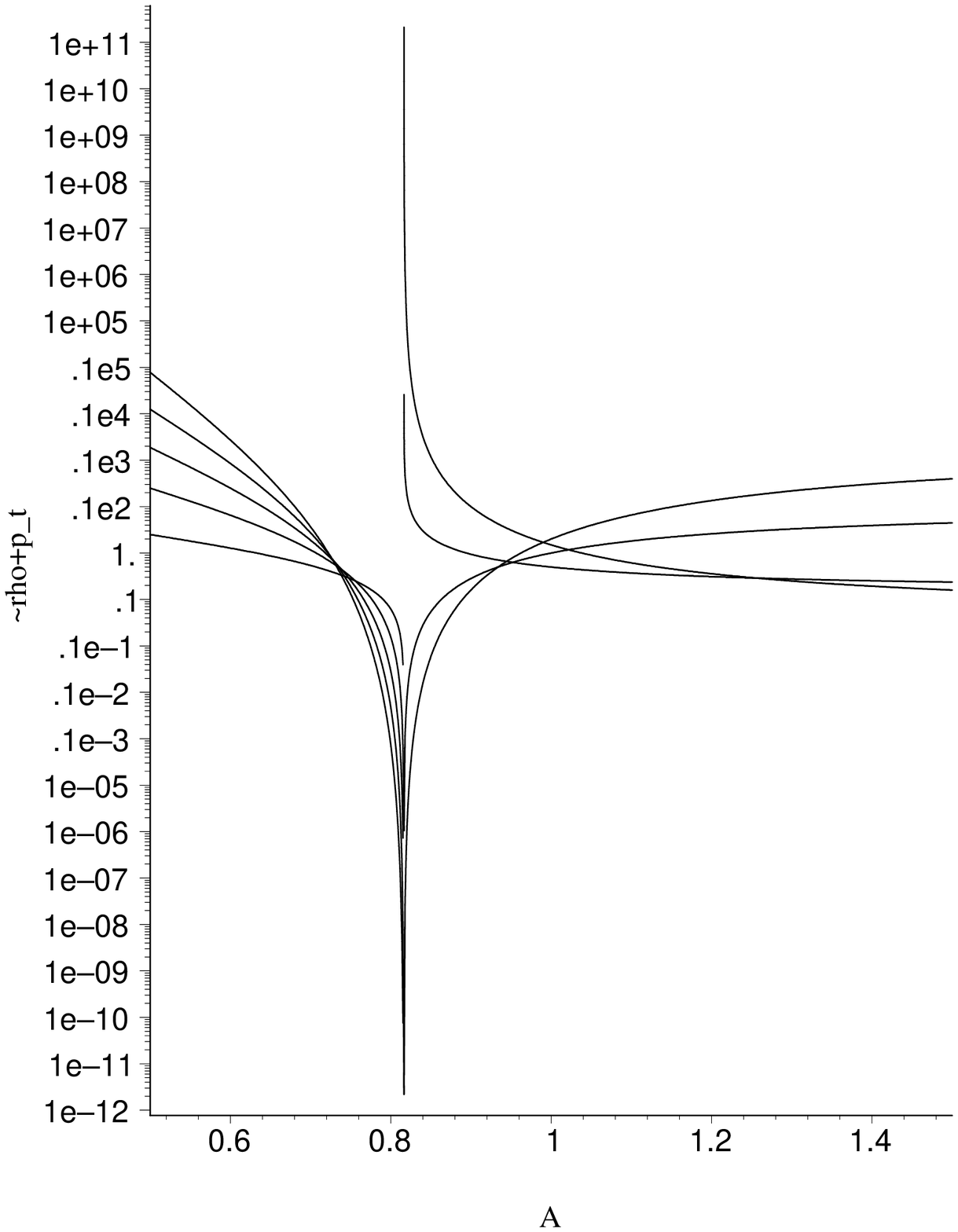}}
\end{tabular}
\end{center}
\vspace{-0.1cm}
\caption{\small{$R^{n}$ contribution at $x=0$ for $x_{0}=1$ and $\alpha_{n}=1$, $g_{tt}$ \emph{increasing}, and varying throat radius. In the far-left graph, the graphs correspond to $n=-3\rightarrow +5$ from bottom to top. In the middle graph, the plots correspond to $n=-3,\,-2,\,-1,\,2,\,3,\,4,\,5$ from bottom to top (the others being either negative or too close to zero to show on the scale). In the far-right plot, the curves intersecting the vertical-axis correspond to $n=1\rightarrow 5$ from bottom to top. The curves that do not approach the vertical-axis corresponds to $n=-1$ (bottom) and $n=-3$ (top).}}
\label{fig:rn_gtt_inc_A}
\end{framed}
\end{figure}

For the non-zero tidal force scenarios we present the following results in order to show the trends as $n$ increases. The graphs (\ref{fig:rn_gtt_dec_xo})-(\ref{fig:rn_gtt_inc_A}) are plots of the energy conditions \emph{exactly at the throat} $x=0$ for $\gtt$ concave-down and $\gtt$ concave-up respectively. Note that from the analyticity of $Q(x)$ in a neighborhood of the throat, if a quantity is positive at the throat then there exists a non-zero neighborhood of the throat where this quantity is positive, and hence the particular energy condition can be respected. In all the following graphs, only results where the energy conditions are positive are shown. (Refer to figure captions for details.)

\vspace{-0.0cm}
\vspace{-1.25cm}

%%\clearpage
%%\begin{figure}[!ht]
\FloatBarrier

\section{{\small CONCLUDING REMARKS}}
We have studied the existence of wormhole throats in modified gravity theories with gravitational actions of the form $S=\int_{M^{4}} {\underset{{\hspace{-0.05cm}\vspace{-0.1cm}n}}{\sum}} \alpha_{n}\,R^{n}\sqrt{g}\,d^{4}x$. This action includes the $n=1$ contribution of Einstein gravity. Of specific interest to our studies are energy condition violation and the degree of anisotropy required to support the throat, since in Einstein gravity alone the existence of throats implies both violation of energy conditions as well as the presence of anisotropy. We studied various terms contributing to the action separately, in order to discern which terms aid and which terms hinder the energy conditions. It would also be worthwhile to consider all terms together, but this would be more computationally involved. We have studied both the zero and non-zero tidal force solutions. {In general, we find that the parameter space for energy condition respecting solutions is much larger for positive $n$ than negative $n$, and that larger $n$ increases the energy conditions. This is due to the fact that $f(R)$ gravity has an equivalent description as a scalar-tensor theory, where the ``scalar field'' component may violate energy conditions. The positive $n$ sector also allows one to minimize the anisotropy required to support a wormhole throat, whereas the negative $n$ sector tends to make the required anisotropy much larger when compared to Einstein gravity.} It may be then that if gravitation is governed by such a Lagrangian, or one which is expandable as such, throats may exist which obey the energy conditions violated by Einstein gravity alone and require little or perhaps no anisotropy. In light of this, it may be possible that if the true gravitational action contains higher positive powers of $R$, then throats in space-time may be more likely than with no such augmentation or with negative powers. Since it is unknown (though perhaps unlikely) whether or not the topology of space-time can change, we cannot say whether throats would be more likely in the early universe (where higher powers may contribute to inflationary expansion) than in the late universe (where inverse powers may contribute to late-time acceleration) (see, for example \cite{ref:no3} for an analysis of modified gravity accommodating early and late-time accelerated expansion). However, it is interesting to speculate on this.

\section*{{\small ACKNOWLEDGMENTS}}
A.D. is grateful for the kind hospitality of the University of Zagreb, where some of this work was carried out. D.H. is partially supported by the Croatian Ministry of Science under the project No. 036-0982930-3144. This work was also partially supported by CompStar, a Research Networking Programme of the European Science Foundation.

\linespread{0.6}
\bibliographystyle{unsrt}

\end{document}